\documentclass[11pt,a4paper]{article} 

\usepackage{adjustbox}
\usepackage{amsfonts}
\usepackage{amssymb}
\usepackage{amsmath}
\usepackage{amsthm}
\usepackage{authblk}
\usepackage{array}
\usepackage{bbm}
\usepackage{booktabs}
\usepackage{caption}
\usepackage{subcaption}
\usepackage{dsfont}
\usepackage{enumerate}
\usepackage{framed,multirow}
\usepackage{graphics}
\usepackage{graphicx}
\usepackage{hyperref}
\usepackage{inputenc}
\usepackage{latexsym}
\usepackage{mathrsfs}  
\usepackage[table]{xcolor}
\usepackage{tikz}
\usepackage[normalem]{ulem}

\theoremstyle{definition}

\title{SHREC 2021: Retrieval and classification of protein surfaces equipped with physical and chemical properties}

\author[1,\thanks{Track organizer},\thanks{Corresponding author}]{Andrea Raffo}
\author[1,$^*$]{Ulderico Fugacci}
\author[1,$^*$]{Silvia Biasotti}
\author[2,$^*$]{Walter Rocchia}
\author[3]{Yonghuai Liu}
\author[4]{Ekpo Otu}
\author[4]{Reyer Zwiggelaar}
\author[4]{David Hunter}
\author[5]{Evangelia I. Zacharaki}
\author[5]{Eleftheria Psatha}
\author[5]{Dimitrios Laskos}
\author[5]{Gerasimos Arvanitis}
\author[5]{Konstantinos Moustakas}
\author[6]{Tunde Aderinwale}
\author[6]{Charles Christoffer}
\author[7]{Woong-Hee Shin}
\author[8,6]{Daisuke Kihara}
\author[9]{Andrea Giachetti}
\author[10,12]{Huu-Nghia Nguyen}
\author[10,12]{Tuan-Duy Nguyen}
\author[10,12]{Vinh-Thuyen Nguyen-Truong}
\author[10,12]{Danh Le-Thanh}
\author[10,12]{Hai-Dang Nguyen}
\author[10,11,12]{Minh-Triet Tran}

\affil[1]{Istituto di Matematica Applicata e Tecnologie Informatiche  ``E. Magenes" CNR, Genova, Italy}
\affil[2]{CONCEPT Lab, Istituto Italiano di Tecnologia, Genova, Italy}
\affil[3]{Department of Computer Science, Edge Hill University, Ormskirk, UK}
\affil[4]{Department of Computer Science, Aberystwyth University, Aberystwyth, UK}
\affil[5]{Department of Electrical and Computer Engineering, University of Patras, Greece}
\affil[6]{Department of Computer Science, Purdue University, West Lafayette, USA}
\affil[7]{Department of Chemical Science Education, Sunchon National University, Suncheon, Republic of Korea}
\affil[8]{Department of Biological Sciences, Purdue University, West Lafayette, USA}
\affil[9]{Department of Computer Science, University of Verona, Verona, Italy}
\affil[10]{University of Science, VNU-HCM, Vietnam}
\affil[11]{John von Neumann Institute, VNU-HCM, Vietnam}
\affil[12]{Vietnam National University, Ho Chi Minh City, Vietnam}

\date{}                     %% if you don't need date to appear
\begin{document}
\maketitle

\begin{abstract}
This paper presents the methods that have participated in the SHREC 2021 contest on retrieval and classification of protein surfaces on the basis of their geometry and physicochemical properties. The goal of the contest is to assess the capability of different computational approaches to identify different conformations of the same protein, or the presence of common sub-parts, starting from a set of molecular surfaces. We addressed two problems: defining the similarity solely based on the surface geometry or with the inclusion of physicochemical information, such as electrostatic potential, amino acid hydrophobicity, and the presence of hydrogen bond donors and acceptors. Retrieval and classification performances, with respect to the single protein or the existence of common sub-sequences, are analysed according to a number of information retrieval indicators.  
\end{abstract}

\section{Introduction}
Automatically identifying the different conformations of a given set of proteins, as well as their interaction with other molecules, is crucial in structural bioinformatics.
The well-established shape-function paradigm for proteins \cite{2002molecular} states that a protein of a given sequence has one main privileged conformation, which is crucial for its function. However, every protein during its time evolution explores a much larger part of the conformational space. The most stable conformations visited by the protein can be experimentally captured by the NMR technique; this is because the hydrogen atoms are already included in the atomic model, thus giving less ambiguities in the charge assignment. 

Recognising a protein from an ensemble of geometries corresponding to the different conformations it can assume means capturing the features that are unique to it and is a fundamental step from the structural bioinformatics viewpoint. It is preliminary to the definition of a geometry-based notion of similarity, and, subsequently, complementarity, between proteins. 
From the application standpoint, the identification of characteristic features can point to protein functional regions and to new target sites for blocking the activity of pathological proteins in the drug discovery field.
These features can become more specific if one adds to the geometry of the molecular surface also the information related to the main physicochemical descriptors, such as local electrostatic potential \cite{Delphi2001}, residue hydrophobicity \cite{kyte1982simple}, and the location of hydrogen bond donors and acceptors \cite{kortemme2003orientation}.

The aim of this track is to evaluate the performance of retrieval and
classification of computational methods for protein surfaces characterized by physicochemical properties.
Starting from a set of protein structures in different conformational states generated via NMR experiments and deposited in the PDB repository \cite{PDB2000}, we build their Solvent Excluded Surface (SES) by the freely available software NanoShaper \cite{DeCherchi2013,Decherchi2019}. Differently from previous SHREC tracks \cite{shrec2017,shrec18,shrec2019,shrec2020} we enrich the protein SES triangulations with scalar fields representing physicochemical properties, evaluated at the surface vertices.

The remainder of this paper is organized as follows. Section \ref{sec:star} overviews the previous benchmarks that were aimed at protein shape retrieval aspects. Then, in Section \ref{sec:bench} we detail the dataset, the ground truth and the retrieval and classification metrics used in the contest. The methods submitted for evaluation to this SHREC are detailed in Section \ref{sec:methods}, while their retrieval and classification performances are presented in Section \ref{sec:results}. Finally, discussions and concluding remarks are in Section \ref{sec:obs_conlusion}.

\section{Related benchmarks}
\label{sec:star}
The interest of recognising proteins and other biomolecules solely based on their structure is a lively challenge in biology and the scientific literature is seeing the rise of datasets and methods for surface-based retrieval of proteins. 
The Protein Data Bank (PDB) repository \cite{PDB2000} is the most widely known public repository for experimentally determined protein and nucleic acid structures.
The PDB collects over $175,000$ biological macromolecular 3D structures of proteins, nucleic acids, lipids, and corresponding complex assemblies. A rather small number of proteins in the PDB dataset are captured with the NMR technique, which is very favourable for characterizing the protein also with respect to physicochemical properties.
The PDB offers also a number of visualization tools of the contained structures but is not intended to perform either sequence or structure similarity tasks. 

Previous benchmarks on protein retrieval based on the shape of their molecular surfaces were provided within the SHape REtrieval Contest (SHREC). In these cases, the molecular surfaces correspond to the protein solvent-excluded surface as defined by Lee and Richards \cite{Richards1977} and firstly implemented by Connolly \cite{Connolly}.
To the best of our knowledge, the first contest on protein shape retrieval solely based on molecular surfaces was launched in 2017 with $10$ query models and a dataset of $5,854$ proteins \cite{shrec2017}. A second contest considered a dataset of $2,267$ protein structures, representing the conformational space of 107 proteins \cite{shrec18}. There, the task was to retrieve for each surface the other conformations of the same protein from the whole dataset. In 2019, the SHREC track on protein retrieval \cite{shrec2019} envisioned the classification of $5,298$ surfaces representing the conformational space of 211 individual proteins. The peculiarity of this contest was in the classification of the dataset, which took into account two levels of \emph{similarity}. In addition to the mere retrieval of the different conformers of a given protein, the evaluation also took into account the retrieval of orthologous proteins (proteins having the same function in different organisms, e.g., human and murine haemoglobin protein) based on their surfaces. Finally, in 2020 the aim of the SHREC track on protein retrieval \cite{shrec2020} moved to the retrieval of related multi-domains protein surfaces. Similarly to the 2019 edition, a $2$-level classification (grouping different conformations of the same protein and grouping orthologs together) of the dataset was considered; in contrast to previous years, the 2020 edition included the evaluation of partial similarity, i.e., limited to sub-regions or domains of the entire molecule, thus moving towards a problem of partial correspondence between the proteins.

Compared to the 2019 track on protein retrieval, 
our benchmark differs in multiple points: 
\begin{itemize}
    \item Our data set does not limit to geometric information, but takes also into account physicochemical properties. Moreover, the data set is already split into a training set and a test set.
    \item The set of retrieval evaluation measures is considerably extended; a set of classification measures is also introduced. 
    \item We consider two novel ground truths, based on the domain of bioinformatics.
\end{itemize}

\section{The benchmark}
\label{sec:bench}
During years, we witness the consolidation of the idea that to have a more satisfactory answer to the protein shape retrieval problem it is necessary to combine geometry with patterns of chemical and geometric features \cite{gainza2020deciphering}. For this reason, we move from the previous SHREC experiences to build a dataset equipped of both characteristics.

\subsection{The dataset}
\label{sec:dataset_creation}
\begin{figure}[htb!]
    \begin{center}
    \begin{tabular}{cccc}         
        \includegraphics[scale=0.05, trim={30cm 16cm 30cm 15cm}, clip]{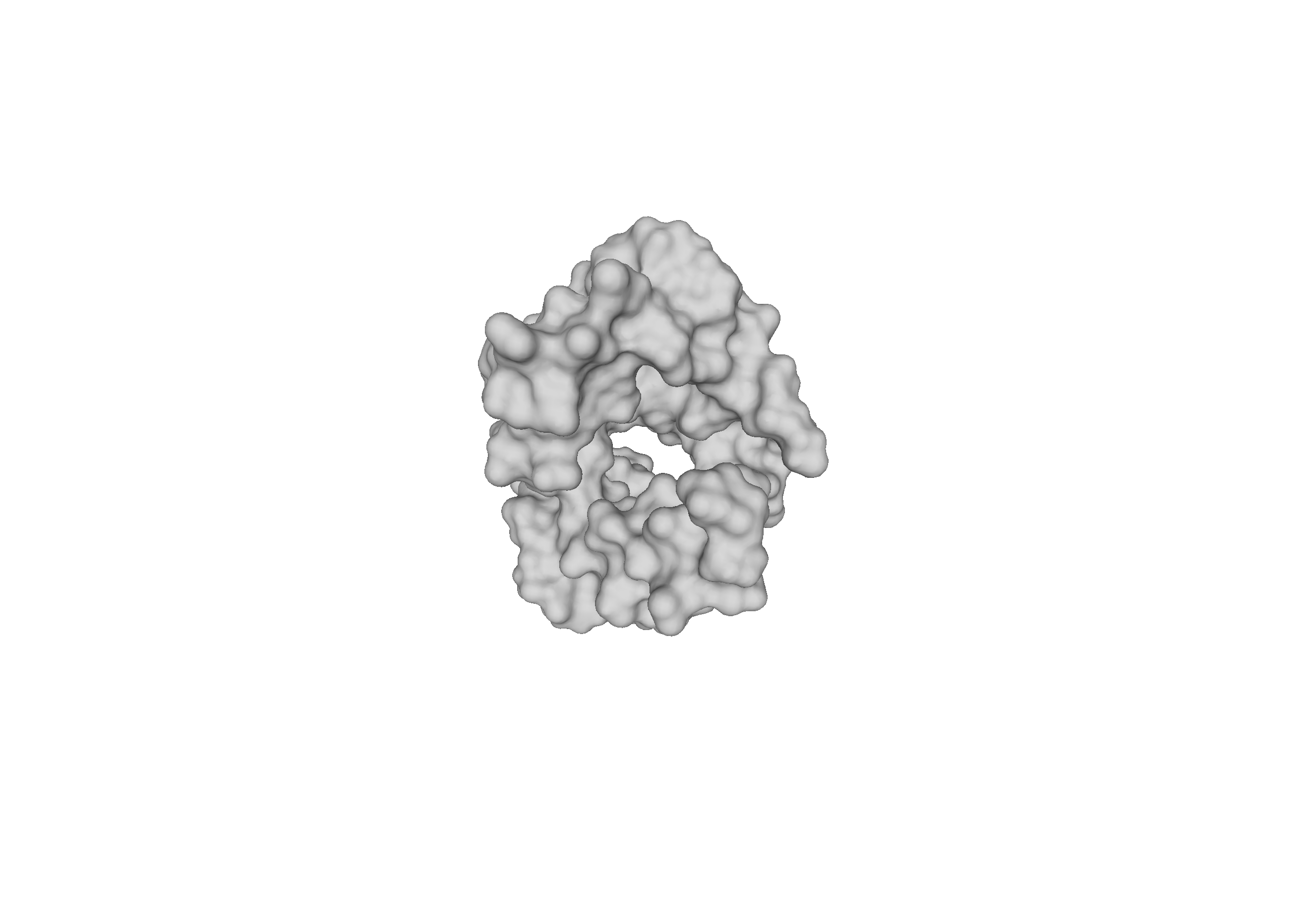}
        &
        \includegraphics[scale=0.045, trim={30cm 16cm 30cm 15cm}, clip]{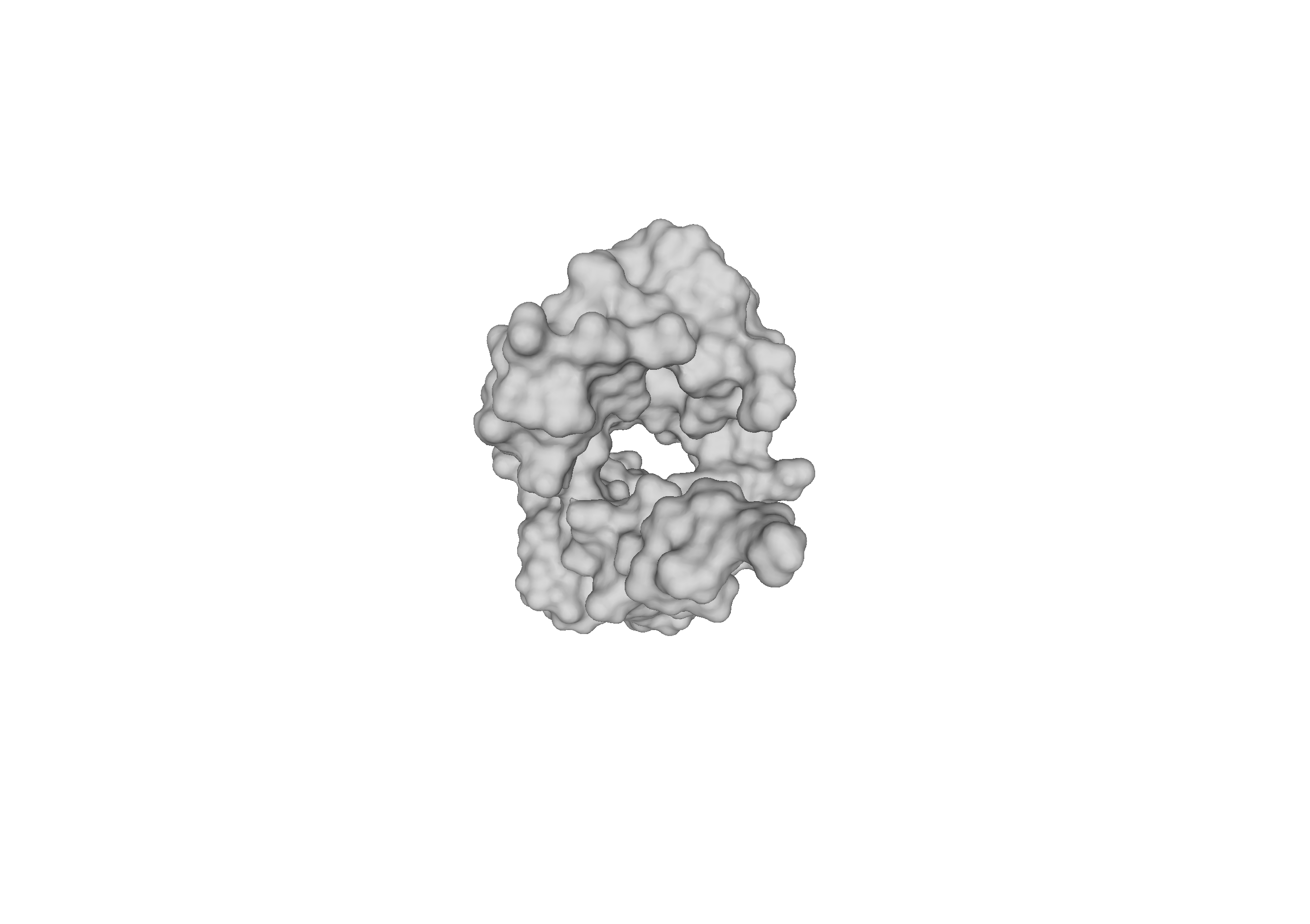}
        &
        \includegraphics[scale=0.045, trim={30cm 16cm 30cm 15cm}, clip]{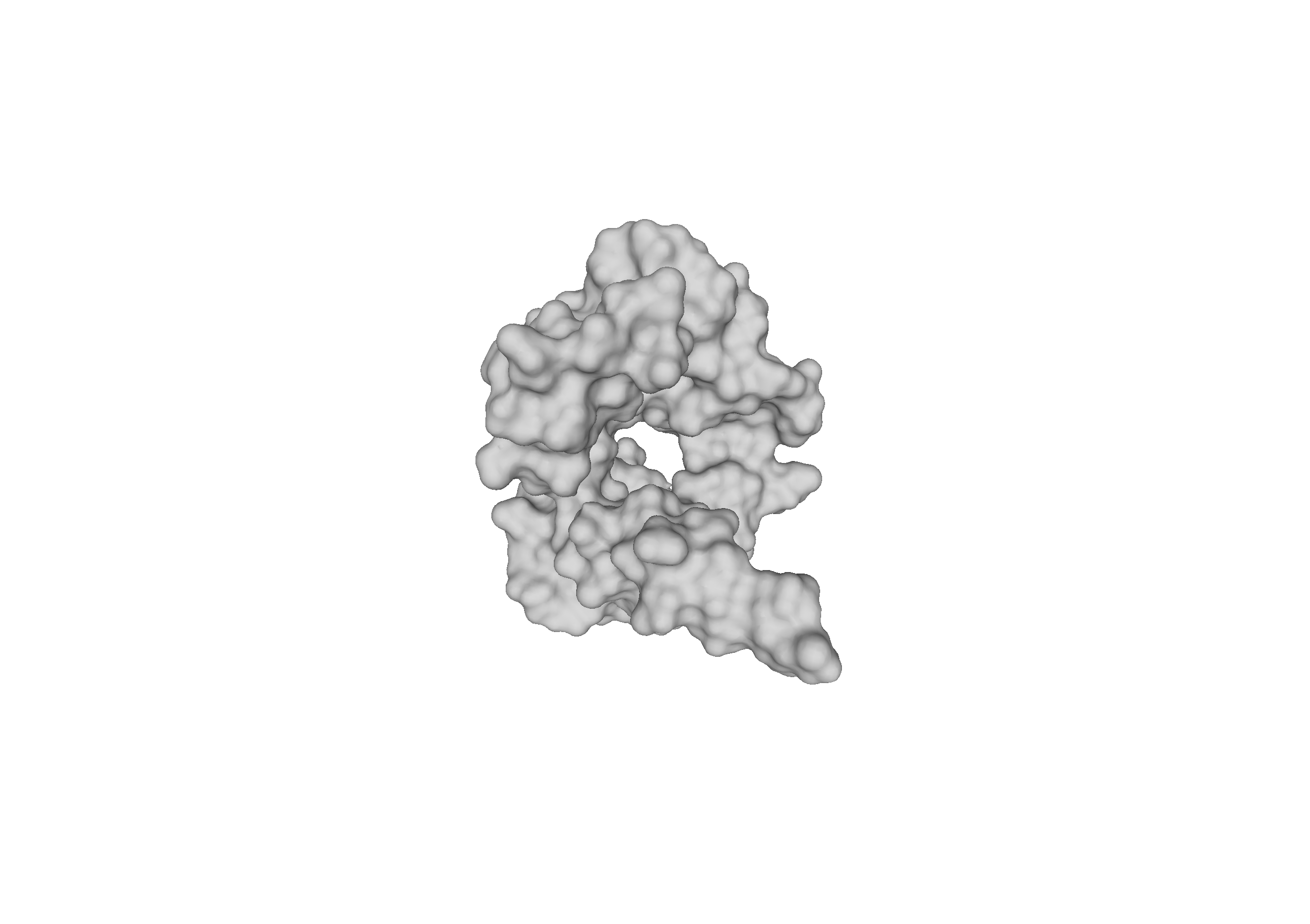}
        &
        \includegraphics[scale=0.045, trim={30cm 16cm 30cm 15cm}, clip]{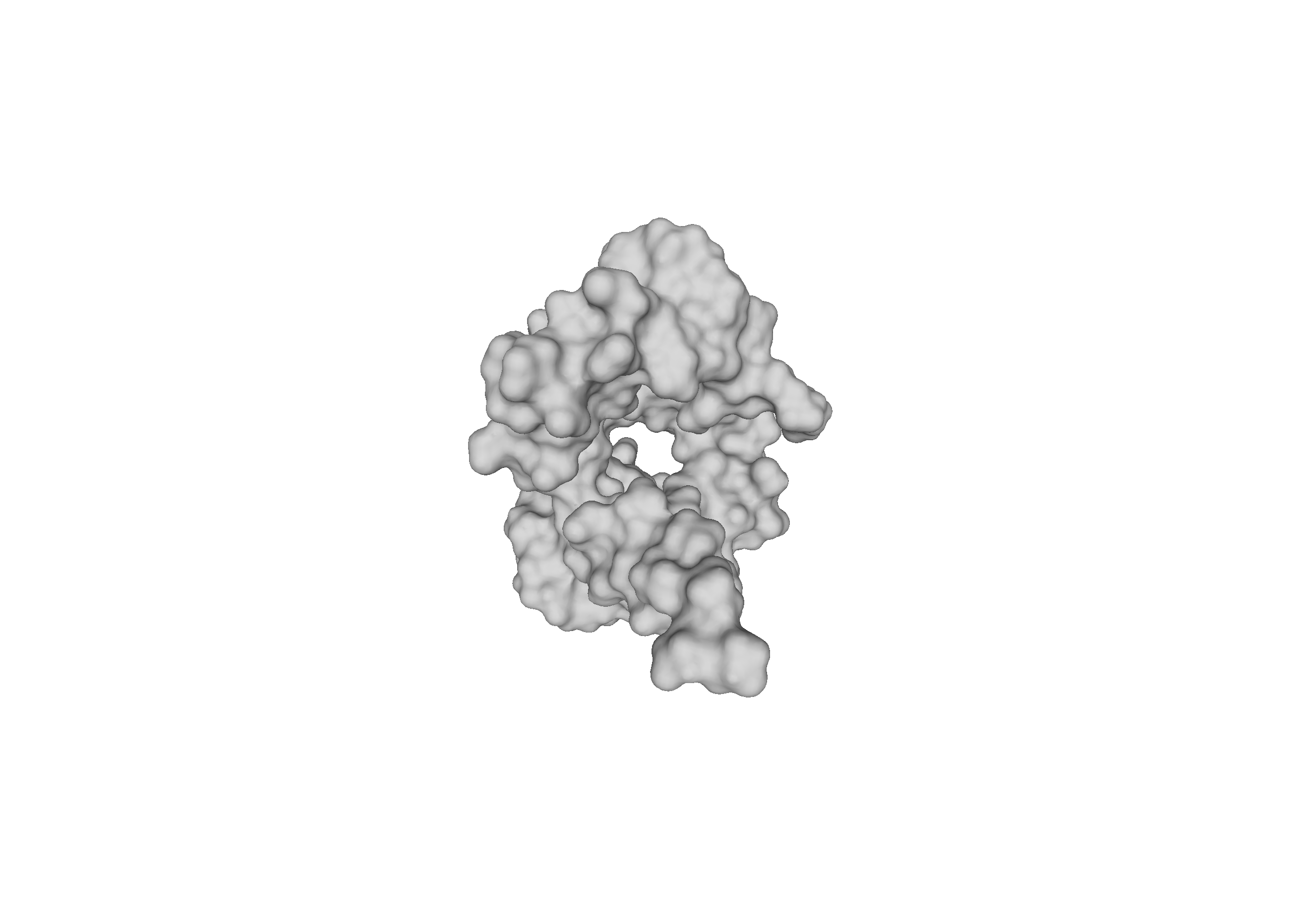}
        \\
        \includegraphics[scale=0.045, trim={27.5cm 11.5cm 27.5cm 13.5cm}, clip]{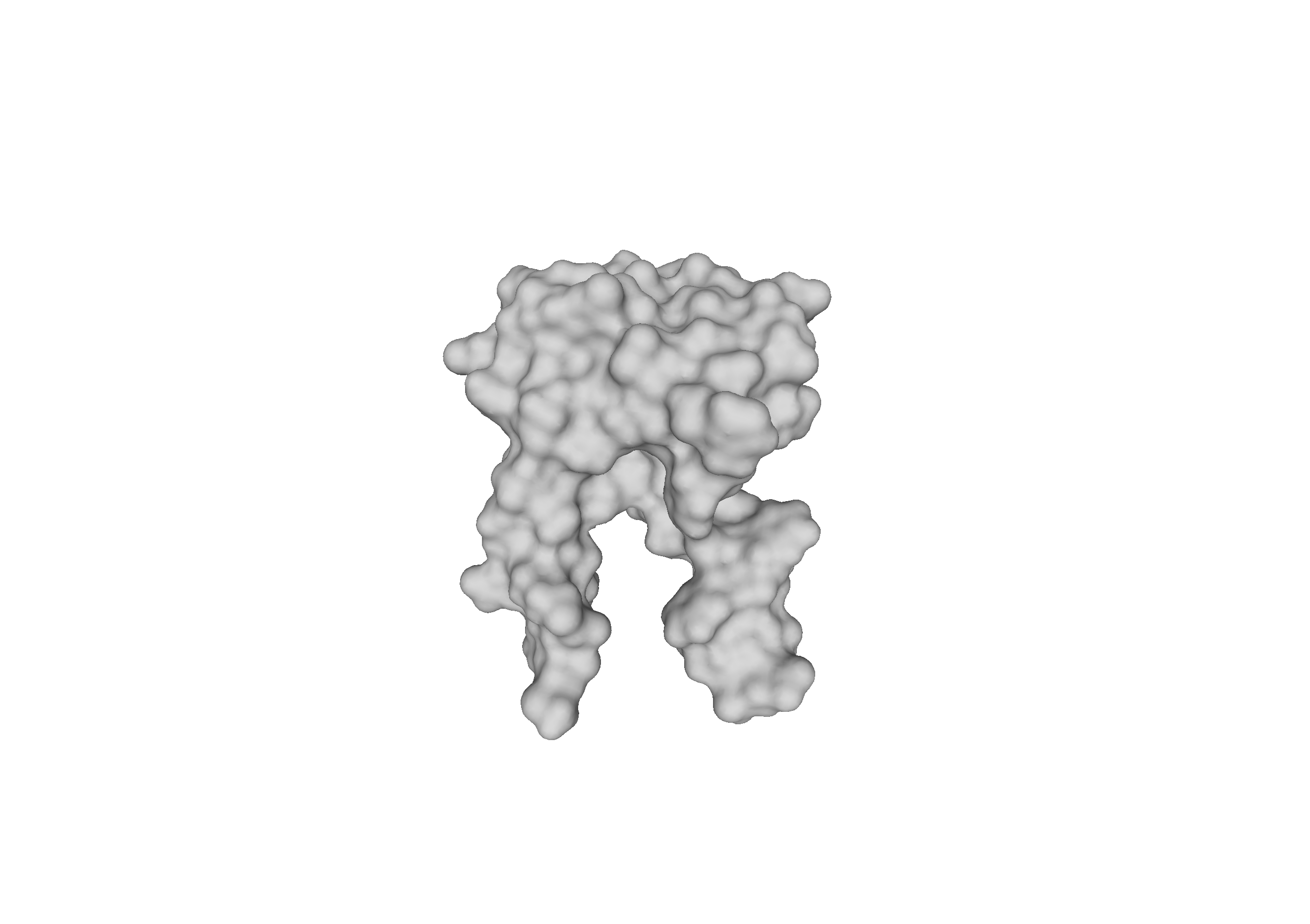}
        &
        \includegraphics[scale=0.045, trim={27.5cm 11.5cm 27.5cm 13.5cm}, clip]{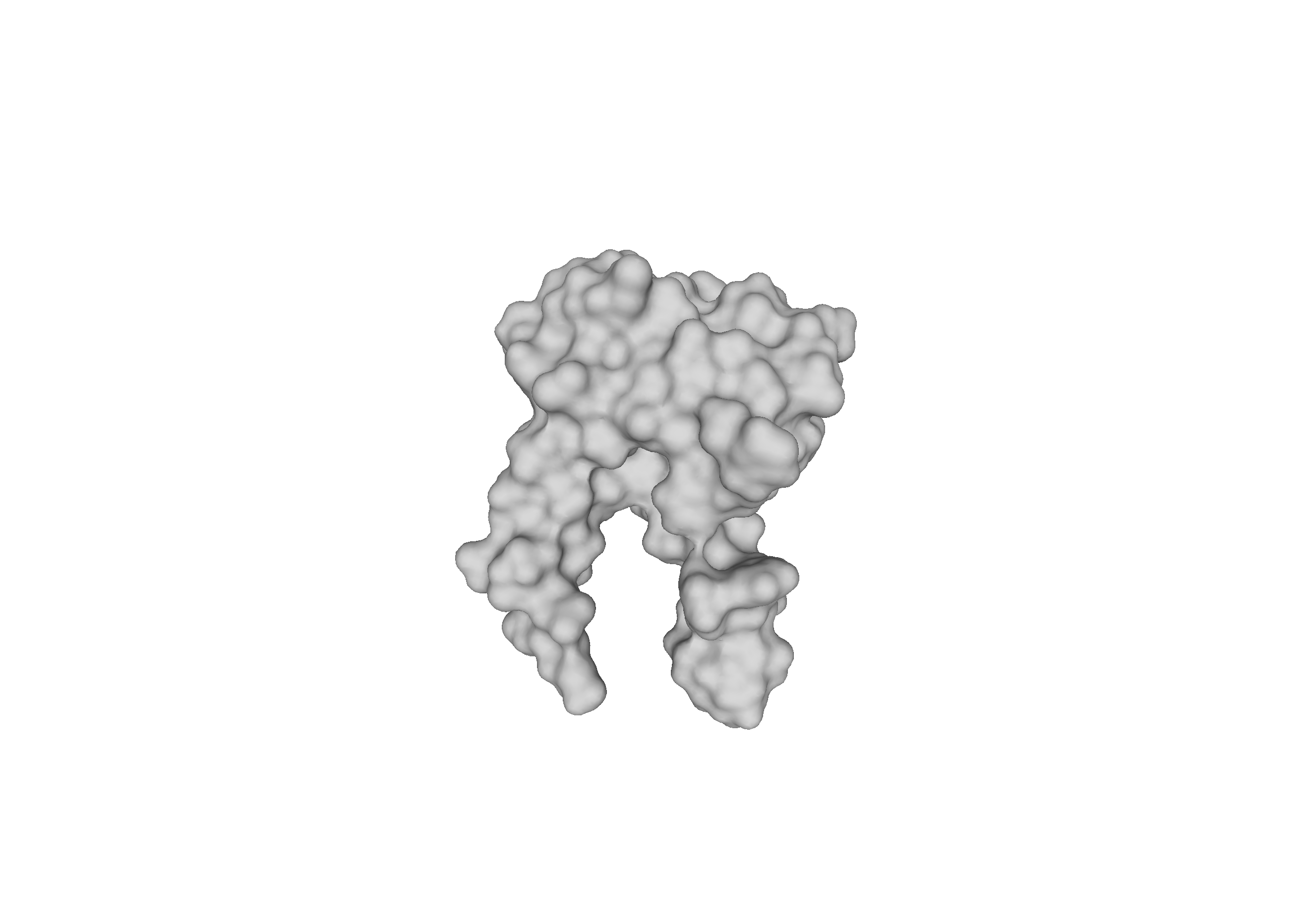}
        &
        \includegraphics[scale=0.045, trim={27.5cm 11.5cm 27.5cm 13.5cm}, clip]{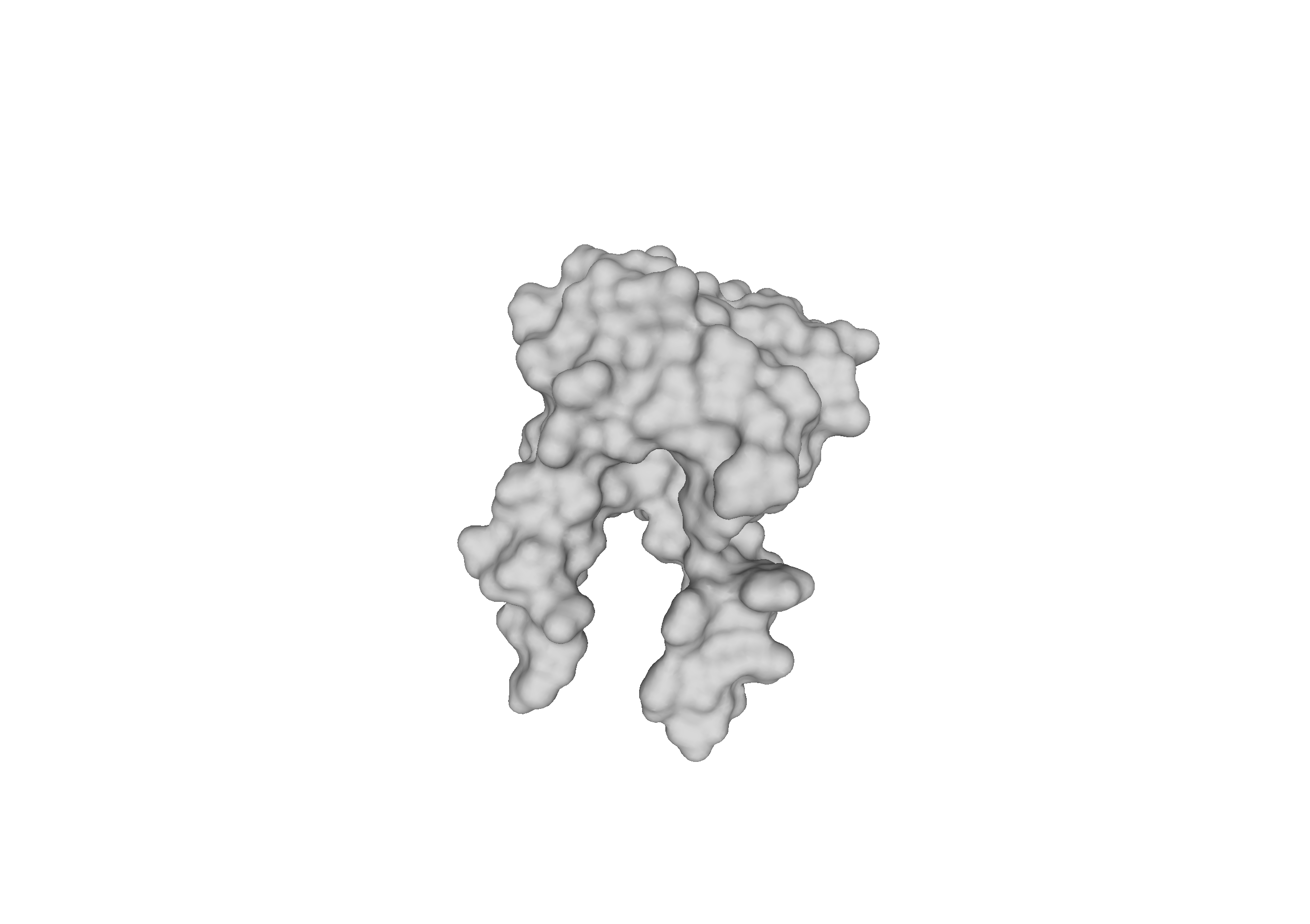}
        &
        \includegraphics[scale=0.045, trim={27.5cm 11.5cm 27.5cm 13.5cm}, clip]{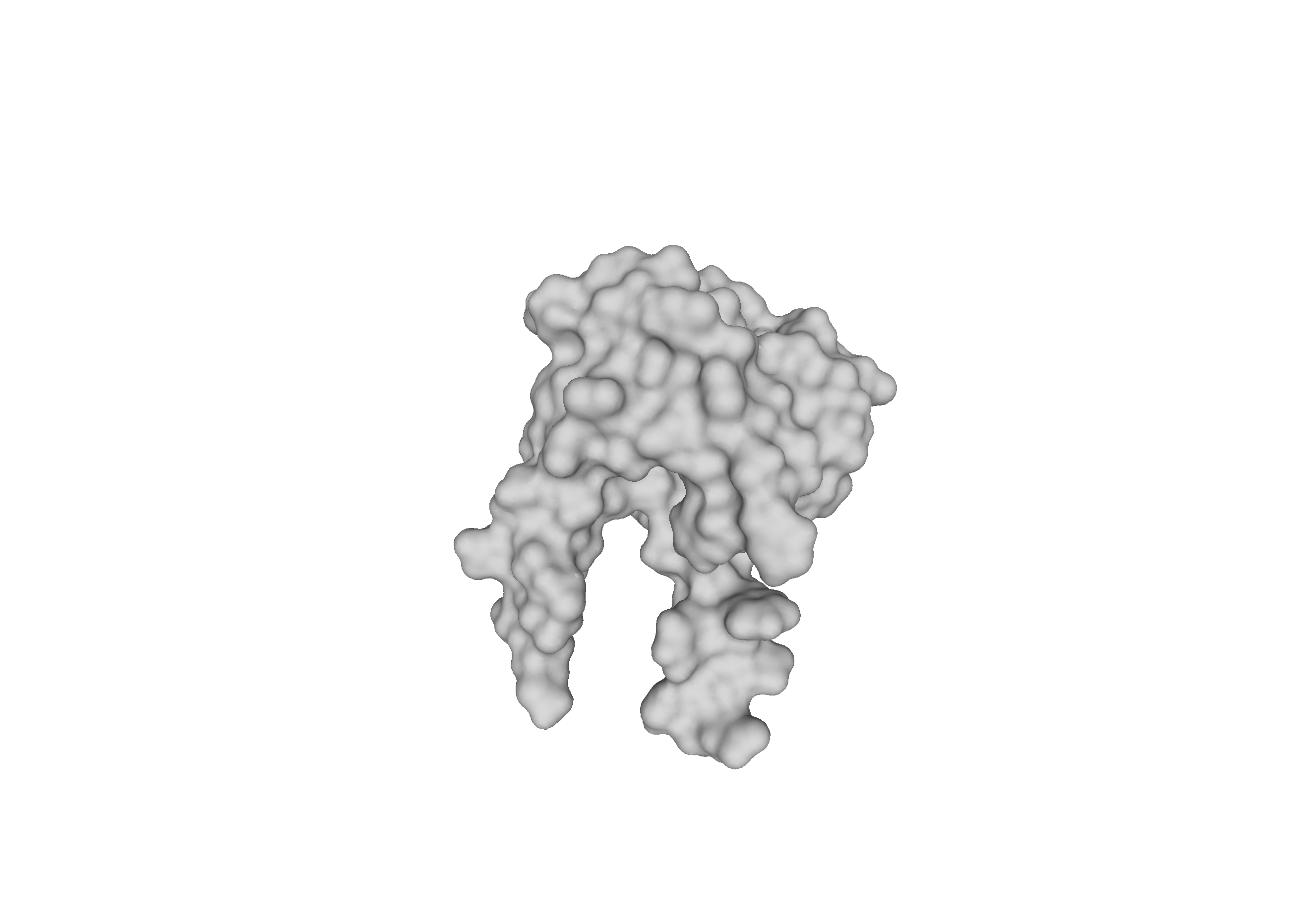}
        \\
        \includegraphics[scale=0.04, trim={25cm 10cm 25cm 12.5cm}, clip]{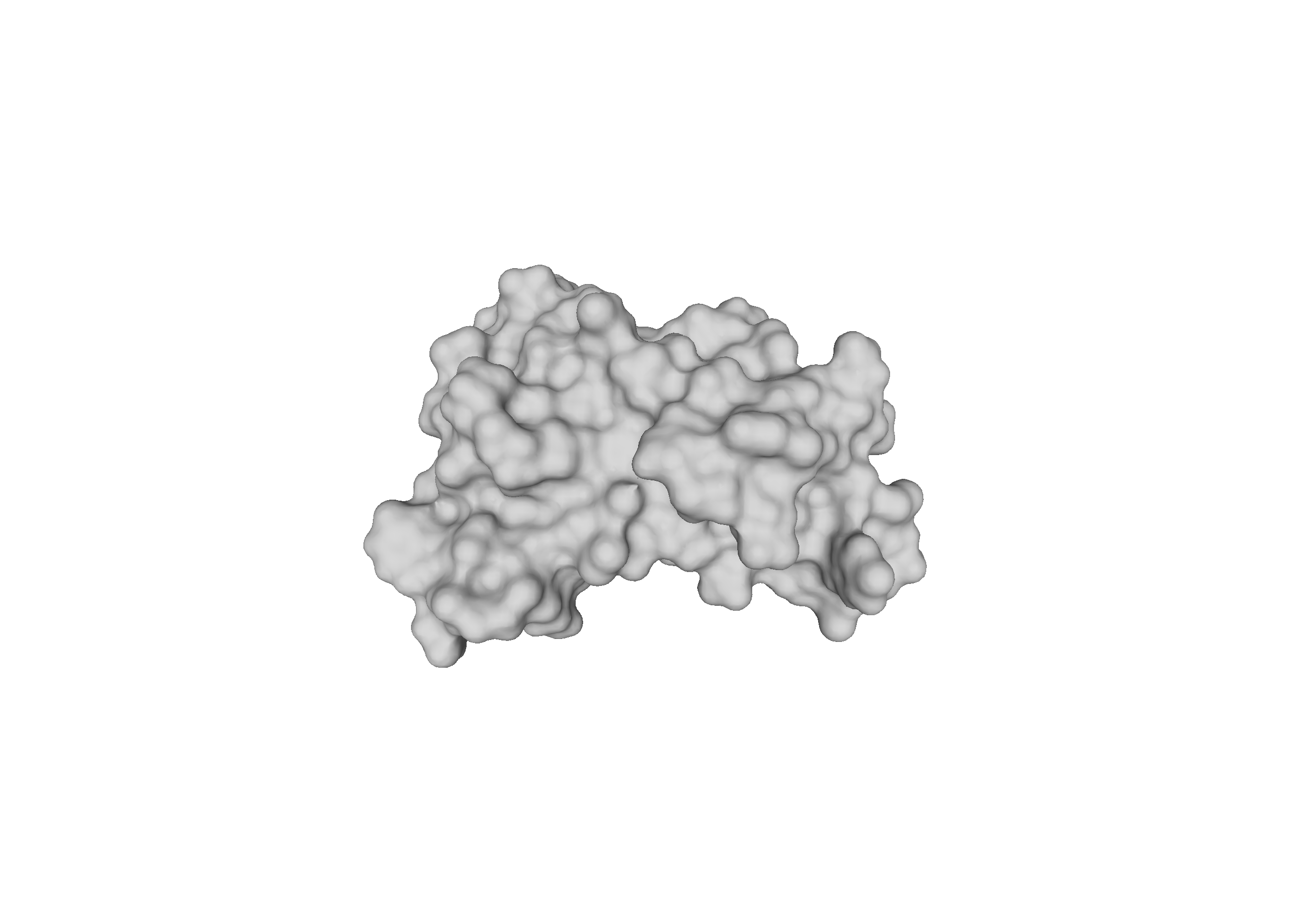}
        &
        \includegraphics[scale=0.04, trim={25cm 10cm 25cm 12.5cm}, clip]{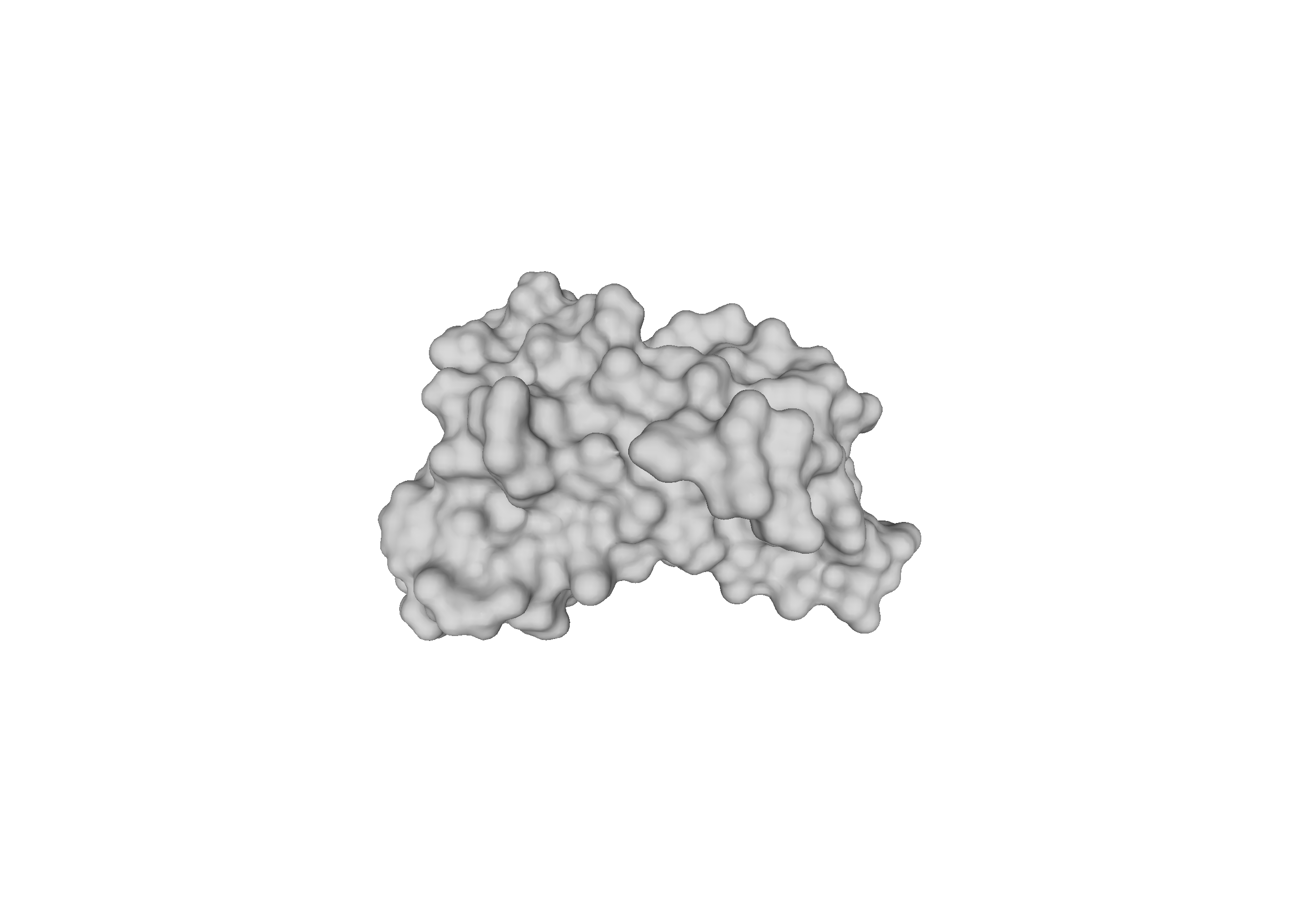}
        &
        \includegraphics[scale=0.04, trim={25cm 10cm 25cm 12.5cm}, clip]{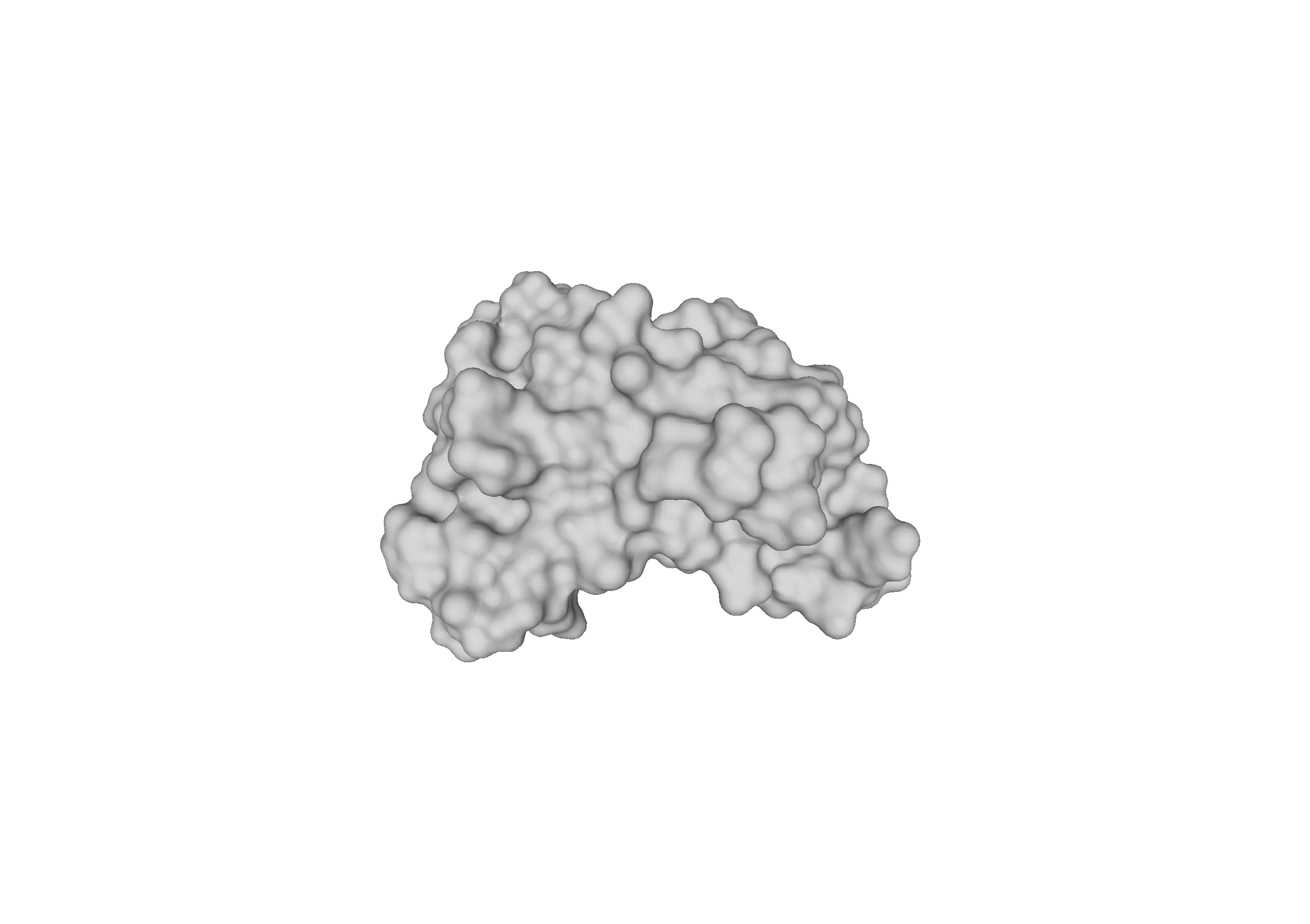}
        &
        \includegraphics[scale=0.04, trim={25cm 10cm 25cm 12.5cm}, clip]{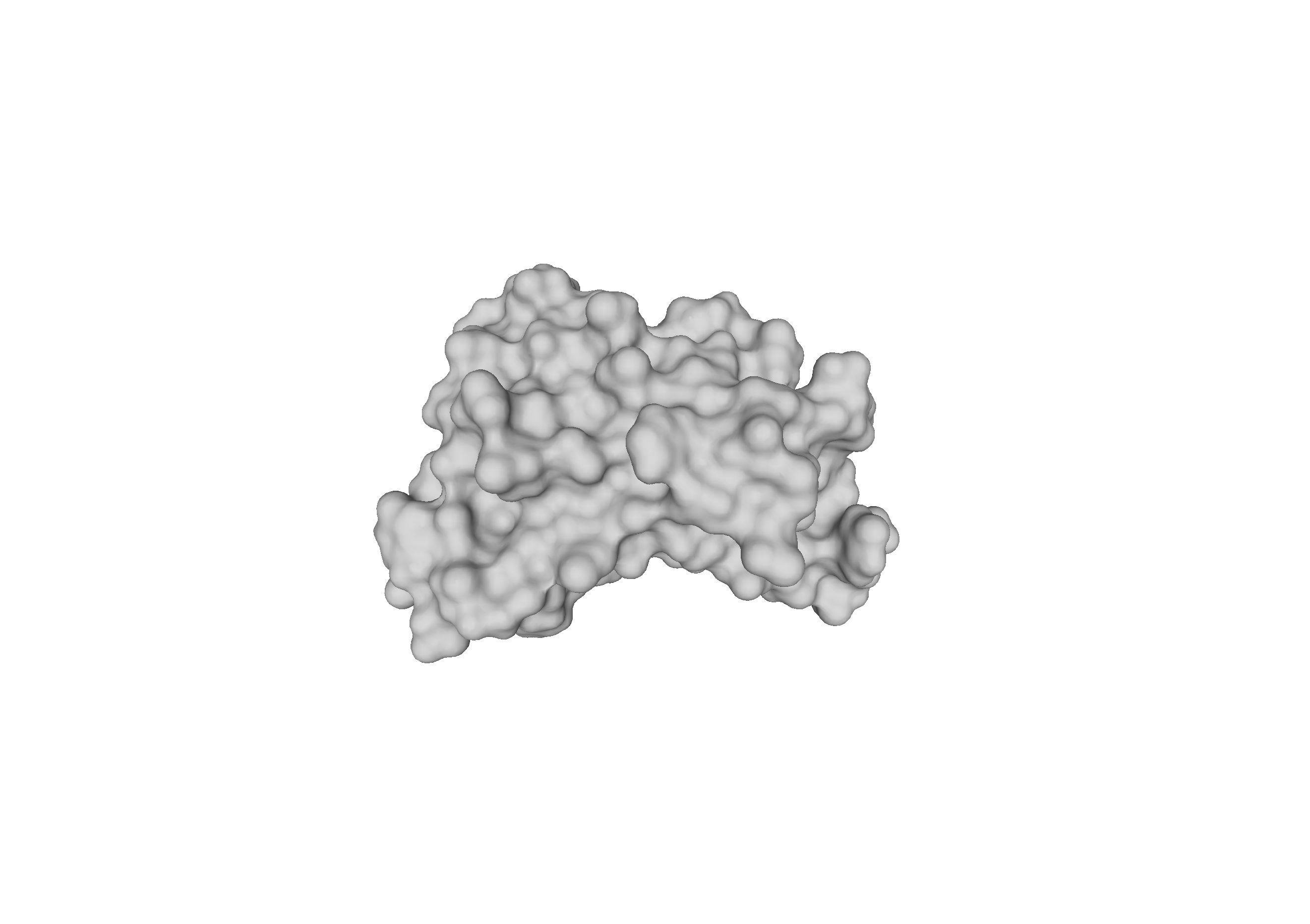}
        \\
        \includegraphics[scale=0.045, trim={27.5cm 15cm 27.5cm 15cm}, clip]{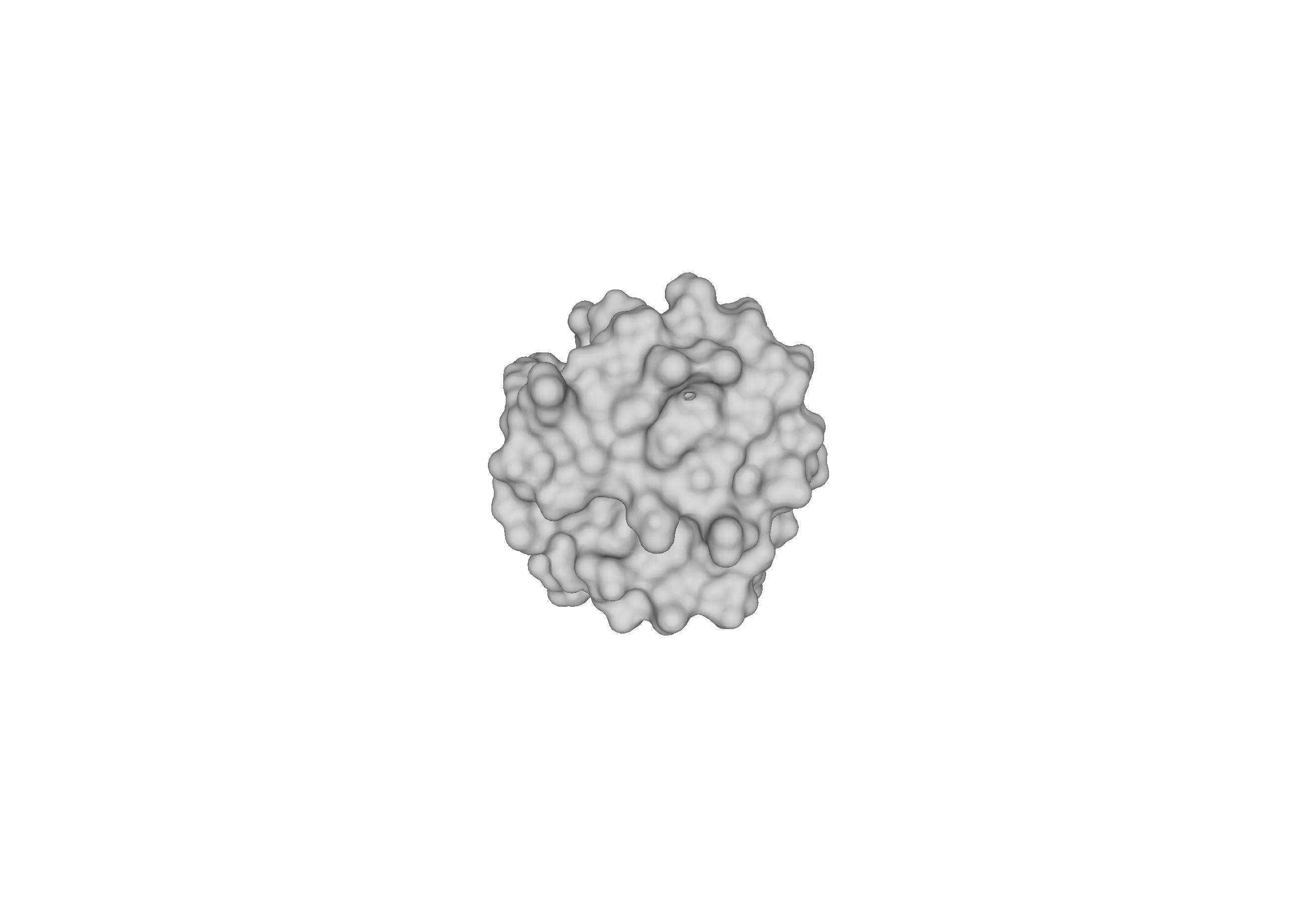}
        &
        \includegraphics[scale=0.045, trim={27.5cm 15cm 27.5cm 15cm}, clip]{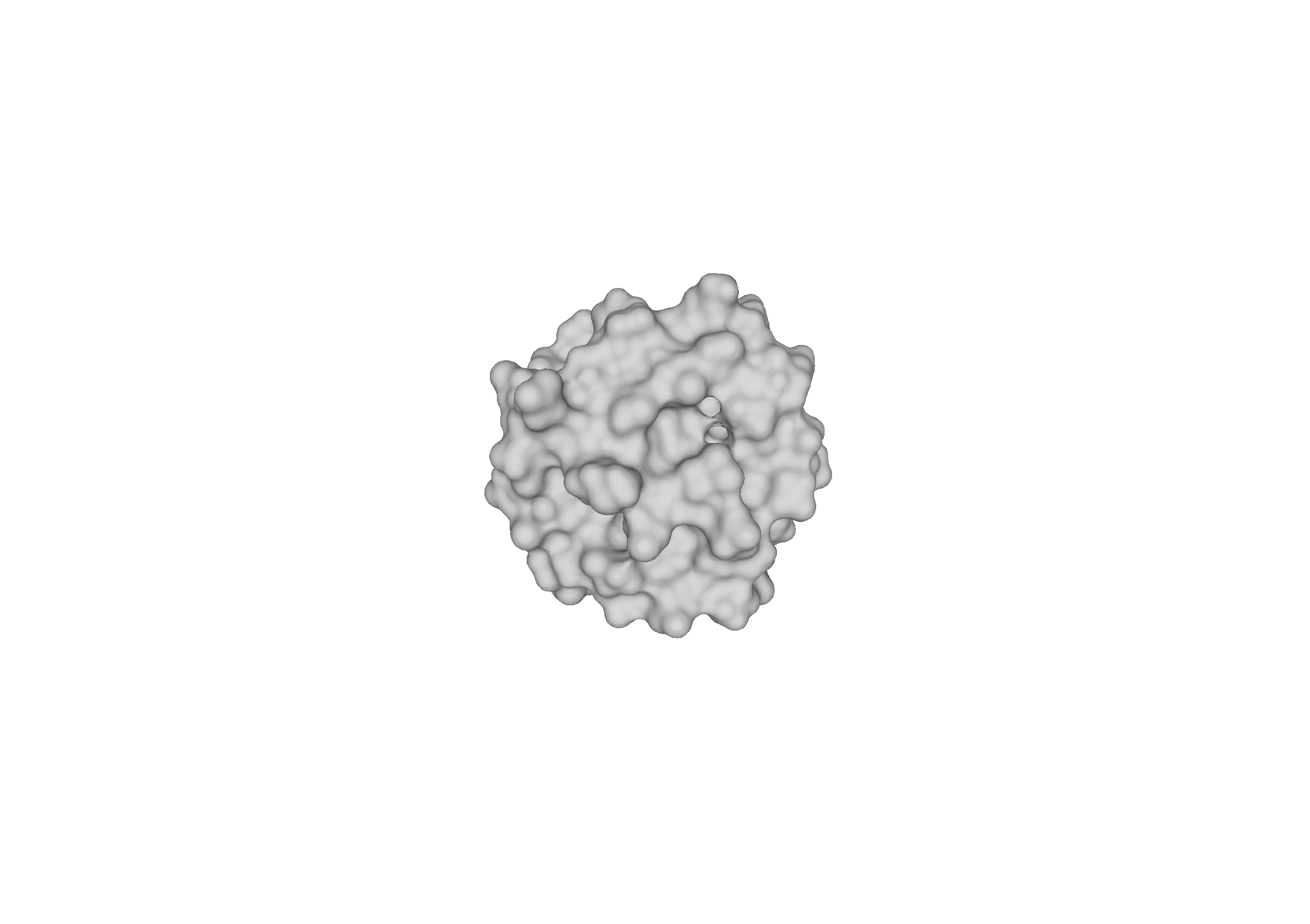}
        &
        \includegraphics[scale=0.045, trim={27.5cm 15cm 27.5cm 15cm}, clip]{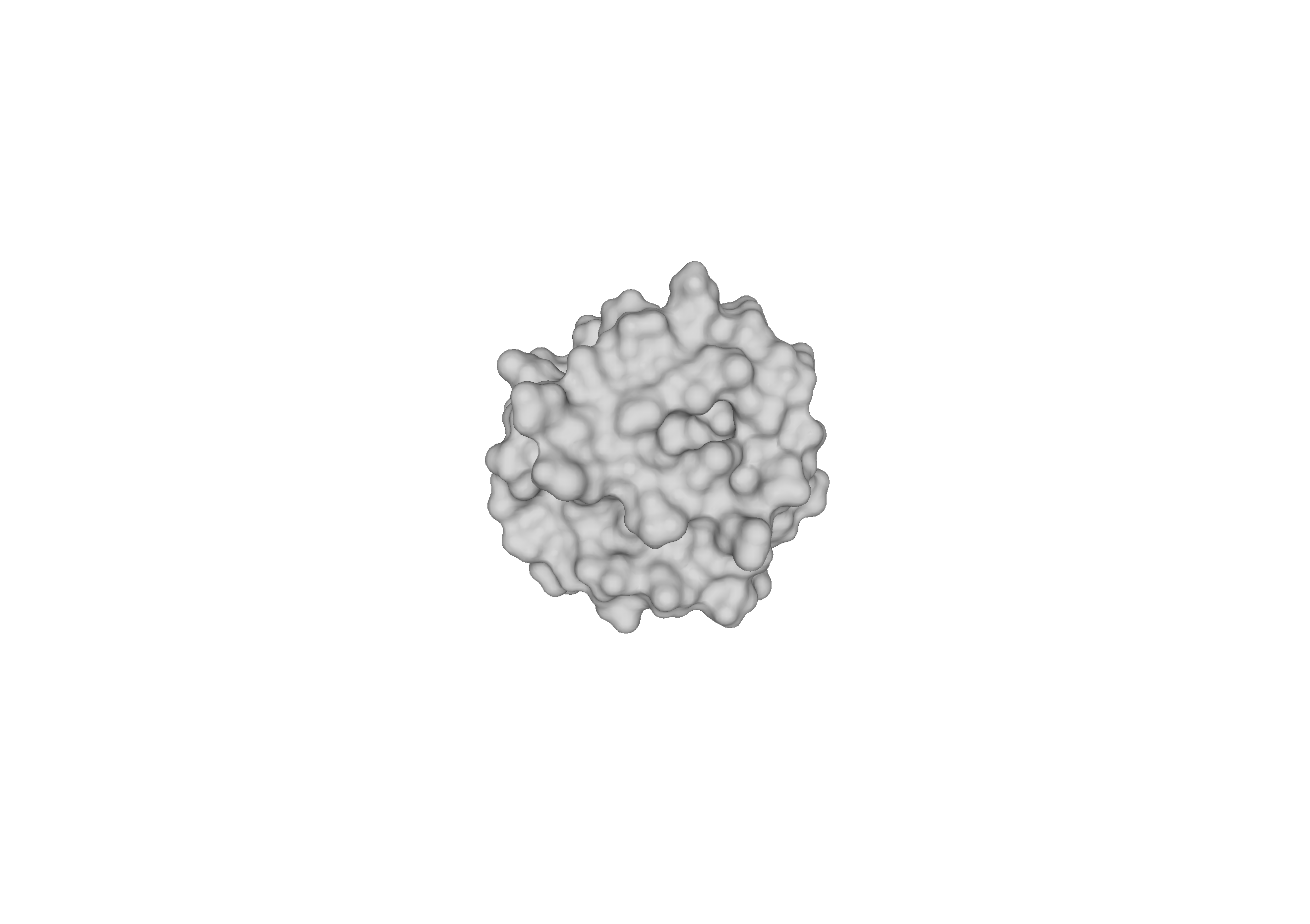}
        &
        \includegraphics[scale=0.045, trim={27.5cm 15cm 27.5cm 15cm}, clip]{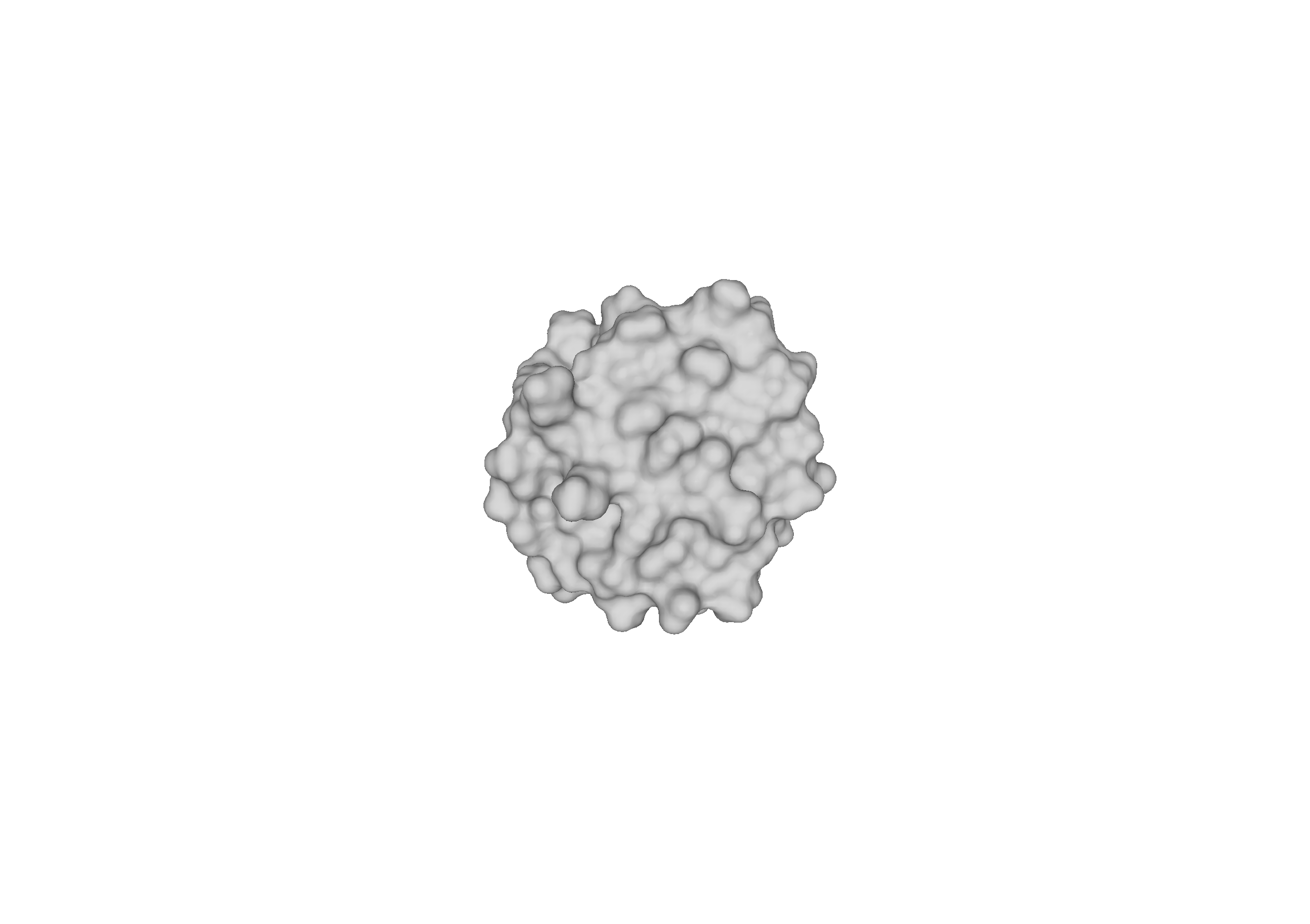}
        \\
        
    \end{tabular}
    \end{center}
    \caption{Example of $4$ proteins in $4$ different conformations (each row identifies a protein). Visualization obtained by using MeshLab \cite{Cignoni:2008}. \label{fig:dataset_conformations}}
\end{figure}

The dataset proposed for this challenge consists of  209 PDB entries, each one containing a protein in different conformations experimentally determined via NMR measurements.
This leads to about $5,000$ surfaces, annotated with physicochemical properties. Some example of proteins in different conformational states are provided in Figure \ref{fig:dataset_conformations}.

Model surfaces were built starting from the PDB files of proteins used in the 2019 SHREC track \cite{shrec2019}. These proteins were experimentally captures with the NMR technique and contain also orthologous structures thus making possible to consider multiple levels of similarity. NMR structures natively include hydrogen atoms, which do not need to be modelled. Importantly, these structures encompass a number of energetically favourable conformations of the same protein, representing important regions of the corresponding conformational space.
Each individual conformation structure was first separated into a unique PDB file. Then, its molecular surface (MS) was calculated and triangulated by means of the NanoShaper computational tool, choosing the Connolly Solvent Excluded Surface model \cite{Connolly}, and default parameters \cite{DeCherchi2013}. The vertices of the triangulated surfaces were stored in OFF\footnote{https://segeval.cs.princeton.edu/public/off\_format.html} format. 

\begin{figure}[htb!]
    \begin{center}
    \begin{tabular}{cccc}         
        \includegraphics[scale=0.07, trim={12.5cm 7cm 12.5cm 8cm}, clip]{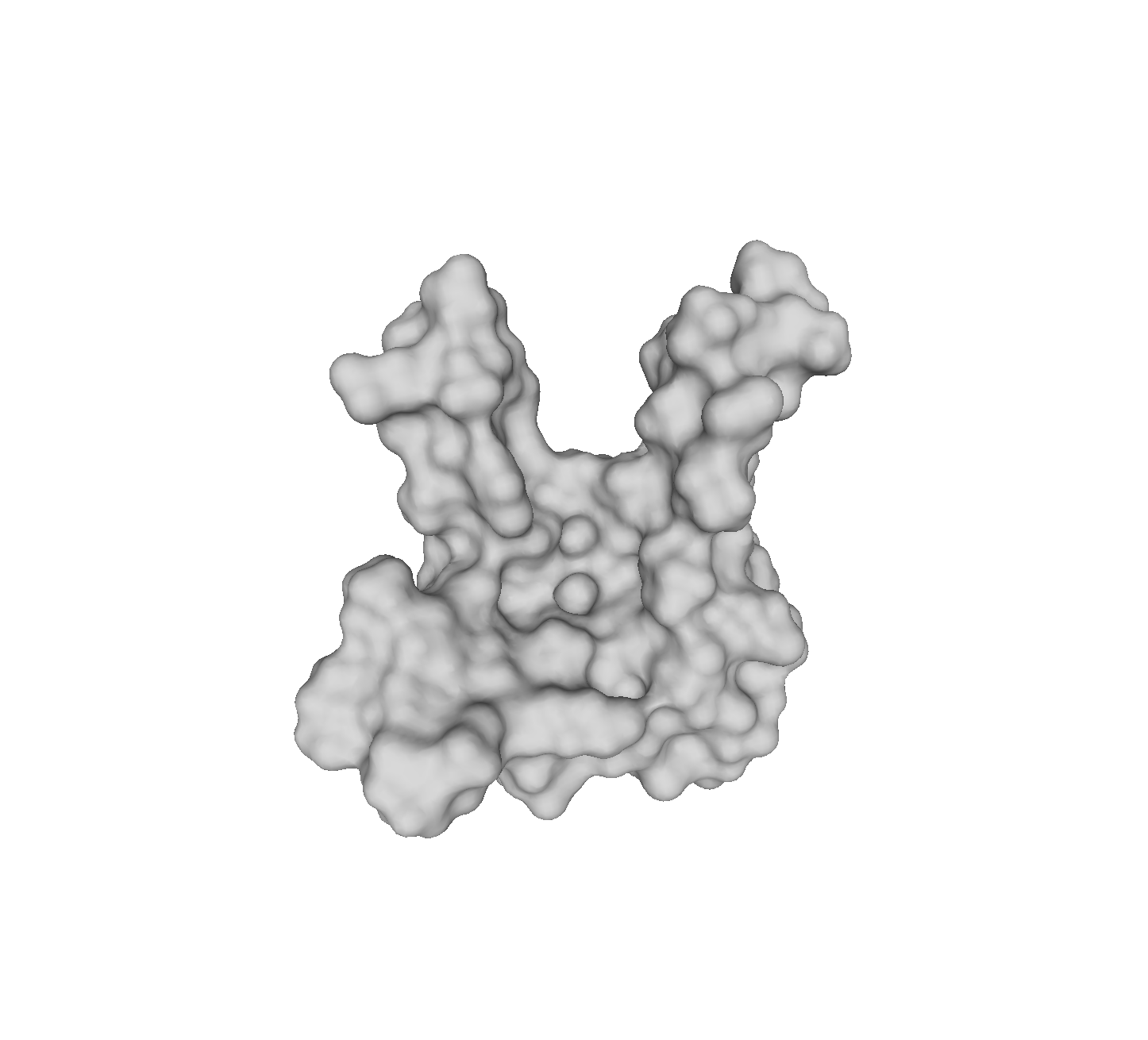}
        &
        \includegraphics[scale=0.07, trim={12.5cm 7cm 12.5cm 8cm}, clip]{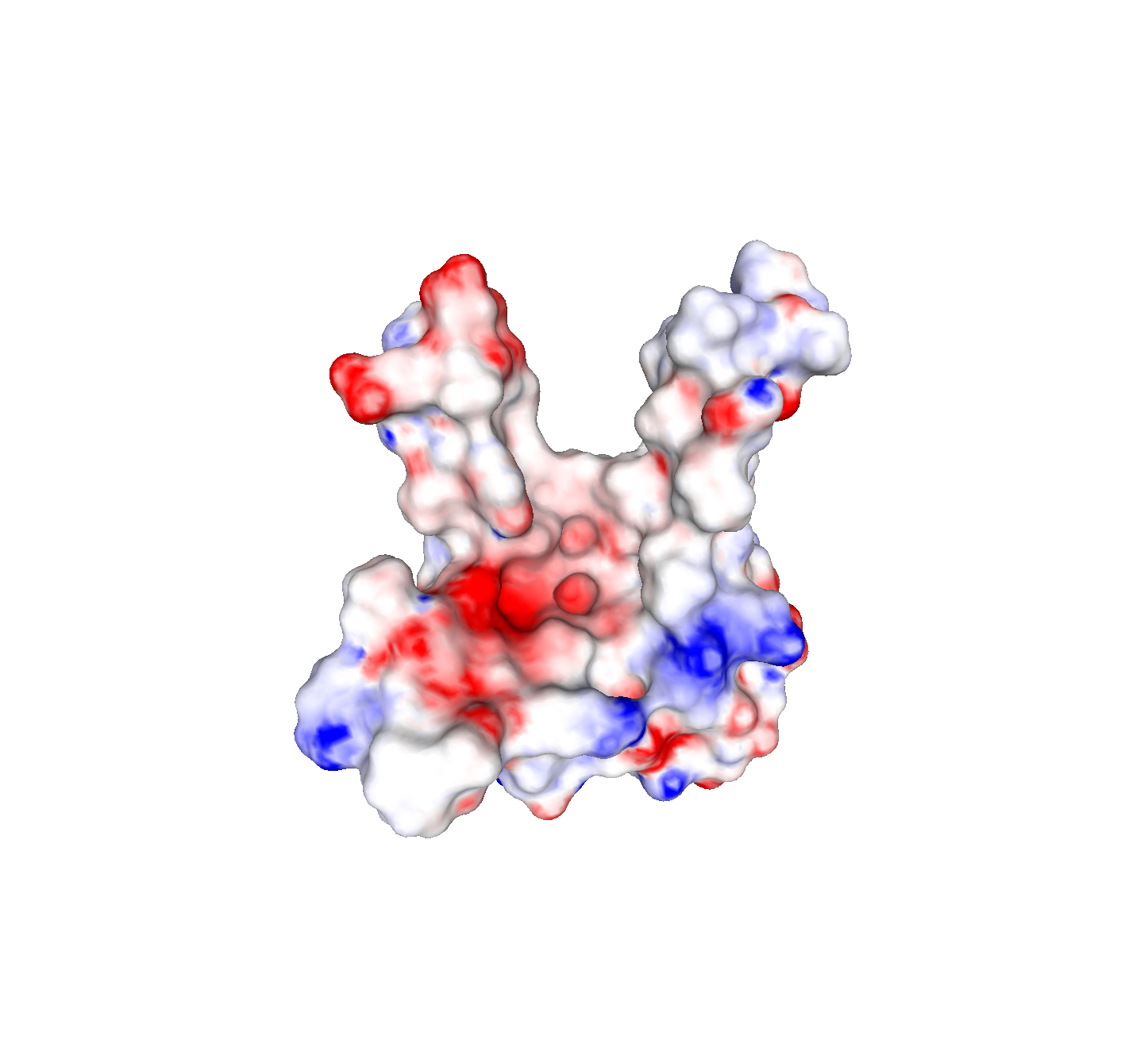}
        &
        \includegraphics[scale=0.07, trim={12.5cm 7cm 12.5cm 8cm}, clip]{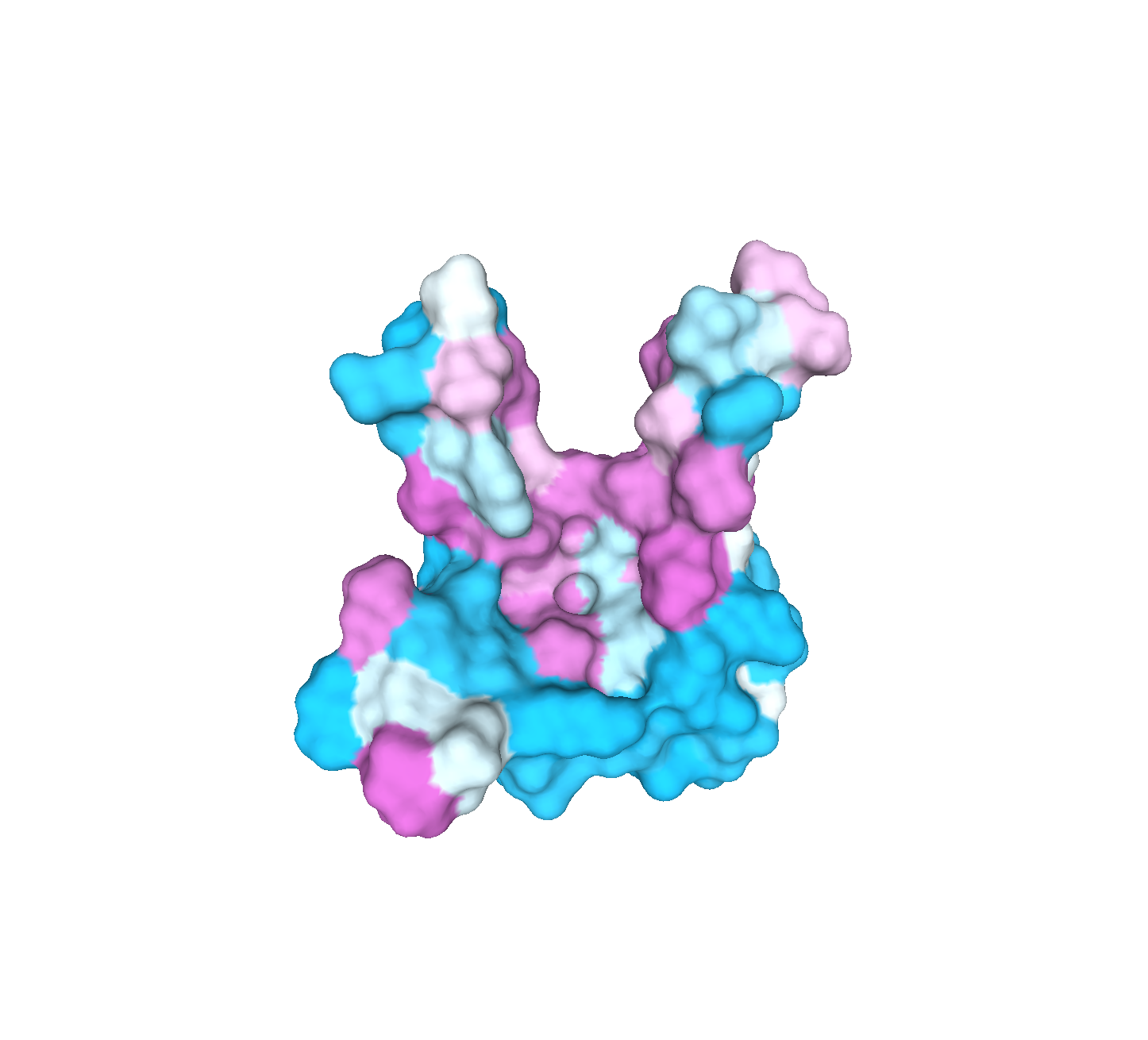}
        &
        \includegraphics[scale=0.07, trim={12.5cm 7cm 12.5cm 8cm}, clip]{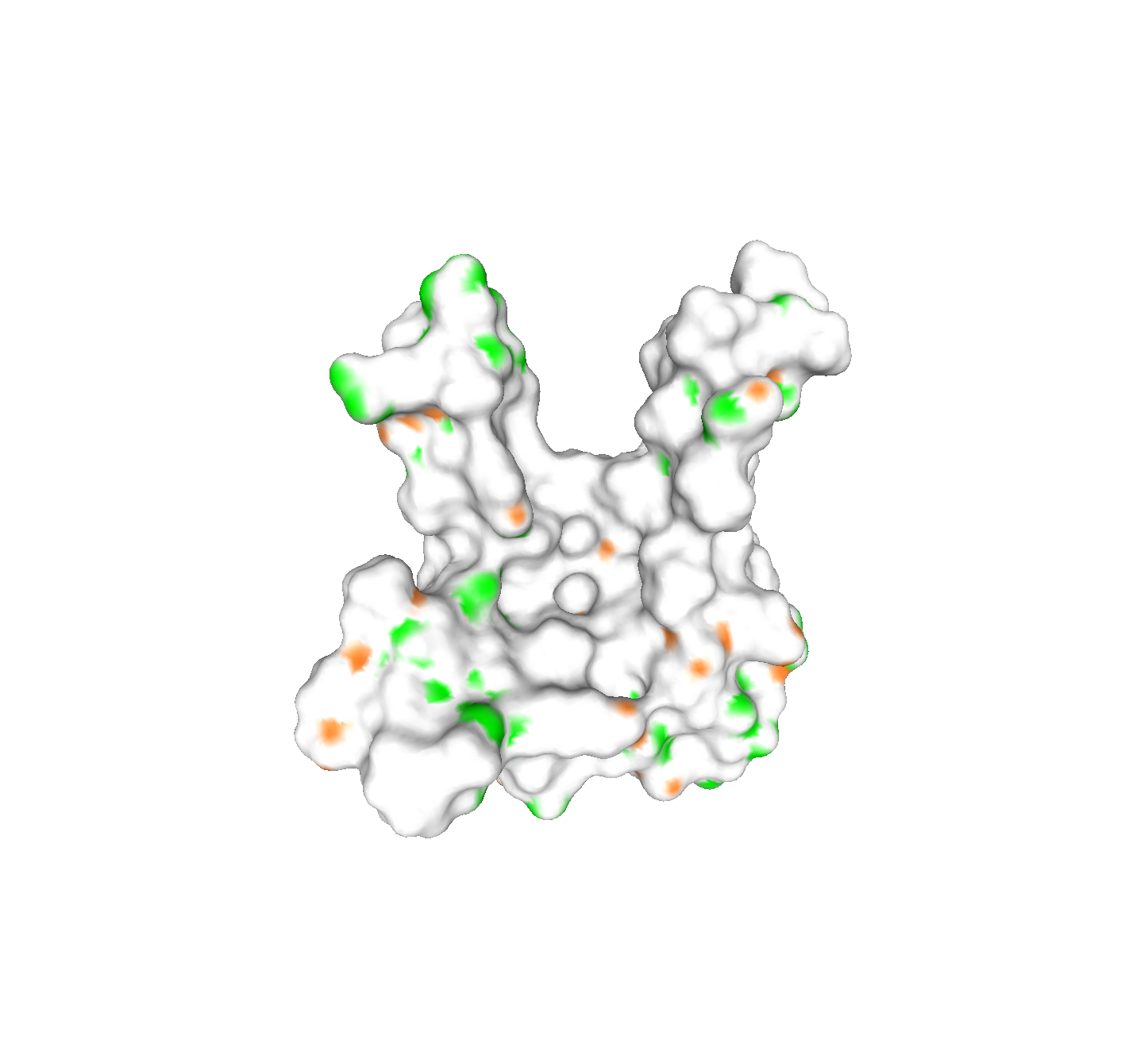}
        \\
        (a) & (b) & (c) &(d)
    \end{tabular}
    \end{center}
    \caption{Example of protein surface (a) equipped with different physicochemical properties: electrostatic potential (b), hydrophobicity (c) and presence of hydrogen bond donors and acceptors (d). Visualization obtained by using MeshLab \cite{Cignoni:2008}. \label{fig:dataset}}
\end{figure}

Each surface model was accompanied by a file with physicochemical information, in TXT format. Each row of the TXT file corresponds to a vertex of the triangulation in the OFF file (in the same order); each row in the TXT file contains the physicochemical properties evaluated at the corresponding vertex in the OFF file. An example of protein surface equipped with physicochemical properties is provided in Figure \ref{fig:dataset}: more specifically, Figure \ref{fig:dataset}(a) exhibits the original triangulated surface, while Figures \ref{fig:dataset}(b-d) represent the three provided physicochemical properties as scalar values on the protein surface. 

\begin{figure}[h!]
     \begin{center}
     \begin{tabular}{cc}
          \includegraphics[scale=0.425, trim={0.25cm 0cm 0.5cm 0.25cm}, clip]{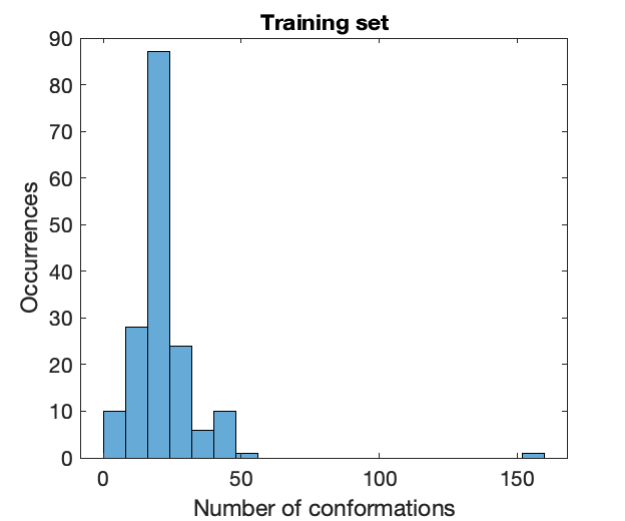}
          &
          \includegraphics[scale=0.425, trim={0.25cm 0cm 0.5cm 0.25cm}, clip]{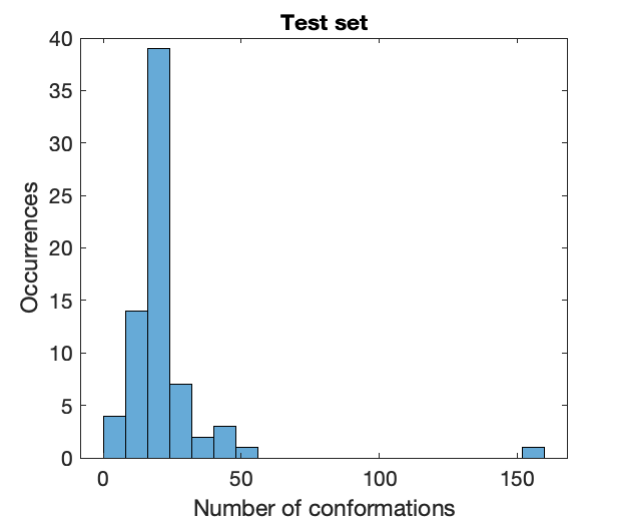}
     \end{tabular}
     \end{center}
     \caption{Distribution of the number of conformations. The two histograms show the distributions for the training set (left) and the test set (right).}
     \label{fig:barchart_distributions}
 \end{figure}

The dataset has been subdivided into a training and a test set (in the proportion of 70\%-30\%). The distribution of the number of conformations per PDB through the training set and the test set is shown in Figure \ref{fig:barchart_distributions}.

 To enrich the MS information we used the electrostatic potential, which we computed by solving the Poisson-Boltzmann equation (PBE) via the DelPhi finite-differences-based solver \cite{Delphi2001,Delphi2013}. 
One of the essential ingredients for the solution of the PBE is a good definition of the MS, which is used to separate the high (solvent) from the low (solute) dielectric regions. In order to guarantee the perfect consistency of the approach, we adopted a DelPhi version integrated with NanoShaper \cite{DeCherchi2013}, so as that the potential is evaluated on the same exact surface that separates the solute from the solvent. 
Other necessary ingredients are atom radii and partial charges, which have been assigned using the PDB2PQR tool \cite{pdb2pqr}.

A different kind of additional information mapped on the MS was the hydrophobicity \cite{kyte1982simple} of the residues exposed to the solvent. In our setting, we assign to each vertex of the MS the hydrophobicity of the residue of the closest atom, on the basis of the scale given in  \cite{kyte1982simple}; this scale ranges from $-4.5$ (hydrophilic) to $4.5$ (hydrophobic).

Lastly, we have computed the location of potential hydrogen bond donors and acceptors in the MS. Firstly, vertices of the MS whose closest atom is a polar hydrogen, a nitrogen or an oxygen were identified. Then, a value between $-1$ (optimal position for a hydrogen bond acceptor) and $1$ (optimal position for a hydrogen bond donor) was assigned to such vertices depending on the orientation between the corresponding heavy atoms (see \cite{kortemme2003orientation}).

\subsection{The ground truth\label{sec:ground}}
The performances of the methods that participated to this SHREC contest are evaluated on the basis of two classifications:

\begin{itemize}
    \item   \emph{PDB-based classification}.
    In the dataset selected, there is a number of entries having different PDB codes that contain the structures of the same protein, possibly interacting with different molecules or having a limited number of point mutations. In these cases it can be expected that the specific condition in which the protein system has been observed impacts on the identified conformations and on the corresponding physicochemical properties. The first classification rewards the techniques that are particularly good at spotting minor differences between similar candidates; in this classification, a class is made by all the conformations corresponding to the same PDB code. For reference, we refer to this ground truth as PDB-based classification.
    
    \item \emph{BLAST-based classification}. Protein sequences fold into unique 3-dimensional (3D) structures and proteins with similar sequences adopt similar structures \cite{Rost1999}. Therefore, on the basis of the similarity among the amino acids sequences, we decided to relax the strict relationship that two surfaces are similar only if they correspond to some conformation of the same PDB code. This choice is based on observations coming from the domain of bioinformatics, where a sequence similarity beyond a value of about $30\%$, and of sufficient length, has a high likelihood of giving rise to the same fold \cite{Rost1999}. We derive a second classification and name it BLAST-based classification, since BLASTP is the tool that we used to perform the sequence alignment and to calculate the sequence similarity \cite{BLASTP}. The BLAST-based classification represents a classification less fine than the PDB-based one, because it is simply based on the similarity between conformations; in this way, not only the different NMR conformations found in the same PDB file, but also these of the same protein in different PDB files or these pertaining to its mutated isoform(s) may be grouped together. The BLAST-based classification presents four levels. In this setting, two structures are: 
    \begin{itemize}
        \item \emph{Extremely similar (similarity level 3)}, i.e. corresponding to the same protein or very closely related protein isoforms: when they have a sequence similarity greater than 95\% on at least the 95\% of both sequences.
        \item \emph{Highly related (similarity level 2)}, i.e. they are expected to have a similar fold as a whole or in a sub-domain (above what in the bioinformatics jargon is called the ``twilight zone"): when they have a sequence similarity greater than 35\% and at least $50$ aligned residuals, but they do not satisfy the conditions of the previous point.
        \item \emph{Similar (similarity level 1)}, i.e. loosely related proteins: when they have a sequence similarity in $[28\%, 35\%]$ and at least $50$ aligned residuals.
        \item \emph{Dissimilar (similarity level 0)}, i.e. unrelated proteins: when none of the previous conditions holds.
    \end{itemize}    
\end{itemize}

To compare the performance of the methods that make use of the physicochemical properties against the simple geometric models, we asked the participants to perform two tasks:
\begin{description}
\item{Task A:}
only the OFF files of the models are considered (i.e. only the geometry is considered);
\item{Task B:} in addition to the geometry, the participant is asked to also consider the TXT files (physicochemical matching).
\end{description}

For a given query, the goal of this SHREC track is twofold: for each Task (A and B), to retrieve the most similar objects (retrieval problem) and to classify the query itself (classification problem). The closeness of the retrieved structures with the ground truth might be evaluated a-priori on the basis of their PDB code or of their sequence similarity ($4$-level BLAST classification) \cite{Rost1999}.

\paragraph{Retrieval problem} Each model is used as a query against the rest of the dataset, with the goal of retrieving the most relevant surface. For the retrieval problem, a dissimilarity $1,543 \times 1,543$ matrix was required, each element $(i,j)$ recording the dissimilarity value between models $i$ and $j$ in the whole dataset. The relevance with respect to the query of a retrieved surface is evaluated with both the PDB and BLAST classifications previously described. 

\paragraph{Classification problem}
PDB-based and BLAST-based classifications define on the training and on the test sets a decomposition into subsets (that will be referred as communities) consisting of conformations grouped together on the basis of their similarity.
The goal of the classification problem is to assign each query of the test set to the correct community with respect to the decompositions induced by the PDB-based and the BLAST-based classifications, respectively.

In the case of the PDB-based classification, each community consists of all the conformations corresponding to the same PDB code.

In the case of the BLAST-based classification, different community decompositions are obtained depending on the choice of the previously described similarity levels.
For each level $\ell$ (with $\ell=0, 1, 2, 3$), it is possible to retrieve a decomposition into communities referred as BLAST-based community decomposition of level $\ell$.
Independently from the chosen level $\ell$, each community of the BLAST-based decomposition of level $\ell$ is an aggregation of communities induced by the PDB-based classification. 

Having fixed a level $\ell$, the communities of the BLAST-based decomposition of level $\ell$ are computed as it follows.
Let us consider a graph $G_\ell$ for which each node represents a PDB-based community (i.e. models corresponding to the same PDB code) and such that there exists an edge $(u,v)$ whenever the structures $u$ and $v$ have a similarity level greater than or equal to $\ell$. Moreover, each edge $(u,v)$ is endowed with a weight $w(u,v)$ coinciding with the percentage of sequence similarity between $u$ and $v$.
The clustering technique for retrieving the BLAST-based decomposition of level $\ell$ adopts the following recursive strategy which has been specifically designed for the considered framework but it is inspired by classic methods for community detection \cite{fortunato2010community}.
Given $G_\ell$, compute the connected components of $G_\ell$ obtained after the removal the edge of $G_\ell$ of minimum weight (and so representing a low similarity score between models). A connected component $C$ is declared a BLAST-based community of level $\ell$ if $C$ is a complete graph (i.e. given any two of its nodes there is an edge connecting them). Otherwise, keep removing edges (prioritising the ones with the lowest weight), compute the connected components and denote them as BLAST-based community of level $\ell$ whenever they are complete.
The procedure ends when all the nodes have been inserted in a community.

\begin{figure}[htb!]
\centering
  \includegraphics[width=0.50\columnwidth]{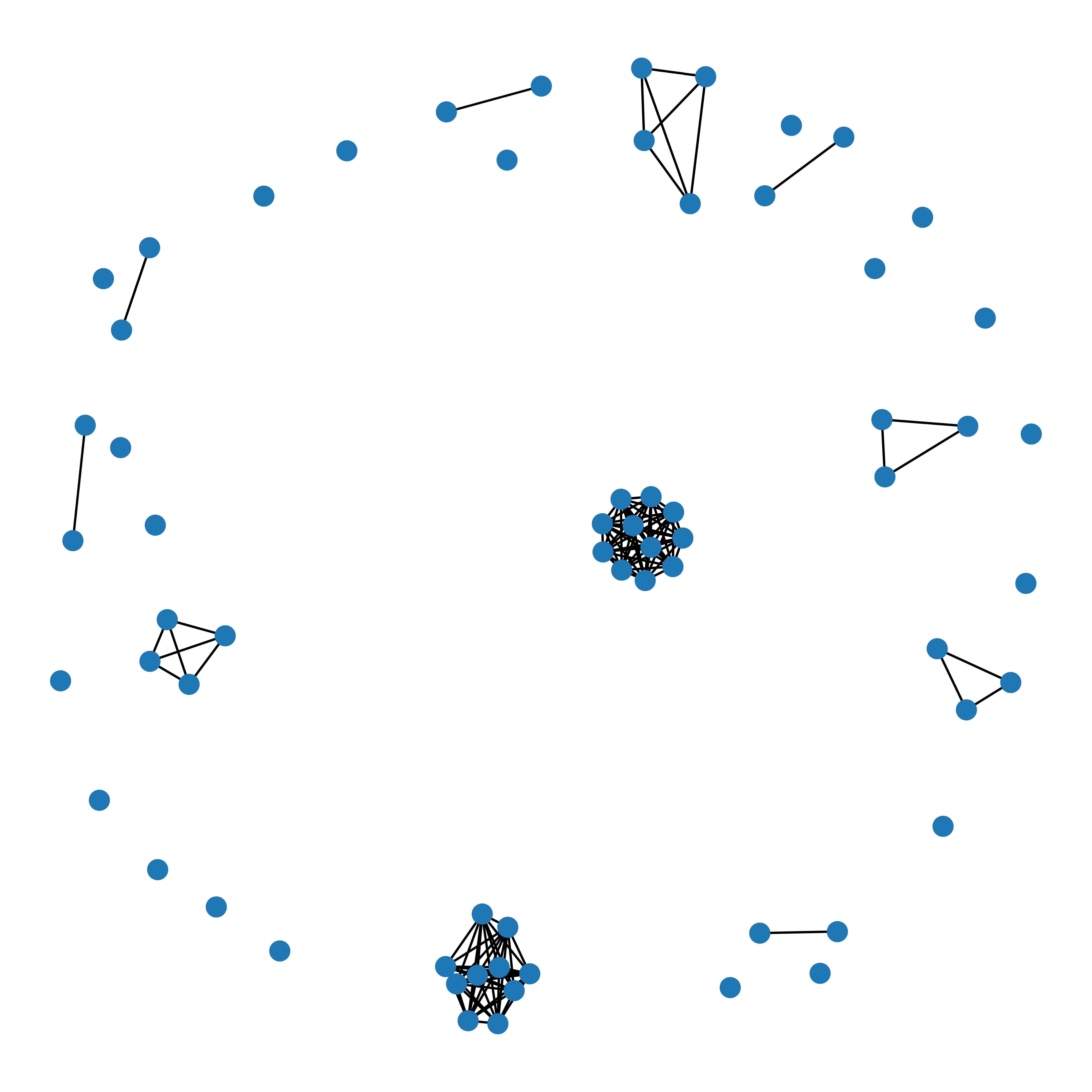}
  \caption{The graph $G_3$ associated with the test set. Each node represents a community induced by the PDB-based classification, while an edge between two nodes occurs whenever the corresponding structures are extremely similar (i.e. they have similarity level equal to 3). Since they are all complete graphs, each connected component of $G_3$ represents a community of the BLAST-based decomposition of level 3.}\label{fig:BLAST-based communities of level 3}
\end{figure}

It is worth to be noticed that the completeness condition has been imposed in order to obtain transitive BLAST-based communities. In this way, we have the theoretical guarantee that any two structures belonging to the same BLAST-based community of level $\ell$ have necessarily a similarity level greater than or equal to $\ell$.
Another relevant aspect to be mentioned is related to the fact that, for $\ell=3$, the proposed algorithm does not remove any edge since the connected components of the graph $G_\ell$ are already complete (see Figure \ref{fig:BLAST-based communities of level 3}). Trivially, the same happens also for $\ell=0$ since the BLAST-based decomposition of level 0 produces just a unique ``giant'' community consisting of the entire dataset.
Finally, please notice that, by increasing the value $\ell$, one obtains BLAST-based decompositions consisting of finer communities.

Given the PDB and BLAST ground truth for the classification problem, the classification performance of each run is obtained through the nearest neighbour (1-NN) classifier derived from the dissimilarity matrices used in the retrieval problem. 
For each run, the output consists of two classification arrays for the test set, with 65 labels/communities for the PDB-based decomposition and with 31 labels/communities for the BLAST-based decomposition of level 3 (see Figure \ref{fig:BLAST-based communities of level 3}). 
In these arrays, the element $i$ is set to $j$ if $i$ is classified in class $j$ (that is, the nearest neighbour of the surface $i$ belongs to class $j$).

\subsection{Evaluation measures}
\label{sec:eval_meas}

We here presents the retrieval and classification evaluation measures that were selected for this benchmark.

\subsubsection{Retrieval evaluation measures}\label{rMeasures} 
3D retrieval evaluation has been carried out according to standard measures, namely precision-recall curves, mean Average Precision (mAP), Nearest Neighbour (NN), First Tier (1T), Second Tier (2T), Normalized Discounted Cumulated Gain (NDCG) and Average Dynamic Recall (ADR) ~\cite{shrec06,Rijsbergen1979,Baeza-Yates:1999}. 

\paragraph{Precision-recall curves and mean average precision} 
Precision and recall measures are commonly used in information retrieval \cite{Rijsbergen1979}. Precision is the fraction of retrieved items that are relevant to the query. Recall is the fraction of the items relevant to the query that are successfully retrieved. Being $A$ the set of relevant objects and $B$ the set of retrieved object,
\begin{equation*}
\textrm{Precision}=\frac{|A\cap B|}{|B|},\quad
\textrm{Recall}=\frac{|A\cap B|}{|A|}.
\end{equation*}
Note that the two values always range from 0 to 1. For a visual interpretation of these quantities we plot a curve in the reference frame recall vs. precision. We can interpret the result as follows: the larger the area below such a curve, the better the performance under examination. In particular, the precision-recall plot of an ideal retrieval system would result in a constant curve equal to 1. As a compact index of precision vs. recall, we consider the mean Average Precision (mAP), which is the portion of area under a precision recall-curve: the mAP value is always smaller or equal to 1.\\

\paragraph{e-Measure} The e-Measure (eM) derives from the precision and recall values for a pre-defined number of retrieved items (32 in our settings), \cite{Rijsbergen1979,princ}. Given the first 32 items for every query, the e-Measure is defined as $e=\frac{1}{\frac{1}{P}+\frac{1}{R}}$, where $P$ and $R$ represent the precision and recall values over them. 
\\

\paragraph{Nearest Neighbour, First Tier and Second Tier} These evaluation measures aim at checking the fraction of models in the query's class also appearing within the top $k$ retrievals. Here, $k$ can be 1, the size of the query's class, or the double size of the query's class. Specifically, for a class with $|C|$ members, $k=1$ for the Nearest Neighbour (NN), $k =|C|-1$ for the First Tier (1T), and $k = 2(|C| - 1)$ for the Second Tier (2T).  Note that all these values necessarily range from 0 to 1. In our this contest, we estimate the NN, FT, ST, and e values using the tools provided in the Princeton Shape Benchmark \cite{princ}.\\

\paragraph{Average dynamic recall} The idea is to measure how many of the items that should have appeared before or at a given position in the result list actually have appeared. The Average Dynamic Recall (ADR) at a given position averages this measure up to that position. Precisely, we adapt the definition of the ADR to our four level BLAST classification, slightly modifying the definition used in previous datasets equipped with a multi-level classification, such as \cite{shrec14,TVC2016}. For a given query let $A$ be the extremely similar (SR) items, $B$ the set of highly related (HR) items, and let  $C$ be the set of similar (MR) items. Obviously $A\subseteq B \subseteq C$. The ADR is computed as:
\begin{equation*}
\textrm{ADR}=\frac{1}{|C|}\sum_{i=1}^{|C|}r_i,
\end{equation*}
where $r_i$ is defined as:
\small
\begin{equation*}
r_i=\begin{cases}
\frac{|\{\textrm{SR items in the first $i$ retrieved items}\}|}{i},& \textrm{if $i\leq |A|$;}\\
\frac{|\{\textrm{HR items in the first $i$ retrieved items}\}|}{i},& \textrm{if $|A| < i\leq |B|$;}\\
\frac{|\{\textrm{MR items in the first $i$ retrieved items}\}|}{i},& \textrm{if $i> |B|$.}
\end{cases}
\end{equation*}
\normalsize
\paragraph{Normalized discounted cumulated gain} For its definition we assume that items with highest similarity score according to the BLAST classification are more useful if appearing earlier in a search engine result list (i.e. are first ranked); and, the higher their level of similarity (extremely similar, highly related, similar and dissimilar) the higher their contribution, and therefore their gain. As a preliminary concept we introduce the \emph{Discounted Cumulated Gain (DCG)}. Precisely, the DCG at a position $p$ is defined as:
\begin{equation*}
\textrm{DCG}_{p} = \textrm{rel}_{1} + \sum_{i=2}^{p} \frac{\textrm{rel}_{i}}{\log_{2}(i)},
\end{equation*}
with $\textrm{rel}_i$ the graded relevance of the result at position $i$. Obviously, the DCG is query-dependent. To overcome this problem, we normalize the DCG to get the Normalized Discounted Cumulated Gain (NDCG). This is done by sorting elements of a retrieval list by relevance, producing the maximum possible DCG till position $p$, also called \emph{ideal DCG (IDCG)} till that position. For a query, the NDCG is computed as 
\begin{equation*}
\textrm{NDCG}_{p}=\frac{\textrm{DCG}_{p}}{\textrm{IDCG}_{p}}.
\end{equation*}
It follows that, for an ideal retrieval system, we would have $\textrm{NDCG}_{p}=1$ for all $p$.

\subsubsection{Classification performance measures.}\label{cMeasures}
A set of popular performance metrics in statistical classification is derived by the so-called \emph{confusion matrix}~\cite{Kuhn:2018}.  A confusion matrix is a square matrix whose order equals the number of classes in the dataset (in our case, in the test set). The diagonal element $\text{CM}(i,i)$ gives the number of items (i.e. molecular surfaces, in our context) which have been correctly predicted as elements of class $i$. On the contrary, off-diagonal elements count items that are mislabeled by the classifier: in other words, $\text{CM}(i,j)$, with $j\neq i$, represents the number of items wrongly labeled as belonging to class $j$ rather than to class $i$. The classification matrix $\text{CM}$ of an ideal classification system is a diagonal matrix, so that no misclassification occurs.\\

\paragraph{Sensitivity and specificity} These statistical measures are among the most widely used in diagnostic test performance. Sensitivity, also called \emph{True Positive Rate} (TPR), measures the proportion of positives which are correctly identified as such (e.g., the percentage of dogs correctly classified as dogs). Specificity, or \emph{True Negative Rate} (TNR), measures the proportion of negatives which are correctly identified as such (e.g., the percentage of non-dogs correctly classified as non-dogs). A perfect classifier is $100\%$ sensitive and $100\%$ specific. \\

\paragraph{Positive and negative predicted values} Specificity and sensitivity tell how well a classifier can identify true positives and negatives. But what is the likelihood that a test result is a true positive (or true negative) rather than a false-positive (or a false-negative)? \emph{Positive Predictive Rate} (TPR) measures the proportion of \emph{true positives} among all those items classified as positives. Similarly, \emph{Negative Predictive Rate} (NPR) measures the proportion of \emph{true negatives} among all those items classified as negatives.\\

\paragraph{Accuracy} This metric measures how often the classifier is correct: it is the ratio of the total number of correct predictions to the total number of predictions.\\

\paragraph{$F_1$ score} It takes into account both PPV and TPR, by computing their harmonic mean: this allows to consider both false positive and false negatives. Therefore, it performs well on an imbalanced dataset.

\section{Description of methods}
\label{sec:methods}

Eight groups from five different countries registered to this track. Five of them proceeded with the submission of their results. Each participant was allowed to send us up to three runs for each task, in the form of a dissimilarity matrix per run. All but one submitted three runs per task; one participant delivered three runs for Task A and one for Task B. Overall, Task A has gathered $15$ runs, while Task B has $13$ runs. 

In the following, we will denote the methods proposed by the five participants as P1, P2, $\dots$, P5.

Specifically,
\begin{itemize}
    \item method P1 has been proposed by Andrea Giachetti;
    \item method P2 has been proposed by Tunde Aderinwale, Charles Christoffer, Woong-Hee Shin, and Daisuke Kihara;
    \item method P3 has been proposed by Yonghuai Liu, Ekpo Otu, Reyer Zwiggelaar, and David Hunter;
    \item method P4 has been proposed by Evangelia I. Zacharaki, Eleftheria Psatha, Dimitrios Laskos, Gerasimos Arvanitis, and Konstantinos Moustakas;
    \item method P5 has been proposed by Huu-Nghia Nguyen, Tuan-Duy Nguyen, Vinh-Thuyen Nguyen-Truong, Danh Le-Thanh, Hai-Dang Nguyen, and Minh-Triet Tran.
\end{itemize}
Lastly, Andrea Raffo, Ulderico Fugacci, Silvia Biasotti, and Walter Rocchia have been the organizers of the SHREC 2021 track on retrieval and classification of protein surfaces on the basis of their geometry and physicochemical properties.

The remaining part of this section is devoted to describe in detail the five proposed methods.

\subsection{P1: Joint histograms of curvatures, local properties and area projection transform}
\label{subsec:P1}

\subsubsection{Adopted descriptors and overall strategy} 
The proposed approach is based on the estimation of simple surface- and volume-based shape descriptors, and on their joining with the local surface properties. 
In a previous contest \cite{gravel20}, it has been shown that simple joint (2D) histograms of min/max curvatures (JHC) are extremely effective in characterizing patterns of elements with approximate spherical symmetry and variable size. On the other hand, in a past contest on protein retrieval \cite{shrec2019}, we used a volumetric descriptor called the Histograms of Area Projection Transform (HAPT) \cite{giachetti2012radial} to characterize radial symmetries at different scales providing good results.

In method P1, we tested both descriptors and their combination to evaluate the similarity of the shapes included in the test dataset. 
Furthermore, having the local information on the physicochemical properties, we can improve the characterization creating joint (3D) histograms counting elements with selected properties in a space characterized by 2 curvature axes and a ``property'' dimension. 
Finally, having a labelled dataset, we evaluated the possibility of applying to the descriptor a trained dimensionality reduction based on Linear Discriminant Analysis, e.g., projecting the high-dimensional joint histogram descriptors onto a lower dimensional space maximizing the separation of the training set classes.

\begin{figure*}[htb!]
\centering
  \includegraphics[width=0.85\textwidth]{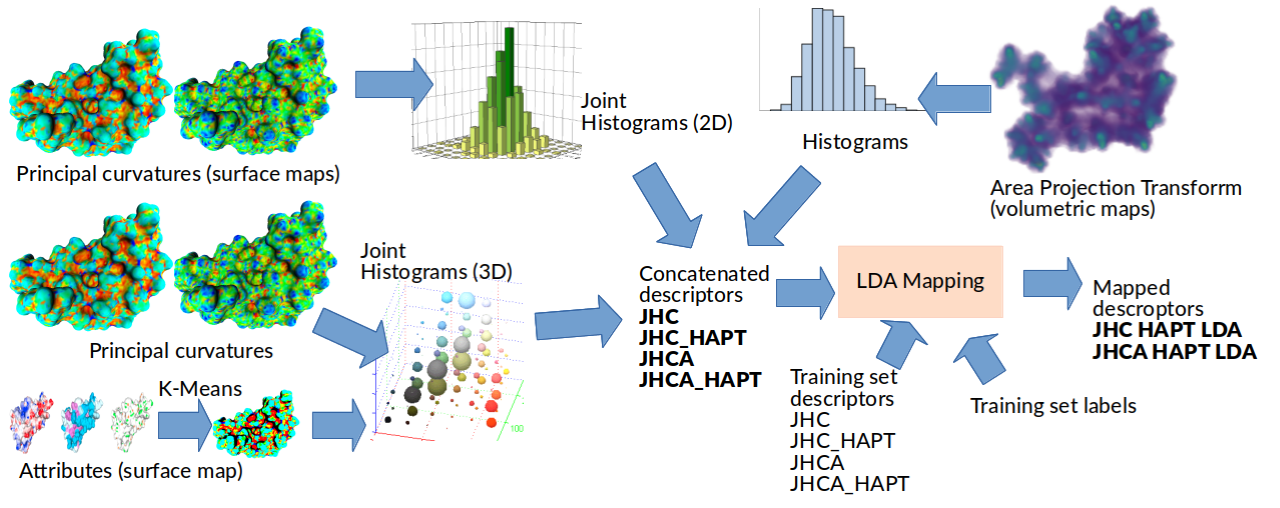}
  \caption{A graphical representation of the strategy adopted by group P1 to obtain untrained and trained descriptors. Joint histograms are obtained from the surface-based descriptors, shape only (principal curvature) and attributes (labels derived from $K$-means clustering of the provided vectors, $K=50$). Simple histograms are obtained from the volumetric symmetry descriptor (APT). Histograms are concatenated in different ways to obtain the different descriptors compared with the Jeffrey Divergence. Trained descriptors are obtained by estimating LDA mappings of the descriptors of the training set shapes with known labels. The mappings are used to perform dimensionality reduction on the test set elements.}\label{fig:P1}
\end{figure*}

A visual description of the pipeline adopted in method P1 is depicted in Figure \ref{fig:P1}.

\subsubsection{Task A}
\paragraph{Joint Histograms of Curvature} 
A basic technique to distinguish surfaces endowed with multiple spherical bumps is to measure curvature values.
Minimum and maximum curvatures have been estimated on the mesh vertices at two different scales. The ranges of min curvature and max curvature have been subdivided in 10 bins estimating the joint histogram (size 100). Concatenating the two joint histograms corresponding to the two smoothing levels a final descriptor with 200 elements is obtained. Histograms are compared with Jeffrey divergence \cite{puz99} to obtain dissimilarity matrices.

\paragraph{Histograms of Area Projection Transform} 
For the mathematical formulation of the technique please refer to the original paper \cite{giachetti2012radial}. In a few words, the internal part of the shape is discretized on a regular grid and for each voxel and for a set of discrete radius values $r$, it is counted how much of the object surface can be considered approximately part of a sphere of radius $r$ centered in the voxel.
Looking at some example protein shapes, we decided to apply the technique with voxel size 0.3 and 9 values of $r$ ranging from 0.6 to 3.0 with step 0.3. Then, we binarized the histograms with 12 bins and concatenated the histograms computed at the 9 scales. 
This resulted in a HAPT descriptor with 108 elements which have been finally concatenated into a 308-element shape descriptor capturing the distribution of curvatures on the surface and radial symmetry inside the volume. The Jeffrey divergence \cite{puz99} was used to obtain from them the dissimilarity matrices.

\paragraph{Trained descriptors}
As the data comes with a labelled training set, it is possible to use it to train the dissimilarity metric to maximally separate elements with different labels. This has been achieved by using the Fisher's Linear Discriminant Analysis \cite{fisher1936use}, projecting the original descriptor onto a $C-1$ dimensional space (where $C$ is the number of classes in the training set) maximizing the ratio of the variance between the classes to the variance within the classes.
LDA mapping for high-dimensional descriptors has been trained on the training set and used on the test data to evaluate the effectiveness of the approach.

For Task A, three different runs adopting method P1 and generated as it follows have been proposed.
\begin{itemize}
    \item {\bf Run 1 (JHC):} generated by the Joint Histograms of Curvature (two different levels of smoothing) compared with the Jeffrey divergence.
    \item {\bf Run 2 (JHC\_HAPT):} generated by concatenating Joint Histograms of Curvature and Histograms of Area Projection Transform compared with Jeffrey divergence.
    \item {\bf Run 3 (JHC\_HAPT\_LDA):} generated by concatenating JHC and HAPT descriptors mapped with trained LDA projection.
\end{itemize}

\subsubsection{Task B}
In method P1, we did not use any prior related to the knowledge of the meaning of the attributes associated with the mesh vertices and we just considered them as generic components of a 3D feature space. The adopted strategy has been to partition this feature space in a set of regions, and estimating for each model histograms counting the number of vertices with features falling in each region. 
To determine a reasonable partitioning, we just applied $K$-means clustering with $K=50$ to the all the dataset vertex attributes and extract corresponding Voronoi cells.
Using 50 cells, a 50-elements histogram to describe shape features is retrieved. Joining the attribute dimension to the two curvature dimensions, a 3D joint histogram per vertex with $10\times 10 \times 50 = 5,000$ elements is obtained.
These Joint Histograms of Curvatures and Attributes JHCA can be directly compared with the Jeffrey divergence.
However, we also tested the combination of JHCA with HAPT and the LDA-based dimensionality reduction.

For Task B, three different runs adopting method P1 and generated as it follows have been proposed.
\begin{itemize}
    \item {\bf Run 1 (JHCA):} generated by the Joint Histograms of Curvature (single smoothing level) and Attributes (50 centroids) compared with the Jeffrey divergence.
    \item {\bf Run 2 (JHCA\_HAPT):} generated by concatenating Joint Histograms of Curvature and Attributes and Histograms of Area Projection Transform compared with Jeffrey divergence.
    \item {\bf Run 3 (JHCA\_HAPT\_LDA):} generated by concatenating JHCA and HAPT descriptors mapped with trained LDA projection.
\end{itemize}

\subsubsection{Computational aspects}
Experiments have been perfomed on a laptop with an Intel\textregistered CoreTM i7-9750H CPU running Ubuntu Linux 18.04.
The estimation of the descriptors JHC and JHCA took on average 1.5 seconds per model using Matlab code, while the estimation of the HAPT descriptor took on average 15 seconds. Other required operations included: the descriptor comparison whose computation time was negligible; the training of the dissimilarity metric and the LDA mapping both implemented using Matlab and requiring 1 minute in the worst case and 0.1 seconds, respectively; the partitioning based on $K$-means clustering whose took approximately 30 minutes.

\subsection{P2: 3D Zernike descriptor}
\label{subsec:P2}

\subsubsection{Adopted descriptors and overall strategy} 
The approach adopted in method P2 is based on the 3D Zernike Descriptor (3DZD). 3DZD is a rotation-invariant shape descriptor derived from the coefficients of 3D  Zernike-Canterakis polynomials \cite{Canterakis1999}.  

3DZD descriptors are adopted as input of a neural network which will return a prediction of the similarity between any pair of proteins.
In a nutshell, neural networks (NN) are a class of tools enabling the estimation of a desired function (in the current case, the similarity between proteins) inspired by the biology of a brain.
NNs can be are typically  represented by a directed weighted graph consisting of nodes, called neurons and subdivided into layers, and edges connecting neurons of different layers.
In a NN, the leftmost layer consists of the so-called input neurons, while the rightmost nodes are called output neurons. In between, there are the hidden layers. In case a NN has at least two hidden layers, it is called a deep neural network.
Each neuron of a layer takes a series of inputs, depending on the edges pointing to it, and transmits an activation value by the edges linking the considered node to a different neuron multiplying this value by the weight of the edge. 
Input neurons receive the features of the input variables and pass them to the next layers while, the activation values of output neurons will form the output of the NN.
The desired function is obtained through a training process of the NN in which the weights are attained minimizing a loss function.

\begin{figure*}[htb!]
\centering
  \includegraphics[width=0.95\textwidth]{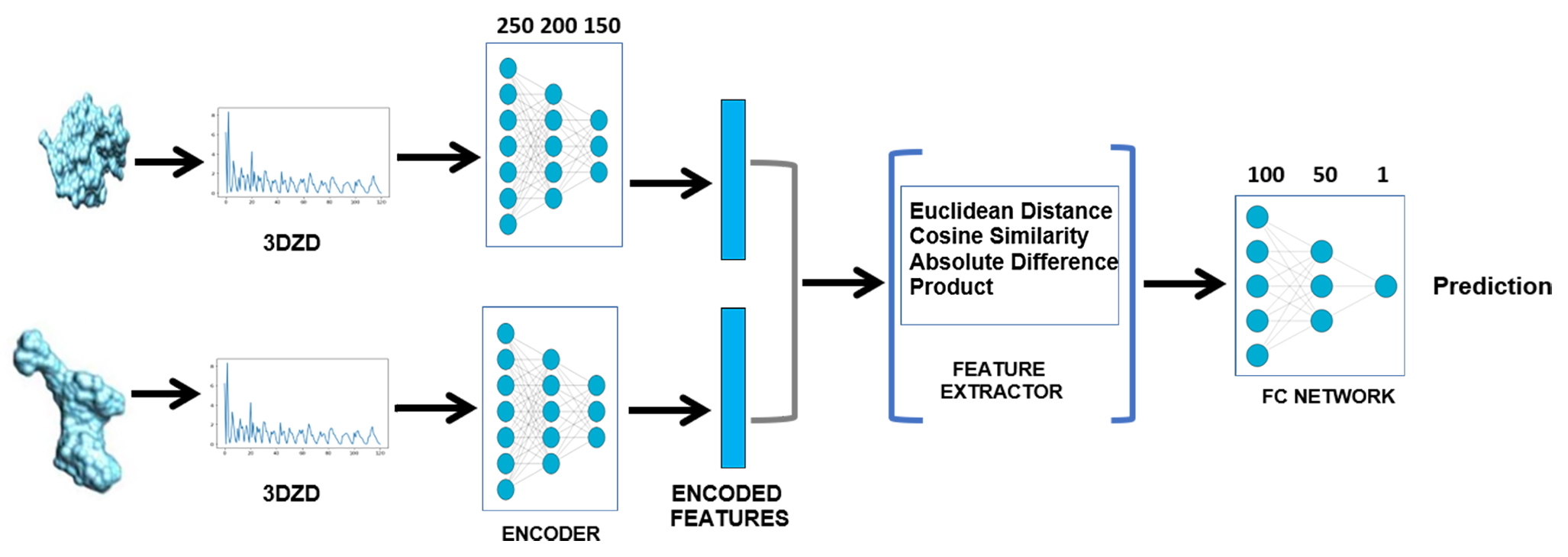}
  \caption{A graphical representation of the two types of neural network adopted in method P2 to measure the dissimilarity of protein shapes encoded via the 3DZDs. The Extractor and the EndtoEnd models differ by presence of the feature comparator layer (depicted inside the blue square brackets).}\label{fig:P2}
\end{figure*}

We trained two types of neural network, visually depicted in Figure \ref{fig:P2}, to output a score that measures the dissimilarity between a pair of protein shapes, encoded via the 3DZDs.

\begin{itemize}
\item The first framework (Extractor model) was previously used in a SHREC track on multi-domain protein shape retrieval, see \cite{shrec2020}. 
The network is structured into multiple layers: 
an encoder layer, which converts 3DZD to a vector of 150 features, has 3 hidden units of size 250, 200, and 150, respectively; a feature comparator layer that computes the Euclidean distance, the cosine distance, the element-wise absolute difference, and product; and a fully connected layer with 2 hidden units of size 100 and 50, respectively.
There are multiple hidden units in each layer. The ReLU activation function is used in all layers, except for the output of the fully connected layer where the sigmoid activation function has been preferred, This choice allows to interpret the output as the probability, for any pair of proteins, to be in the same class.

\item The second framework (EndtoEnd model) is similar to the first one, except for the removal of the feature comparator layer. The output of the encoder layer directly flows into the fully connected layer and the network is trained end-to-end.
\end{itemize}
The network was trained on the training set of $3,585$ protein structures that was provided by the organizers. The training set was further split into a training set and a validation set, by using respectively $80\%$ and $20\%$ of data (i.e. $2,868$ conformations for training and the remaining $717$ for validation). From the $717\times 717$ classification matrix of the validation set, $10,436$ protein pairs were extracted for the purpose of network validation. 

A third attempt is made via a simple Euclidean model, where the Euclidean distance between pairs of proteins has been computed directly from the generated 3DZD of the pairs. 

\subsubsection{Task A}
The performance of the networks on the validation set was used to determine models to use for inference on the test set. Training for Task A was performed on the 3DZDs of shape files only.

For Task A, three different runs adopting method P2 and generated as it follows have been proposed.
\begin{itemize}
    \item {\bf Run 1 (extractor):} generated by the  Extractor model.
    \item {\bf Run 2 (extractor\_e2e):} generated by the average between Extractor and EndtoEnd models.
    \item {\bf Run 3 (extractor\_eucl):} generated by the average between Extractor and Euclidean models.
\end{itemize}

\subsubsection{Task B}
As for Task A, model selection was carried out on the validation set.  Training was performed on input files that concatenate 3DZD of shape with 3DZDs of the three physicochemical properties.

For Task B, three different runs adopting method P2 and generated as it follows have been proposed.
\begin{itemize}
    \item {\bf Run 1 (extractor):} generated by the Extractor model.
    \item {\bf Run 2 (extractor\_e2e):} generated by the average between Extractor and EndtoEnd models.
    \item {\bf Run 3 (extractor\_e2e\_eucl):} generated by the average between Extractor, EndtoEnd, and Euclidean models.
\end{itemize}

\subsubsection{Computational aspects}
For each protein in the dataset, we performed some pre-processing step to convert the OFF and TXT files provided by the organizers. The mesh and property files were converted to a volumetric skin representation (the Situs file) where points within 1.7 grid intervals were assigned with values interpolated from the mesh \cite{sael2008fast}. For the electrostatic features, the interpolated values were the potentials at the mesh vertices. For the shape features, a constant value of 1 was assigned to grids which overlap with the surface.  The resulting Situs files were then fed into the EM-Surfer pipeline \cite{Esquivel2015} to compute 3DZD. It took approximately $12-13$ minutes to pre-process each file. Generating the 3DZD descriptors took averagely $8$ seconds for each protein on an Intel\textregistered Xeon\textregistered CPU E5-2630 0 @ 2.30GHz.

For Task A, training the extractor model took averagely 6 hours and the EndtoEnd model took about 11 hours. For Task B, training the extractor model took about 9 hours and approximately 14 hours for the EndtoEnd model. Training was performed on a Quadro RTX 800 GPU.

The 3DZD model took averagely $0.22$ seconds to predict the dissimilarity between two proteins, using TitanX GPU. The Euclidean model took averagely $0.17$ seconds per prediction. Finally, the averaging of the three matrices was virtually instant and negligible.

\subsection{P3: Hybrid Augmented Point Pair Signatures and Histogram of Processed Physicochemical Properties of Protein molecules}
\label{subsec:P3}

\subsubsection{Adopted descriptors and overall strategy}
Considering the twofold nature of this challenge, in P3 we adopted two separate retrieval strategies for the two different tasks.
For Task A, we used the Hybrid Augmented Point Pair Signature (HAPPS)~\cite{9071111}, a 3D geometric shape descriptor.
For Task B, we adopted the Histogram of Processed Physicochemical Properties of Protein molecules descriptor following an Exploratory Data Analysis (HP4-EDA). Both the strategies rely on traditionally hand-crafted feature extraction from the respective datasets, using the knowledge-based approach (i.e. non-learning nor data-driven approach).

\begin{figure}[htb!]
\centering
  \includegraphics[width=0.98\columnwidth,trim={0.10cm 0 0.10cm 0}, clip]{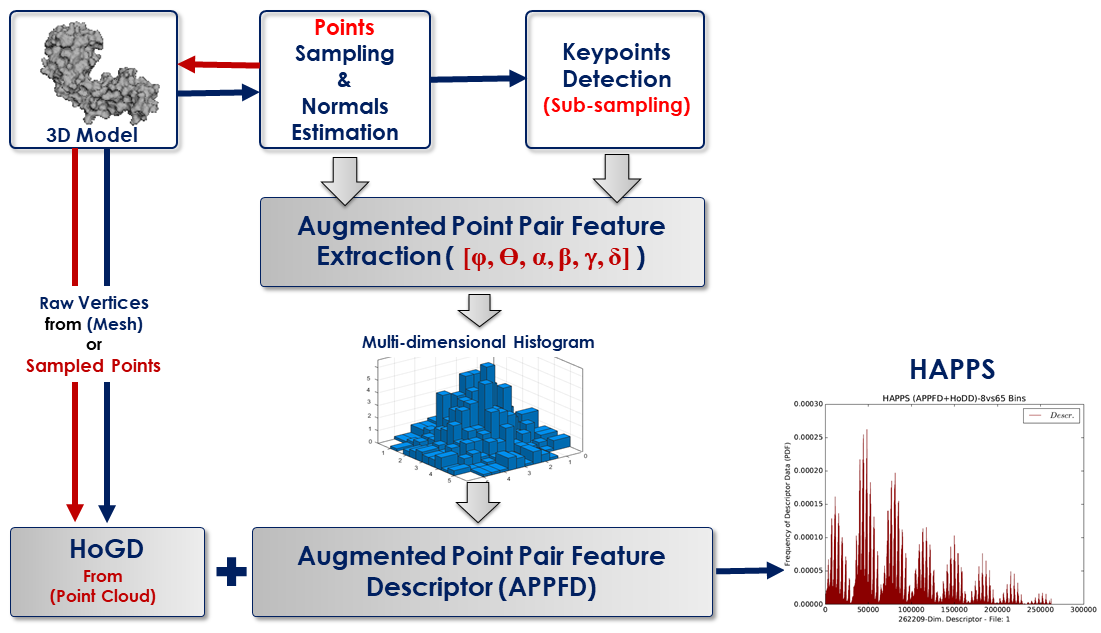}
  \caption{A graphical representation of the strategy adopted in method P3 for Task A.}\label{fig:P3A}
\end{figure}

\begin{figure}[htb!]
\centering
  \includegraphics[width=0.98\columnwidth,trim={0.10cm 0 0.10cm 0}, clip]{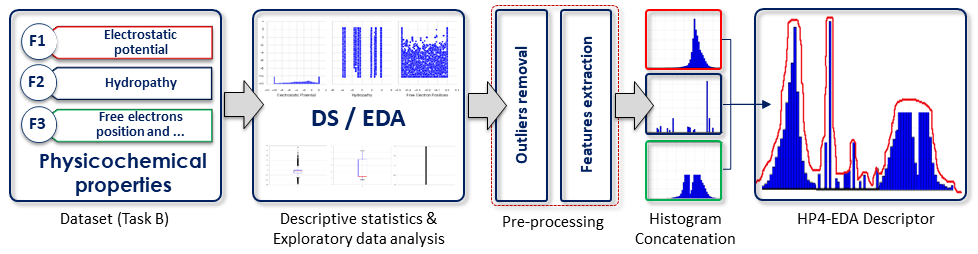}
  \caption{A graphical representation of the strategy adopted in method P3 for Task B.}\label{fig:P3B}
\end{figure}

The goal of the proposed methods (HAPPS and HP4-EDA) is to provide simple, efficient, robust and compact representations, describing both the 3D geometry and physicochemical properties of protein surfaces, using statistically-based descriptors. Visual descriptions of the pipelines adopted in method P3 for Tasks A and B are depicted in Figures~\ref{fig:P3A} and~\ref{fig:P3B}, respectively.

\subsubsection{Task A}
Each 3D geometrical protein surface in this challenge contains an average of $35,000$ vertices and $70,000$ triangular faces. The HAPPS method (first introduced in~\cite{9071111}) involves a combination of local and global descriptors. Specifically, the Augmented Point Pair Feature Descriptor (APPFD), and the Histogram of Global Distances (HoGD). 
% The proposed method is claimed to be particularly interested in providing a robust, compact, and accurate representation of protein structures with as low as $3,500$ to $4,500$ points sampled from each surface.
The proposed HAPPS method is particularly interested in providing a robust, compact, and accurate representation of protein structures with as low as $N$ points (i.e., $[N \times 3]$) sampled from the triangular mesh surface of each input protein model, where $N = 3,500$ and $N = 4, 500$.

\paragraph{Histogram of Global Distances (HoGD)}
This descriptor involves binning a set of normalized vectors $\delta_i = \|P_c - p_i\|$ between the centroid $P_c$ of a given 3D object to all other points, $p_i$ on its surface into a 1D histogram with $\sqrt{P} \approx 65$ bins, normalized to give HoGD, following some pre-processing steps, where $p_i \in P$ and $P$ is a 3D point cloud object, with a number of points, $N = 3,500$ or $N = 4,500$.
Such normalized vectors $\delta_i$ are regarded as global features whose distribution (histogram) is capable of expressing the configuration of the entire shape relative to its centroid, and is a rich description of the global structure of the shape. Pre-processing also involves applying uniform scale, $S$ in all direction to all points in $P$, such that the root mean square (RMS) distance of each point to the origin is 1, and centering $P$ on its centroid, i.e. $P = p_i - P_c$.

\paragraph{Augmented Point Pair Feature Descriptor (APPFD)}
The APPFD describes the  local  geometry around a point, $p_i = [p_{ix}, p_{iy}, p_{iz}]$ in $P$. Its computation involves a 5-step process: (i) pointcloud sampling and normals estimation; (ii) keypoint, $p_{k_i}$ determination; (iii) local surface region (i.e. LSP), $P_i$ selection; (iv) Augmented Point Pair Feature (APPF) extraction per LSP; 
(v) bucketing of locally extracted $6$-dimensional APPF into a multi-dimensional histogram, with the number of bins, $b_{APPFD} = 8$ in each feature-dimension which is then flattened and normalised to give $8^6 = 262,144$-dimensional single local descriptor (APPFD) per 3D shape.

Finally, HoGD is combined with APPFD to give HAPPS, with a final feature vector of dimension $262,144 + 65 = 262,209$ (see Figure~\ref{fig:P3A}). For more details regarding the HoGD, APPFD and HAPPS algorithms, the reader is referred to~\cite{9071111, shrec2020}.

The APPFD is characterised by four key parameters, which are $r$, $vs$, $b_{APPFD}$, and $N$. $r$ is the radius value used by the nearest-neighbour algorithm to determine the number of points in a LSP, and is directly proportional to the size of LSP. The voxel-size parameter, $vs$ determines the size (big or small) of an occupied voxel-grid by the pointcloud down-sampling algorithm~\cite{zhou2018open3d}. It is inversely proportional to the number of sub-sampled points (used as keypoints). $r$ and $vs$ influence the overall performances of the APPFD/HAPPS.  

For Task A, three different runs adopting method P3 and generated as it follows varying the values of $r$, $vs$, and $N$ have been proposed.
\begin{itemize}
    \item {\bf Run 1 (HAPPS):} generated by choosing $r=0.40$, $vs=0.20$, and $N=4,500$.
    \item {\bf Run 2 (HAPPS):} generated by choosing $r=0.50$, $vs=0.30$, and $N=4,500$.
    \item {\bf Run 3 (HAPPS):} generated by choosing $r=0.50$, $vs=0.30$, and $N=3,500$.
\end{itemize}

Parameters $b_{APPFD} = 8$ and $b_{HoGD} = 65$ remained the same for all three runs. Overall, the Cosine distance metric between final vectors gave good approximation of the similarity between the HAPPS for Task A datasets.

\subsubsection{Task B}
The HP4-EDA method involves a descriptive statistics (DS) of the 3-dimensional physicochemical variables or properties, following exploratory data analysis (EDA) of each of these properties.

Let the three physicochemical properties of the dataset in Task B be denoted as $f_1$, $f_2$, and $f_3$, for \textit{Electrostatic Potential}, \textit{Hydrophobicity}, and \textit{Position of potential hydrogen bond donors and acceptors}, respectively.
First, we carried out an in-depth EDA of the physicochemical properties to investigate their values distribution, followed by data pre-processing (majorly outliers detection and removal). Next, we investigated the performances of combining some DS, such as the mean, variance, first and third interquartile values, and correlation coefficients between these variables as a final descriptor, including the construction of histograms of these variables, post-processing (see Figure~\ref{fig:P3B}).

\paragraph{Outliers detection and removal}
Considering that the presence of outliers  would adversely affect the performance of any retrieval system, we checked for the presence of outliers in each of $f_1$, $f_2$, and $f_3$. Unlike $f_2$, the $f_1$ and $f_3$ variables contain lots of outliers with $f_3$ having almost negligible amount of useful data.
Empirically, the presence of outliers in a distribution may not necessarily make the observation a ``bad data''. For outliers detection and removal, we adopted the Interquartile Range (IQR) Score, represented by the formula $IQR = Q_3 - Q_1$, which is a measure of statistical dispersion calculated as the difference between lower ($Q_1$) and upper ($Q_3$)  percentiles.
Here, any observation that is not in the range of $(Q_1 - 1.5 IQR)$ and $(Q_3 + 1.5 IQR)$ is an outlier, and can be removed. We further investigated the effect of using $Q_1 =  10^{th}$  or $25^{th}$, and $Q_3 =  75^{th}$  or $90^{th}$ and recorded better performances with the later option where $Q_1 =  10{th}$ and $Q_3 =  90{th}$ for the training set, which parameter settings were also applied to the test data.

For Task B, three different runs adopting method P3 and generated as it follows by applying statistical description techniques for the extraction of statistical features and/or construction of final descriptors from the pre-processed data have been proposed.
\begin{itemize}
    \item {\bf Run 1 (HP4-EDA):} generated by binning each pre-processed physicochemical variable into a 1D histogram (using $150$ bins) and combining the final histograms as the final descriptor for each input physicochemical surface where matching between two descriptors is done using the Earth Mover's Distance (EMD) metric.
    \item {\bf Run 2 (HP4-EDA):} generated by binning each of the pre-processed values of $f_1$, $f_2$, and $f_3$ into a multi-dimensional histogram, with $5$ bins in each feature dimension, where the flattened and normalised histogram frequencies represent the final descriptor for a single input data and descriptors are matched using the Kullback Liebner Divergence (KLD) metric.
    \item {\bf Run 3 (HP4-EDA):} 
    generated by first normalizing each of the feature (variable) dimensions or columns, and selecting their $mean$, $variance$, $Q_1$, $Q_3$, and some correlation coefficient values between $f_1$, $f_2$, and $f_3$ to represent a single input physicochemical surface, with a total of 14-dimensional feature vector, and combining the outcome of Run 1 to have a feature vector of dimension $14 + (150 \times 3) = 464$ as a final descriptor representing a single input.
\end{itemize}

\subsubsection{Computational aspects}
For Task A, the HAPPS method has been implemented in Python 3.6 and all experiments have been carried out under Windows 7 desktop PC with Intel Core i7-4790 CPU @ 3.60GHz, 32GB RAM. 
It took on average, $0.3$ and $20.0$ seconds to sample point cloud and estimate normals from 3D mesh, and extract features and compute HAPPS, respectively. Matching $1,543 \times 1,543$ testing set HAPPS descriptors took $3,212.3$ seconds using the Cosine metric, which implies an average of $2.1$ seconds to match any two HAPPS.

For Task B, the HP4-EDA method has been implemented in Python 3.6 and all experimental run have been performed on 64-bit Windows 10 notebook, Intel Core(TM) i3-5157U CPU @ 2.50GHz, 8GB RAM.
The extraction of the features and the computation of the HP4-EDA descriptors took an average of $0.01$ seconds.
Additionally, it took $322.5$ seconds to match $1,543 \times 1,543$ HP4-EDA descriptors for the testing dataset with both the \textit{EMD} and \textit{KLD} distance metrics, an average of $0.2$ seconds to match two HP4-EDAs.

\subsection{P4: Global and Local Feature (GLoFe) fit}
\label{subsec:P4}

\subsubsection{Adopted descriptors and overall strategy} 
\begin{figure*}[htb!]
\centering
  \includegraphics[width=0.75\textwidth, trim={1cm 0 1.5cm 0}, clip]{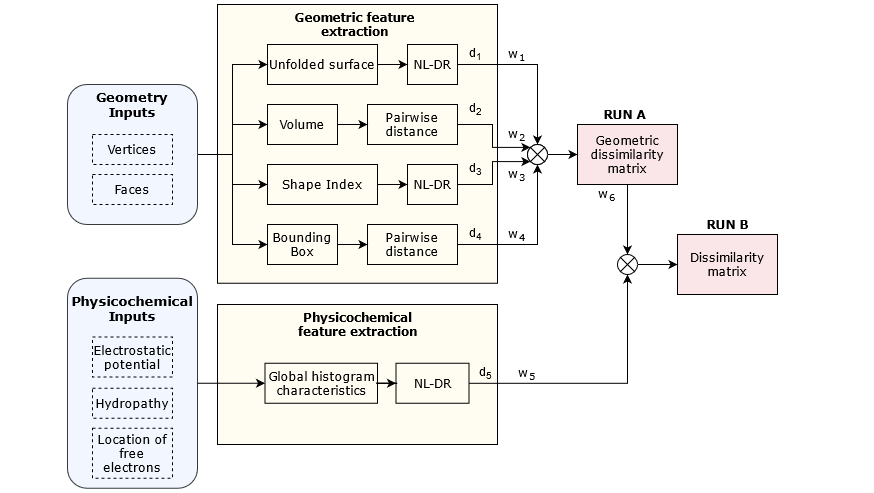}
  \caption{Processing pipeline of the strategy adopted in method P4. It includes extraction of multiple global and local shape descriptors, non-linear dimensionality reduction (NL-DR), calculation of pairwise distances, and fusion of the obtained dissimilarity matrices.}\label{fig:P4}
\end{figure*}

The strategy adopted in method P4 is based on a direct approach. Depending on the track, a collection of local and global features have been calculated. For each feature vector $f$, a dissimilarity matrix $d_f$ has been computed, while their weighted combination produced the total dissimilarity matrix. The pipeline of the adopted strategy is depicted in Figure \ref{fig:P4}.

The features used for Task A have been 3D shape descriptors from surface unfolding, a shape index describing local curvature, the volume of each protein, and the size of an encompassing bounding box.
Differently, the ones taken into account for Task B have been global histogram characteristics of the three provided physicochemical properties.

Using the Euclidean distance normalized by the standard deviation, we calculated the matrices $d_f$ that indicate the pairwise distance between pairs of observations, for each geometric or physicochemical feature $f$. For some of them, we slightly modified the distance value by examining the inverse consistency of mapping. Specifically, for each protein structure $j$, we identified the first $k$-Nearest Neighbours, i.e. $k \in N(j)$, where $N$ denotes the set of neighbours. For each $k$, we examined whether $j \in N(k)$. If true, the forward and inverse retrieval was consistent, thus the certainty in estimation of pairwise similarity between $j$ and $k$ was considered high. In this case, we increased by a factor, $\alpha$, the dissimilarity value in $d(j,k)$. The values used in the experiments were $|N| = 5$ for both forward and inverse mapping and $\alpha = 0.3$.
The final distance matrix $d$ was constructed from a linear combination of the individual dissimilarity matrices, i.e. $d=\sum w_f \cdot d_f$. where $w_f$ is a weight determining the contribution of each feature $f$. The weights have been empirically estimated by optimizing the classification performance on the training set.

The classification performance was assessed on the training set in order to allow the optimization of the several (hyper)parameters of the methodology. As evaluation criterion, the percentage of proteins for which the first (or correspondingly the second) closest neighbour belonged to the same class has been adopted.

\subsubsection{Task A}
For Task A, four local and global geometric features have been extracted.

\paragraph{3D shape descriptors from surface unfolding}
In order to remove translation and rotation differences across the protein structures, for each data matrix $V^{(j)}$ associated with protein $j$, we performed Principal Component Analysis (PCA) on the mean centered data and replace them by their projection in the principal component space. This results in globally normalized surface data. Since the vertex coordinates belong to 2D surfaces lying in the 3D space, a dimensionality reduction technique to ``unfold'' the manifold and embed into a 2-dimensional space has been applied. 
For this purpose, we used Locality Preserving Projections (LPP) \cite{he2004locality} due to the algorithm’s stability, high performance, and mainly its capability to preserve local (neighbourhood) structure. The embedded data $Y^{(j)}$ were calculated as $Y^{(j)} = V^{(j)} \cdot W^{(j)}$
where $W^{(j)}$ is the transformation matrix that maps the set of vertices of protein $j$ from $\mathbb{R}^3$ to $\mathbb{R}^2$. Scores in $Y^{(j)}$ cannot be directly used for protein retrieval because of their high number and variable length across structures. Thus, we used as data representation the multi-variate kernel density estimate, $p_{Y^{(j)}}\in \mathbb{R}^{b_1\times b_2}$ , where $b_1$ and $b_2$ are the number of bins (common for all proteins) for the two columns of $Y^{(j)}$, respectively. Then, since the obtained kernel density maps were sometimes
anti-symmetric, we augmented the whole dataset by horizontally and vertically flipping $p_{Y^{(j)}}$ resulting in 4 replicates for each protein. Finally, the 4 replicates of $p_{Y^{(j)}}$ for all proteins were linearized and concatenated in a big data matrix, in order to learn the manifold of different proteins and their conformations. The non-linear dimensionality reduction technique t-distributed Stochastic Neighbour Embedding (t-SNE) \cite{hinton2002stochastic} was used to calculate 5 scores that model each protein structure in the lower dimensional space by retaining data similarity as much as possible. For indexing purposes, the distance of protein structure $j$ to some other protein structure was defined as the smallest distance across the 4 replicates.

\paragraph{Shape index}
The third incorporated geometric feature is the shape index which was part of the first pre-processing phase of MaSIF \cite{gainza2020deciphering}, a network that combines geometric and physicochemical properties into a single descriptor.
The shape index describes the shape of the surface around each vertex with respect to the local curvature, which is calculated in a neighbourhood of geodesic radius $12$\r{A} around it. The proposed neighbourhood size was chosen empirically. The shape index of $v_i$ is defined according to
$$\frac{2}{\pi}\tan^{-1}\Big(\frac{k_1 + k_2}{k_1- k_2}\Big)$$
where $k_1, k_2$ (with $k_1\geq k_2$) are the principal curvature values in $v_i$'s neighbourhood. High shape index values imply that $v_i$'s neighbourhood is highly convex while lower values indicate that is highly concave.

\paragraph{Volume}
Since the volume of a protein does not significantly change when the protein obtains a different conformation, it has been also used as a feature. Although this feature helps to reduce some possible matches, it has low specificity because the range of volumes across different protein classes overlaps for many of them. Moreover, volume cannot be accurately calculated for the very few protein structures which contain holes.

\paragraph{Global scale}
In order to characterize global scale of the protein, we fitted a bounding box on the protein surface defined by 8 vertices in the 3D space. We used as global scale descriptor the 3 eigenvalues of the square matrix produced by multiplying the $8 \times 1$ vector by its transpose.

For Task A, three different runs adopting method P4 and generated as it follows by a linear combination of the individual dissimilarity matrices of four separate geometric descriptors with weights have been proposed.
Specifically, $w_1$, $w_2$, $w_3$, and $w_4$ refer to unfolded surface, volume, shape index, and bounding box, respectively.
\begin{itemize}
    \item {\bf Run 1 (GLoFe):} generated by choosing $w_1 = 0.22$, $w_2 = 0.67$, $w_3= 0.055$, $w_4 = 0.055$.
    \item {\bf Run 2 (GLoFe):} generated by choosing $w_1 = 0.22$, $w_2 = 0.67$, $w_3= 0.055$, $w_4 = 0.055$ (without using inverse consistency).
    \item {\bf Run 3 (GLoFe):} generated by choosing $w_1 = 0.25$, $w_2 = 0.725$, $w_3= 0$, $w_4 = 0.025$ (without using inverse consistency).
\end{itemize}

\subsubsection{Task B}
For each of the three provided physicochemical properties, we extracted global histogram characteristics (first order statistics) that included mean intensity, standard deviation, mode of histogram (i.e. the most frequent intensity value), kurtosis, skewness, and energy. These six features for each of the three physicochemical properties provide global information on the distribution of gray-level intensities. Among these 18 features, the mode for the location of hydrogen bond donors and acceptors assumes the same value for all proteins and so it has been discarded. Differently, the remaining features has been merged into a 17-dimensional vector for each protein.

For Task B, three different runs adopting method P4 and generated as it follows by a linear combination of the geometric and physicochemical dissimilarity matrices with weights have been proposed.
Specifically, $w_5$ refers to the physicochemical dissimilarity matrix and $w_6$ to the geometric dissimilarity matrix produced by the corresponding run of Task A.
\begin{itemize}
    \item {\bf Run 1 (GLoFe):} generated by choosing $w5 = 0.065$, $w6 = 0.935$ (geometric dissimilarity matrix produced by Run 1 in Task A).
    \item {\bf Run 2 (GLoFe):} generated by choosing $w5 = 0.065$, $w6 = 0.935$ (geometric dissimilarity matrix produced by Run 2 in Task A).
    \item {\bf Run 3 (GLoFe):} generated by choosing $w5 = 0.075$, $w6 = 0.925$ (geometric dissimilarity matrix produced by Run 3 in Task A).
\end{itemize}

\subsubsection{Computational aspects}
The experiments on both tracks have been carried out using an AMD Ryzen 7 3700X 8-core Processor @3.59 GHz PC with 16 GB of RAM, except from the extraction of the surface unfolding and the physicochemical features which have been carried out using an Intel i5-6402P @2.80 GHz CPU with 8 GB of RAM. The software has been written in Matlab 2019b. 

The total time required for obtaining the results for each task in the test set is indicated in Table \ref{table:P4}.
In addition, an exhaustive search procedure (requiring 54 minutes) was followed to optimize the weights for fusion of the different dissimilarity matrices based on the training set. 
Notice that the total inference cost (illustrated in Table \ref{table:P4}) corresponds to the time aggregated due to sequential calculation of the different feature sets, whereas with a multi-threaded implementation it is reasonable to expect that the total computational time would be significantly decreased.

\begin{table}[htb!]
\centering
\caption{Computational times required for Tasks A and B by method P4.}
\label{table:P4}
\begin{adjustbox}{width=0.475\textwidth}
\begin{tabular}{c|c|c|c}
\hline
\textbf{Task} & \multicolumn{2}{c|}{} & \textbf{Time (mins)} \\ \hline
 \multirow{8}{*}{\textbf{A}} & \multirow{3}{*}{Surface unfolding} & $n_S^{test}=1,543$ & 986 \\ \cline{3-4} 
 &  & $n_S^{train}=3,585$ & \multicolumn{1}{c}{2,290} \\ \cline{3-4}
 &  & Augment \& DR & \multicolumn{1}{c}{83} \\ \cline{2-4}
 & \multicolumn{2}{c|}{Volume} & 961 \\ \cline{2-4} 
 & \multicolumn{2}{c|}{Shape index} & 85 \\ \cline{2-4} 
 & \multicolumn{2}{c|}{Global scale} & 284 \\ \cline{2-4} 
 & \multicolumn{2}{c|}{Dissimilarity matrix calculation} & 0.005 \\ \cline{2-4} 
 
 & \multicolumn{2}{c|}{\textbf{Total time}} & \multicolumn{1}{c}{4,689} \\ \hline
\multirow{3}{*}{\textbf{B}} & \multicolumn{2}{c|}{Physicochemical features} & 1 \\ \cline{2-4} 
 & \multicolumn{2}{c|}{Dissimilarity matrix calculation} & 0.001 \\ \cline{2-4} 
 
 & \multicolumn{2}{c|}{\textbf{Total time}} & \multicolumn{1}{c}{4,690} \\ \hline
\end{tabular}
\end{adjustbox}
\end{table}

\subsection{P5: Message-Passing Graph Convolutional Neural Networks (MPGCNNs) and PointNet}
\label{subsec:P5}

\subsubsection{Adopted descriptors and overall strategy} 
For the meshes in each 3D model of a protein surface, in method P5 we first sampled 512 points on the surfaces of the meshes based on the area of the meshes. Because a sampled point might not be an original vertex in the 3D meshes, the original physical and chemical properties are not valid for newly sampled points. To generate physical and chemical properties for a sampled point $p$, trilinear interpolation has been performed from the properties of the three vertices forming the face that $p$ is on. Then, to re-assign the topological structures for sampled points, each node has been connected with their $k$-Nearest Neighbours based on their original coordinates choosing $k=16$.

\begin{figure*}[htb!]
    \centering
    \begin{subfigure}{\linewidth}
    \includegraphics[width=\textwidth]{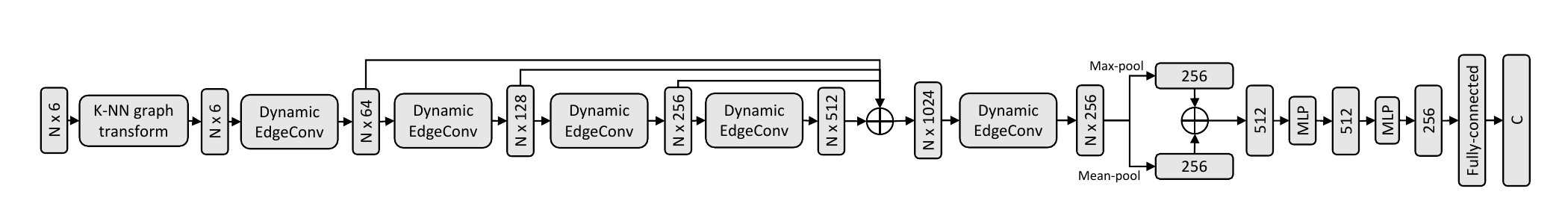}
    \vspace*{-1cm}
    \caption{Dynamic Edge Convolutional Neural Network}
    \end{subfigure}
    \begin{subfigure}{\linewidth}
    \centering
    \includegraphics[width=0.6\textwidth]{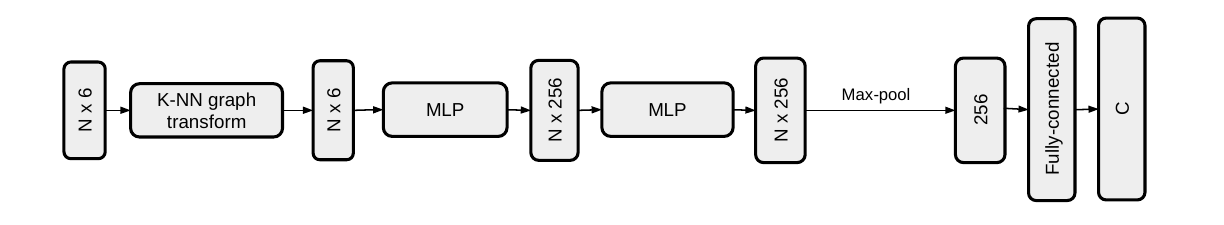}
    \vspace*{-0.5cm}
    \caption{PointNet}
    \end{subfigure}
    \caption{The summary of the graph neural networks employed in method P5 optimized over the classification of the training set. The depicted pipelines are used for the geometry with physicochemical properties tasks, where a graph has $N$ nodes and each node has $6$ initial features. For Task A, there are only $3$ initial node features, which are the spatial coordinates of points. After training, the 256-dimension vector before the fully-connected layer is used for all the tasks.}
    \label{fig:P5}
\end{figure*}

In method P5, we adopted a deep learning strategy by exploiting the availability of protein class labels to optimize the representation of protein surfaces with and without textures. 
The chosen strategy is based on the use of graph neural networks (GNNs). GNNs are deep learning based methods that, unlike classic NNs, operate on a graph domain rather than on Euclidean domains. Remarkable advancements in NNs can be achieved by including in the network hidden layers that perform convolutions. In such a case, a neural network will be called convolutional neural network (CNN) or graph neural network (GNNs) in the specific considered case.
In particular, we designed message-passing graph convolutional neural networks (MPGCNNs) with the Edge Convolution paradigm \cite{wang2019dynamic}. 
A visual description of the pipeline adopted in method P5 is depicted in Figure \ref{fig:P5}.

\paragraph{Edge convolution}
In the framework of Task A, the initial node features are the coordinates of sampled points, while in the setting of Task B, the features are concatenated tuples of coordinates and interpolated physicochemical properties. Each protein surface is represented by a $k$-Nearest Neighbours graph generated in the pre-processing step with 512 vertices. 
The module that performs the graph message-passing function is the Edge Convolution (EdgeConv) layer \cite{wang2019dynamic}. In the EdgeConv layer, the information of a vertex $i$ after layer $l$ is calculated as $x_i^{l+1}=\max_{j \in N} h(x_i^l, x_j^l)$
where $N$ is the neighbouring vertices of vertex $i$ and 
$$h(x_i^l, x_j^l)=ReLU(MLP(x_i^l\oplus x_j^l))$$
where ReLU is Rectified Linear Unit (in the implementation, LeakyReLU, a variant of ReLU, has been used), MLP is a standard multi-layer perceptron (MLP), $\oplus$ is the concatenation operator.
In the implementation of method P5, we adopted a dynamic variant of EdgeConv instead of the standard EdgeConv described above.
At each Dynamic EdgeConv layer, the each vertex's $k$-Nearest Neighbours is re-calculated in the feature space produced by the previous layer, before applying the standard EdgeConv operation. After the graph is recomputed, standard EdgeConv operation is performed.
After the pre-processing phase, the vertex features first go through 4 layers of Dynamic EdgeConv. The dimensions of output features for each vertex after these first 4 layers are 64, 64, 128, and 256, respectively. 
Then, the outputs of these 4 layers are concatenated to become a 512-dimensional vector for each vertex. This 512-dimensional vector is then fed through another Dynamic EdgeConv layer, creating the output vector with 512 dimensions $v$. The feature vector $v$ is pooled using the concatenation of the outputs of a max-pooling and a mean pooling layer to generate the first graph-level feature vector. This vector is passed through two MLP blocks with BatchNorm, Leaky-ReLU, and Dropout layers. 
The retrieval tasks can then be performed by exploiting the L2-distances between the output vectors.

\paragraph{PointNet}
Adopting the same data pre-processing procedure, we also implemented PointNet \cite{qi2017pointnet}, a well-known graph-based learning strategy for 3D data. In this network architecture, the vertex features first pass through 2 message-passing modules. Each messages-passing module contains a MLP block that uses ReLU as activation function. The module captures spatial information between a node and its neighbours by performing subtractions between each pair of the center's position and its neighbour's position. After having performed calculation, the information of a vertex $i$ after layer $l$ is calculated as $x_i^{l+1}=\max_{j \in N} h(x^l_j,p_j,p_i)$
where $N$ is the neighbouring vertices of vertex $i$ and 
$$h(x^l_j,p_j,p_i)=ReLU(MLP(x_j^l \oplus (p_j-p_i)))$$
where $p_j$ is the position of the vertex $j$ and $p_i$ is the position of the vertex $i$.
For a MLP block that the vertex features pass through, its output further goes through a ReLU function. After two MLP blocks, the feature vector is pooled using a single global max-pooling layer. Then, retrieval tasks can be performed by the same way as with the Dynamic EdgeConv strategy.

\subsubsection{Task A}

For Task A, three different runs adopting method P5 and generated as it follows have been proposed.
\begin{itemize}
    \item {\bf Run 1 (EdgeConv):} generated by the Dynamic EdgeConv strategy where distances between output vectors are L2-distances between embeddings.
    \item {\bf Run 2 (PointNet):} generated by the PointNet strategy but choosing the number of mesh surface sample points as 258 instead of 512.
    \item {\bf Run 3 (Ensemble):} generated by taking the weighted average of embedding distances from the above Dynamic EdgeConv and PointNet embedding distances. Specifically, the distances from Dynamic EdgeConv are empirically weighted by 0.6, while those from PointNet are weighted by 0.4.
\end{itemize}

\subsubsection{Task B}
For Task B, one run adopting method P5 and generated as it follows has been proposed.
\begin{itemize}
    \item {\bf Run 1 (EdgeConv):} generated by the Dynamic EdgeConv strategy where the concatenation of spatial coordinates and properties made up of initial vertex features. The distances between output vectors are L2-distances between embeddings.
\end{itemize}

\subsubsection{Computational aspects}
All of the methods have been implemented in Python 3.8, using Pytorch \cite{paszke2019pytorch} and Pytorch Geometric \cite{fey2019fast} libraries.
The experiments have been carried out a machine with an Intel Core i7-8700K 6-core CPU Processor @3.70 GHz PC with 32 GB of RAM and an NVIDIA TITAN V with 12 GB of VRAM. The training and test set's embedding extraction used both the CPU and the GPU, whose time is represented in Table \ref{tab:emb-time}. The computation of distance matrix only used the CPU and it required approximately 15 minutes for Run 3 of Task A while just approximately 7 minutes for the other runs. 

\begin{table}[htb!]
    \centering
     \caption{The (approximated) training and extraction times of employed strategies in method P5.}
    \label{tab:emb-time}
    \begin{tabular}{@{}c|lcc@{}} \toprule
        \textbf{Task} & \textbf{Strategy} & \textbf{Training} & \textbf{Test Set Extr.}\\
        \midrule
        \multirow{2}{*}{A} & PointNet & 720 mins &  0.5 mins \\
         & Dyn. EdgeConv &  1,100 mins &  3 mins \\
         \midrule
         B & Dyn. EdgeConv &  1,100 mins &  3 mins \\
         \bottomrule
    \end{tabular}
\end{table}

\section{Comparative analysis}
\label{sec:results}
The performances of each run presented in Section \ref{sec:methods} are here quantitatively evaluated on the basis of the measures described in Section~\ref{sec:eval_meas}. We remind the reader that: Task A refers to the mere use of geometry, while Task B includes both geometry and physicochemical properties; for any run, the method name and its specific settings are given in Section \ref{sec:methods}. The performance measures are presented for both the PDB and BLAST classifications detailed in Section \ref{sec:ground}.

An additional analysis, reported in the supplementary material in  \ref{sec:suppl_mat}, is performed on a $3$-level BLAST-based classification: we introduce a further relaxation by merging the classes containing the same proteins or their isoforms with structures of proteins that have a significant sequence similarity, according to what introduced in Section \ref{sec:ground}. In this case, the BLAST-based decomposition of level 2 consists of 25 communities.

\subsection{Retrieval evaluation measures}
\begin{figure*}[t!]
    \begin{center}
    \begin{tabular}{|cc|cc|}
    \hline
        \rowcolor{blue!12}\multicolumn{2}{|c|}{Geometry} & \multicolumn{2}{c|}{Geometry and physicochemical characterization} \\
        \hline
         \includegraphics[scale=0.45, trim={0.1cm 0 0.7cm 0}, clip]{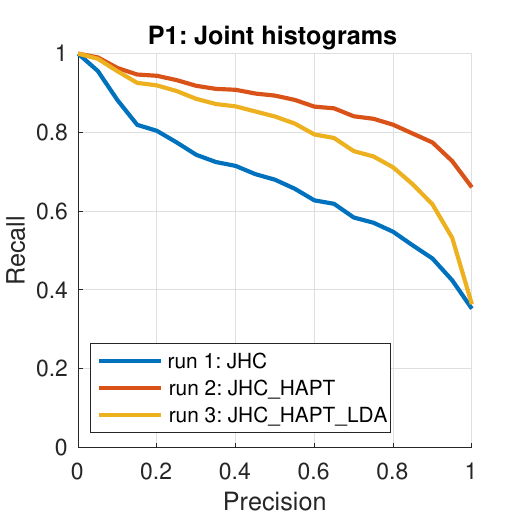}
         &
          \includegraphics[scale=0.45, trim={0.1cm 0 0.7cm 0}, clip]{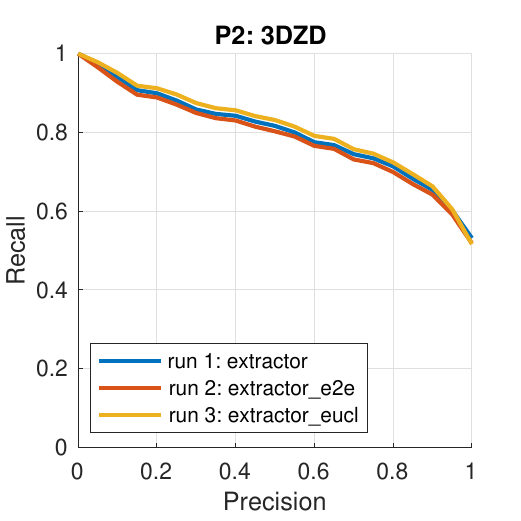}
         &
         \includegraphics[scale=0.45, trim={0.1cm 0 0.7cm 0}, clip]{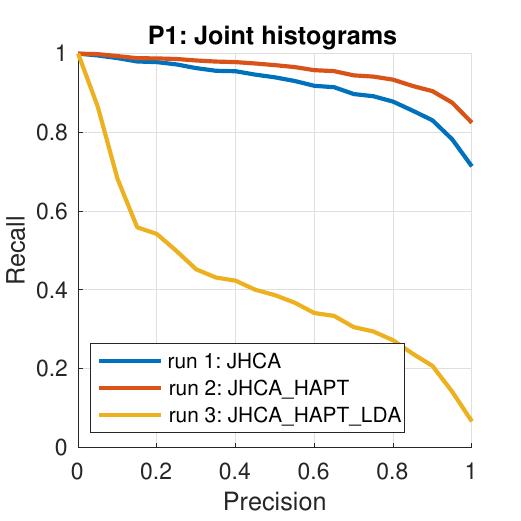}
         &
         \includegraphics[scale=0.45, trim={0.1cm 0 0.7cm 0}, clip]{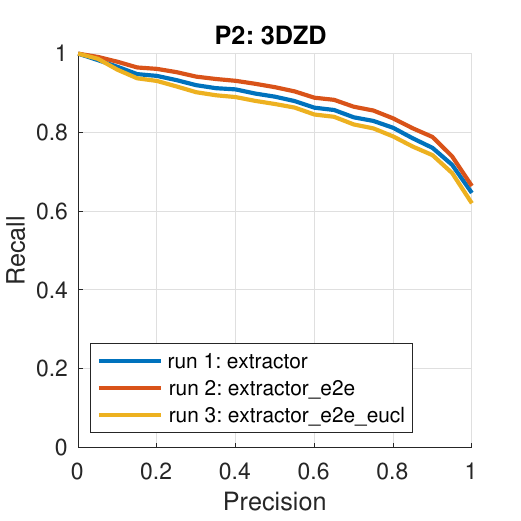}\\
         
         \includegraphics[scale=0.45, trim={0.1cm 0 0.7cm 0}, clip]{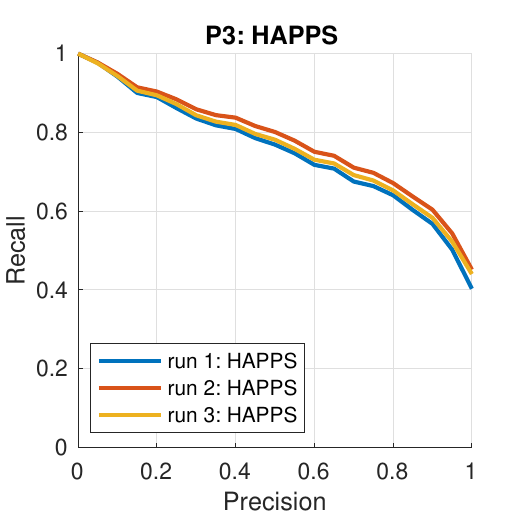}
         &
         \includegraphics[scale=0.45, trim={0.1cm 0 0.7cm 0}, clip]{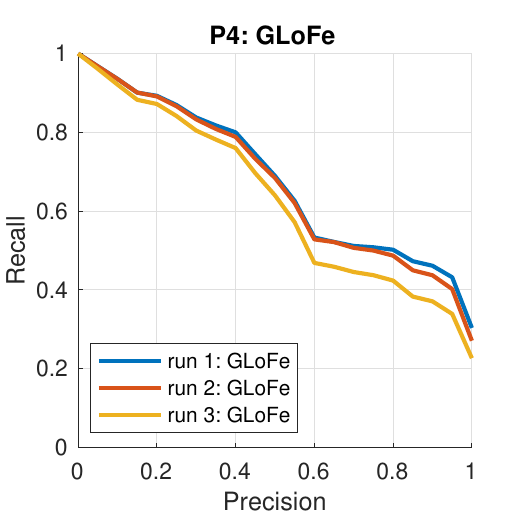}
         &
         \includegraphics[scale=0.45, trim={0.1cm 0 0.7cm 0}, clip]{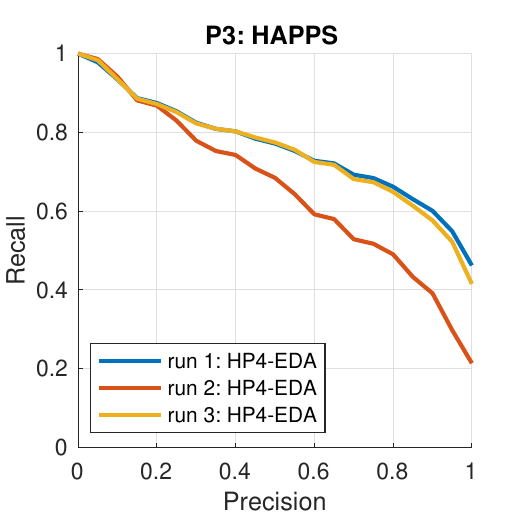}
         &
         \includegraphics[scale=0.45, trim={0.1cm 0 0.7cm 0}, clip]{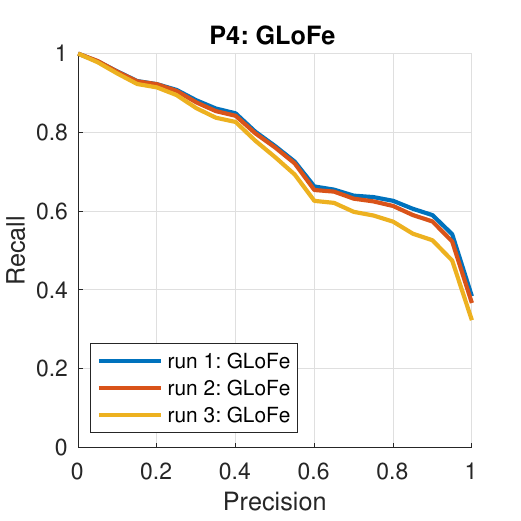}
         \\
         \includegraphics[scale=0.45, trim={0.1cm 0 0.7cm 0}, clip]{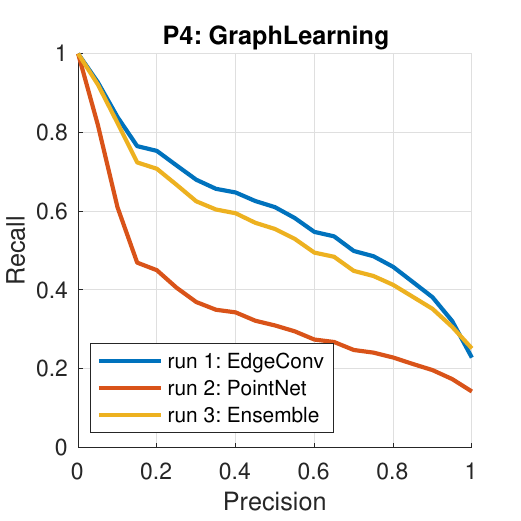}
         &
         \includegraphics[scale=0.45, trim={0.1cm 0 0.7cm 0}, clip]{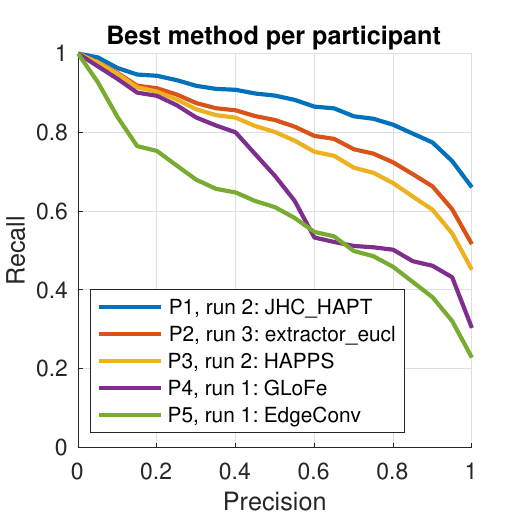}
         &
         \includegraphics[scale=0.45, trim={0.1cm 0 0.7cm 0}, clip]{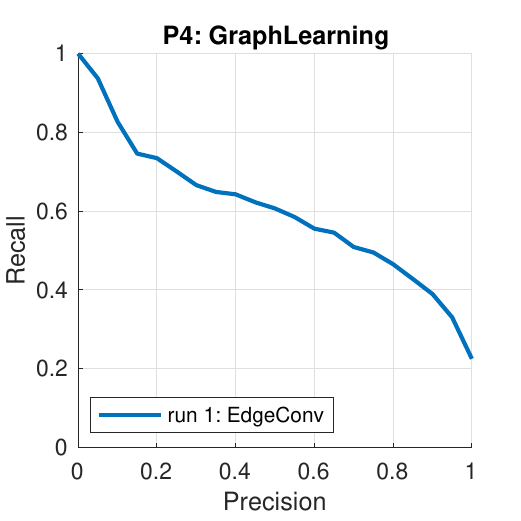}
         &
         \includegraphics[scale=0.45, trim={0.1cm 0 0.7cm 0}, clip]{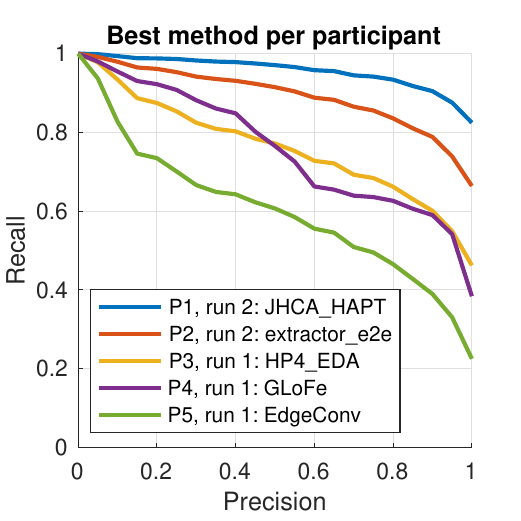}\\
         \hline
    \end{tabular}
    \end{center}  
    \caption{Precision-recall curves for Task A (geometry only) and Task B (geometry and physicochemical properties), with respect to the PDB classification.}
    \label{fig:prerec_geom_geomchem_SCOPe}
\end{figure*}

Table \ref{table:summary_retrieval_SCOPe} summarizes the retrieval performances of all the runs submitted for evaluation with respect to the PDB classification. More specifically, the table provides the following information: the Nearest Neighbour (NN), the First Tier (1T), the Second Tier (2T), the e-measure (eM), the Discounted Cumulated Gain (DCG), and the mean Average Precision (mAP). For each task, method and retrieval measure, the best performance is highlighted in bold; for each task and retrieval measure, the best performance among all methods is highlighted in red. All values are averaged for all queries.
Many methods achieve great or excellent performances. For instance: 
\begin{itemize}
    \item For Task A, $10$ out of $15$ runs have an NN value above $0.9$, i.e. their classification rate is above $90\%$.
    \item For Task B, $11$ out of $13$ runs have the NN value above $0.9$.
\end{itemize}
The same methods have mAP and DCG values above, respectively, $0.6$ and $0.8$. Precision-Recall plots are provided in Figure \ref{fig:prerec_geom_geomchem_SCOPe}. 

%  RETRIEVAL MEASURES --- SCOPe CLASSIFICATION
\begin{table*}[!h]
    \centering
    \caption{Summary of results by method and property type (only geometry vs. geometry and physicochemical properties) for the PDB classification. Here: NN = Nearest Neighbour, 1T = First Tier, 2T = Second Tier, eM = e-Measure, DCG = Discounted Cumulated Gain, mAP = mean Average Precision. For each task and for each measure, the best value for each method is in bold. The best among them is highlighted in red. \label{table:summary_retrieval_SCOPe}}
    \begin{adjustbox}{width=1\textwidth}
    \begin{tabular}{l l c c c c c c l c c c c c c}
    
    \toprule
    & \multicolumn{7}{c}{Geometry} & \multicolumn{7}{c}{Geometry and physicochemical characterization} \\
    \cmidrule(l){2-8}
    \cmidrule(l){9-15}

    & method & NN &  1T & 2T  & eM & DCG & mAP & method & NN &  1T & 2T  & eM & DCG & mAP \\[1.5ex]

    %FIRST PARTICIPANT
    \midrule
    \multirow{3}{*}{P1} & \multicolumn{1}{|l}{run 1} &   0.837 &  0.605 & 0.778  & 0.504 & 0.845 & 0.675 & \multicolumn{1}{|l}{run 1} & 0.982 & 0.873 & 0.951 & 0.685 & 0.971 & 0.922  \\

    & \multicolumn{1}{|l}{run 2} & \textbf{\textcolor{red}{0.947}} & \textbf{\textcolor{red}{0.815}} & \textbf{\textcolor{red}{0.940}}  & \textbf{\textcolor{red}{0.654}} & \textbf{\textcolor{red}{0.947}} & \textbf{\textcolor{red}{0.877}} & \multicolumn{1}{|l}{run 2} & \textbf{\textcolor{red}{0.989}} & \textbf{\textcolor{red}{0.922}} & \textbf{\textcolor{red}{0.979}} & \textbf{\textcolor{red}{0.714}} & \textbf{\textcolor{red}{0.985}} & \textbf{\textcolor{red}{0.958}} \\

    & \multicolumn{1}{|l}{run 3} & 0.927 & 0.729 & 0.884 & 0.597 & 0.921 & 0.806 & \multicolumn{1}{|l}{run 3} & 0.585 & 0.364 & 0.518 & 0.306 & 0.670 & 0.414 \\

    %SECOND PARTICIPANT
    \midrule
    \multirow{3}{*}{P2} & \multicolumn{1}{|l}{run 1} & 0.914 &  0.735 & 0.888  & 0.607 & 0.916 & 0.802 & \multicolumn{1}{|l}{run 1} & 0.951 & 0.815 & \textbf{0.938} & 0.653 & 0.949 & 0.874 \\

    & \multicolumn{1}{|l}{run 2} & 0.894 &  0.723 & 0.880  & 0.605 & 0.908 & 0.791 & \multicolumn{1}{|l}{run 2} & 0.947 & 0.800 & 0.927 & 0.649 & 0.942 & \textbf{0.895} \\

    & \multicolumn{1}{|l}{run 3} &  \textbf{0.924} & \textbf{0.748}  & \textbf{0.889} & \textbf{0.613} & \textbf{0.921} & \textbf{0.813} & \multicolumn{1}{|l}{run 3} & \textbf{0.979} & \textbf{0.839} & 0.937 & \textbf{0.665} & \textbf{0.962} & 0.858 \\
 
    %THIRD PARTICIPANT
    \midrule
    \multirow{3}{*}{P3} & \multicolumn{1}{|l}{run 1} & 0.920 & 0.683 & 0.836 & 0.562 & 0.897 & 0.756 & \multicolumn{1}{|l}{run 1} & 0.902 & \textbf{0.696} & \textbf{0.848} & \textbf{0.572} & \textbf{0.893} & \textbf{0.764} \\

    & \multicolumn{1}{|l}{run 2} & \textbf{0.930} & \textbf{0.711} & \textbf{0.858} & \textbf{0.586} & \textbf{0.911} & \textbf{0.782} &  \multicolumn{1}{|l}{run 2} & \textbf{0.923} & 0.592 & 0.720 & 0.486 & 0.847 & 0.663 \\

    & \multicolumn{1}{|l}{run 3} & 0.922  & 0.692 & 0.846 & 0.572 & 0.903 & 0.767 &  \multicolumn{1}{|l}{run 3} & 0.903 & 0.683 & 0.820 & 0.560 & 0.887 & 0.757 \\

    %FOURTH PARTICIPANT
    \midrule
    \multirow{3}{*}{P4} & \multicolumn{1}{|l}{run 1} & \textbf{0.927} & \textbf{0.593} & \textbf{0.716} & \textbf{0.493} & \textbf{0.865} & \textbf{0.684} & \multicolumn{1}{|l}{run 1} & \textbf{0.941} & \textbf{0.684} & \textbf{0.791} & \textbf{0.550} & \textbf{0.901} & \textbf{0.761} \\

    & \multicolumn{1}{|l}{run 2} & \textbf{0.927} & 0.586 & 0.705 & 0.487 & 0.862 & 0.675 & \multicolumn{1}{|l}{run 2} & \textbf{0.941} & 0.676 & 0.785 & 0.544 & 0.899 & 0.755 \\

    & \multicolumn{1}{|l}{run 3} & 0.907 & 0.549 & 0.672 & 0.453 & 0.840 & 0.634 & \multicolumn{1}{|l}{run 3} & 0.933 & 0.653 & 0.758 & 0.529 & 0.888 & 0.730 \\

    %FIFTH PARTICIPANT
    \midrule
    \multirow{3}{*}{P5} & \multicolumn{1}{|l}{run 1} & \textbf{0.755} & \textbf{0.537} & \textbf{0.734} & \textbf{0.468} & \textbf{0.806} & \textbf{0.603} & \multicolumn{1}{|l}{run 1} & \textbf{0.718} & \textbf{0.532} & \textbf{0.731} & \textbf{0.466} & \textbf{0.798} & \textbf{0.602} \\

    & \multicolumn{1}{|l}{run 2} & 0.437 & 0.300 & 0.477 & 0.263 & 0.633 & 0.358 & \multicolumn{1}{|l}{} \\

    & \multicolumn{1}{|l}{run 3} & 0.713 & 0.494 & 0.699 & 0.435 & 0.783 & 0.563 & \multicolumn{1}{|l}{ } \\
 
    \bottomrule
    \end{tabular}
    \end{adjustbox}
\end{table*}

% PRECISION RECALLS --- SCOPe CLASSIFICATION

For sake of conciseness, we do not list the analogous values of Table \ref{table:summary_retrieval_SCOPe} and plots in Figure \ref{fig:prerec_geom_geomchem_SCOPe} for the BLAST classification, rather we focus on multi-level indicators, such as the ADR values and the NDCG plots. 
 Interpreting the elements of a query queue in terms \emph{identity or isoform}, \emph{highly related}, \emph{similar} and \emph{dissimilar} surfaces with respect to the BLAST classification, Table \ref{table:summary_ADR_FASTA} reports the average dynamic recall (ADR) values for the runs of all methods.  All the ADR scores, which range from 0 (worst case) to 1 (ideal performance), are averaged over all
the models in the dataset.

% ADR MEASURE --- BLAST CLASSIFICATION
\begin{table*}[!h]
    \centering
    \caption{Summary of average dynamic recalls (ADRs) for the $4$-level BLAST classification. For each task, the best ADR for each method is in bold. The best among them is highlighted in red. \label{table:summary_ADR_FASTA}}
    \begin{tabular}{l c c c c c  l c c c c c}
    
    \toprule
    \multicolumn{6}{c}{Geometry} & \multicolumn{6}{c}{Geometry and physicochemical properties} \\
    \cmidrule(l){1-6}
    \cmidrule(l){7-12}
    & P1 & P2 & P3 & P4 & P5 & & P1 & P2 & P3 & P4 & P5 \\[1.5ex]

    \midrule
    \multicolumn{1}{l}{run 1} &   0.640 &  \textbf{0.700} & 0.697  & \textbf{0.681} & 0.631 & \multicolumn{1}{|l}{run 1} & \textcolor{red}{\textbf{0.809}} & \textbf{0.770} & 0.756 & \textbf{0.719} & \textbf{0.733}  \\

    \multicolumn{1}{l}{run 2} & 0.721 & 0.688 & \textbf{0.706}  & 0.676 & 0.543 & \multicolumn{1}{|l}{run 2} & 0.755 & 0.738 & \textbf{0.804} & 0.715 & - \\

    \multicolumn{1}{l}{run 3} &  \textbf{\textcolor{red}{0.729}} & 0.670 & 0.699 & 0.660 & \textbf{0.641} & \multicolumn{1}{|l}{run 3} & 0.635 & \textbf{0.770} & 0.751 & 0.704 & - \\

    \bottomrule

    \end{tabular}
\end{table*}

\begin{figure*}[t!]
    \begin{center}
    \begin{tabular}{|cc|cc|}
    \hline
        \rowcolor{blue!12}\multicolumn{2}{|c|}{Geometry} & \multicolumn{2}{c|}{Geometry and physicochemical properties} \\
        \hline
         \includegraphics[scale=0.45, trim={0 0 0 0}, clip]{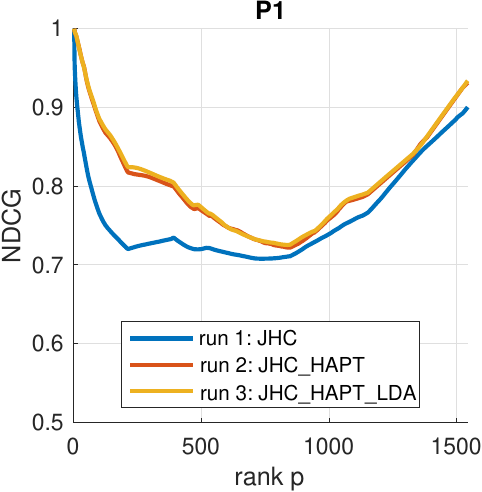}
         &
         \includegraphics[scale=0.45, trim={0 0 0 0}, clip]{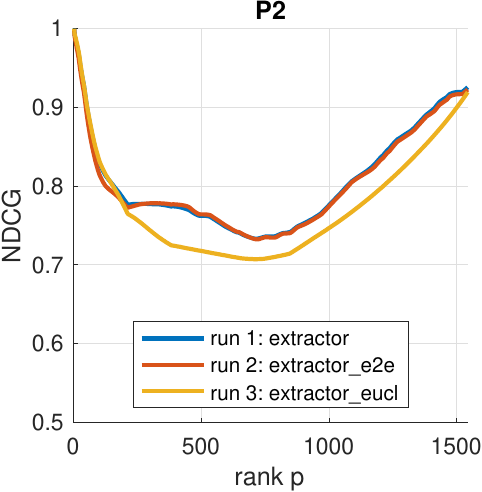}
         &
         \includegraphics[scale=0.45, trim={0 0 0 0}, clip]{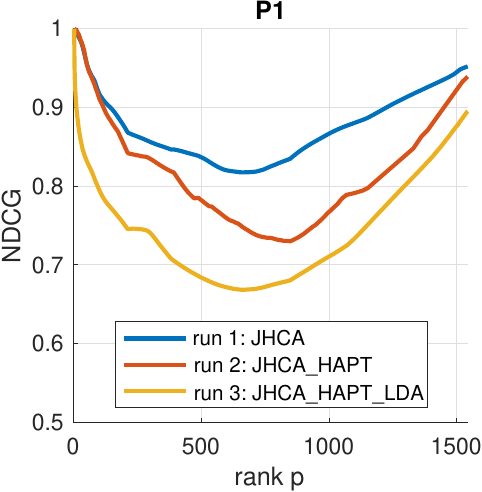}
         &
         \includegraphics[scale=0.45, trim={0 0 0 0}, clip]{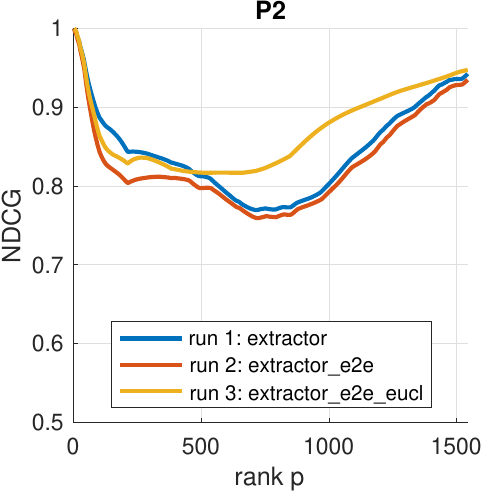}
         \\
         
         \includegraphics[scale=0.45, trim={0 0 0 0}, clip]{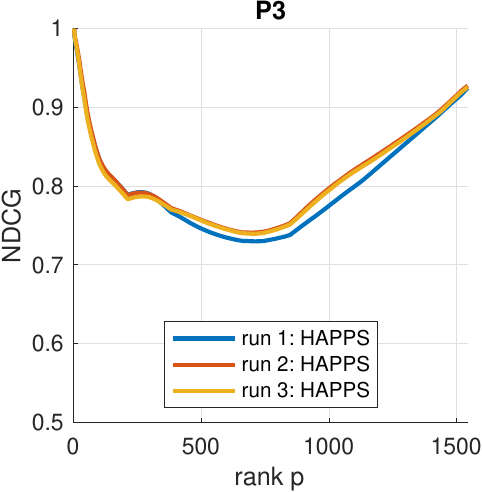}
         &
         \includegraphics[scale=0.45, trim={0 0 0 0}, clip]{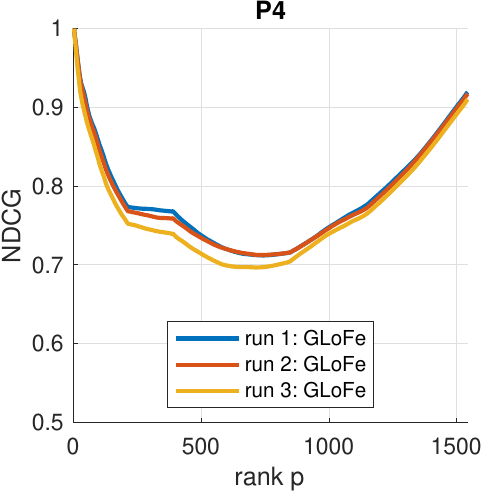}
         &
         \includegraphics[scale=0.45, trim={0 0 0 0}, clip]{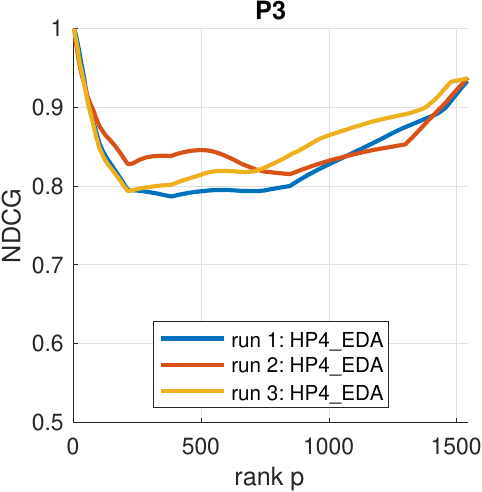}
         &
         \includegraphics[scale=0.45, trim={0 0 0 0}, clip]{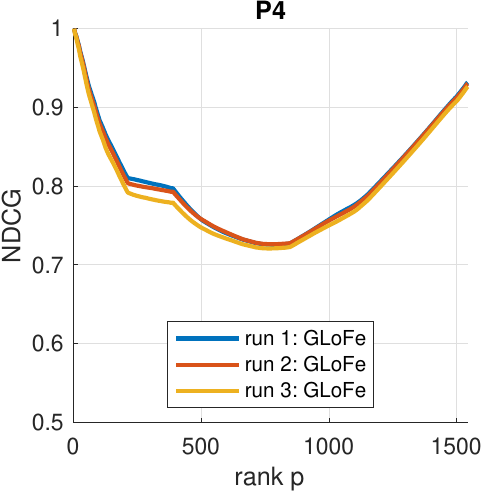}
         \\
         
         \includegraphics[scale=0.45, trim={0 0 0 0}, clip]{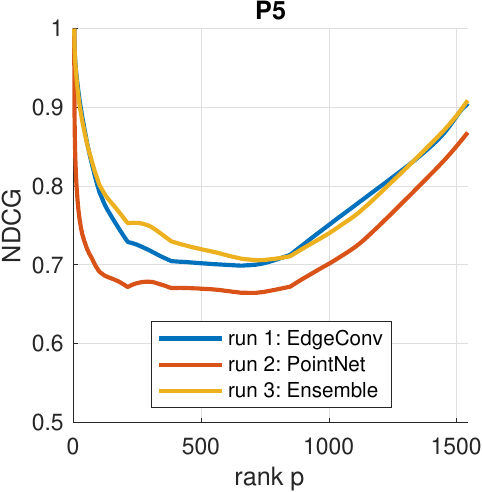}
         &
         \includegraphics[scale=0.45, trim={0 0 0 0}, clip]{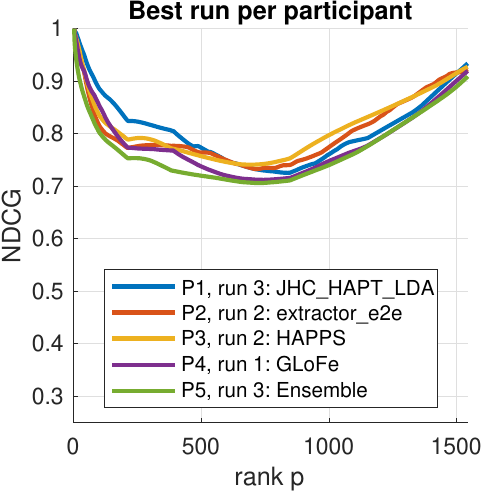}
         &
         \includegraphics[scale=0.45, trim={0 0 0 0}, clip]{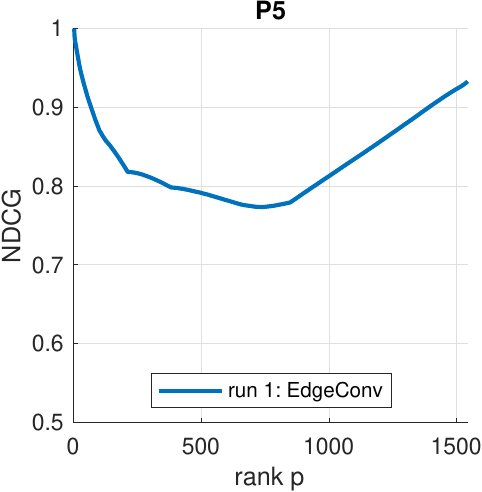}
         &
         \includegraphics[scale=0.45, trim={0 0 0 0}, clip]{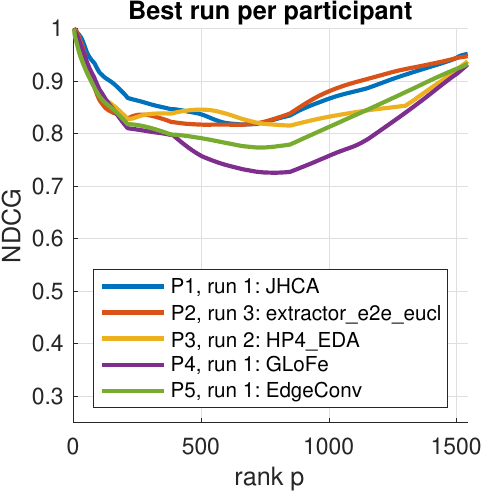}\\
         \hline
    \end{tabular}
    \end{center}  
    \caption{Normalized discounted cumulated gain (NDCG) for Task A (geometry only) and Task B (geometry and physicochemical properties), with respect to the $4$-level BLAST classification.}
    \label{fig:NDCG_geom_geomchem_FASTA}
\end{figure*}

A more comprehensive analysis of the retrieval queue with respect to the BLAST classification is provided by the normalized discounted cumulative gain plots in Figure \ref{fig:NDCG_geom_geomchem_FASTA}.
The NDCG measure is represented as a function of the rank $p$.  The NDCG values for all queries are averaged to obtain a measure of the average performance for each submitted run. Remind that, for an ideal run, it would be NDCG $\equiv$ 1. The NDCG measure takes BLAST classification performances into larger account than PDB one, as all surfaces corresponding to the same PDB code are the same protein for the BLAST classification.

In \ref{sec:suppl_mat}, we include the same multi-level indicators, namely ADR and NDCG plots, for the 3-level BLAST classification. Since this classification aggregates communities that are extremely similar and highly related, the ADR scores slightly increase but the overall relationships between the different methods and runs are confirmed.

%  NDCG --- CLASTP CLASSIFICATION

\subsection{Classification performance measures}

Table \ref{table:summary_class_meas_SCOPe} summarizes the classification performances for both tasks A and B, when considering the PDB classification; the confusion matrices originating such values are shown, for the sake of completeness, in Figures \ref{fig:conf_mat_geom_SCOPe} and \ref{fig:conf_mat_geomchem_SCOPe}. More precisely, the table contains the following information: True Positive Rate (TPR), True Negative Rate (TNR), Positive Predicted Values (PPV), Negative Predicted Values (NPV), Accuracy (ACC) and $F_1$ score (F1). One fact we can immediately note is that, maybe not surprisingly, methods that showed robustness when evaluated with the retrieval measures yet exhibit strong performances when tested with classification measures. However, we can additionally notice that:
\begin{itemize}
    \item All methods have TNR higher than TPR, making them more reliable in correctly finding true negatives rather than true positives.
    \item All methods have NPV higher than TPV: it is more likely for these methods to be right when reporting a negative rather than a positive.
    \item All methods have accuracy over $97.5\%$, although they are not equally ``accurate" in finding true/false positives/negatives. This a sign -- rather obvious from the nature of the problem -- that positives and negatives are not in equal proportion; methods showing a greater difference between TPR and TNR (and between PPV and NPV) are more inclined to privilege the negatives (the predominant class) at the expense of the positives (the minor class). In our context, accuracy is therefore an overoptimistic estimation.
    \item $F_1$ score provides a ``better'' metric than the accuracy, in the sense that it suffers more from imbalance.
\end{itemize}
Enriching the protein SES triangulation with physicochemical properties does not always lead to an improvement. For example, in the third run by P1 it dramatically decreases the performances. On the other hand, run 2 from the same method shows a marked improvement to deal with (true) positives.

Table \ref{table:summary_class_meas_FASTA} summarizes the classification performance for both tasks A and B, when considering the ($4$-level) BLAST classification; the confusion matrices originating such values are shown, for the sake of completeness, in Figures \ref{fig:conf_mat_geom_FASTA_4levels} and \ref{fig:conf_mat_geomchem_FASTA_4levels}. As expected, a decrease in the number of classes leads to an improvement of the classification measures. However, it is also worth noting that this improvement is not the same in all methods: by comparing Tables \ref{table:summary_class_meas_SCOPe} and \ref{table:summary_class_meas_FASTA}, one can indeed note changes in the best run per method (and in the best overall run). 

The supplementary material, reported in \ref{sec:suppl_mat}, includes classification measures (see Table \ref{table:summary_class_meas_FASTA_old}) and the corresponding confusion matrices (see Figures \ref{fig:conf_mat_geom_FASTA_3levels} and \ref{fig:conf_mat_geomchem_FASTA_3levels}) in the case of a $3$-level BLAST classification; one can notice that this latter leads to the same considerations as in the $4$-level BLAST classification.

%  CLASSIFICATION MEASURES --- SCOPe CLASSIFICATION
\begin{table*}[!h]
    \centering
    \caption{Summary of statistical measures by method and property type (only geometry vs. geometry and physicochemical properties) for the PDB-based community decomposition. Here: TPR = True Positive Rate, TNR = True Negative Rate, PPV = Positive Predictive Value, NPV = Negative Predictive Value, ACC = ACCuracy, F1 = $F_1$ score. For each task and for each measure, the best value for each method is in bold. The best among them is highlighted in red.  \label{table:summary_class_meas_SCOPe}}
    \begin{adjustbox}{width=1\textwidth}
    \begin{tabular}{l l c c c c c c l c c c c c c}
    
    \toprule
    & \multicolumn{7}{c}{Geometry} & \multicolumn{7}{c}{Geometry and physicochemical properties} \\
    \cmidrule(l){2-8}
    \cmidrule(l){9-15}
    & method & TPR & TNR &  PPV & NPV  & ACC & F1 & method & TPR & TNR &  PPV & NPV  & ACC & F1 \\[1.5ex]

    %FIRST PARTICIPANT
    \midrule
    \multirow{3}{*}{P1} & \multicolumn{1}{|l}{run 1} & 0.8373 & 0.9967 & 0.8401 & 0.9973 & 0.9941 & 0.8354 & \multicolumn{1}{|l}{run 1} & 0.9825 & 0.9997 & 0.9860 & 0.9997 & 0.9994 & 0.9832 \\

    & \multicolumn{1}{|l}{run 2} & \textbf{\textcolor{red}{0.9475}} & \textbf{\textcolor{red}{0.9991}} & \textbf{\textcolor{red}{0.9489}} & \textbf{\textcolor{red}{0.9992}} & \textbf{\textcolor{red}{0.9983}} & \textbf{\textcolor{red}{0.9467}} & \multicolumn{1}{|l}{run 2} & \textbf{\textcolor{red}{0.9890}} & \textbf{\textcolor{red}{0.9999}} & \textbf{\textcolor{red}{0.9921}} & \textbf{\textcolor{red}{0.9999}} & \textbf{\textcolor{red}{0.9998}} & \textbf{\textcolor{red}{0.9893}} \\

    & \multicolumn{1}{|l}{run 3} &  0.9274 & 0.9990 & 0.9304 & 0.9988 & 0.9974 & 0.9239 & 
    \multicolumn{1}{|l}{run 3} & 0.5839 & 0.9885 & 0.6086 & 0.9912 & 0.9807 & 0.5727\\
 
    %SECOND PARTICIPANT
    \midrule
    \multirow{3}{*}{P2} & \multicolumn{1}{|l}{run 1} & 0.9145 & 0.9979 & 0.9159 & 0.9984 & 0.9965 & 0.9119 & \multicolumn{1}{|l}{run 1} & 0.9514 & 0.9989 & 0.9525 & 0.9992 & 0.9981 & 0.9504 \\

    & \multicolumn{1}{|l}{run 2} & 0.8944 & 0.9975 & 0.8977 & 0.9976 & 0.9954 & 0.8931 & \multicolumn{1}{|l}{run 2} & 0.9469 & 0.9989 & 0.9470 & 0.9991 & 0.9981 & 0.9460 \\

    & \multicolumn{1}{|l}{run 3} &  \textbf{0.9242} & \textbf{0.9983} & \textbf{0.9253} & \textbf{0.9985} & \textbf{0.9969} & \textbf{0.9222} &  \multicolumn{1}{|l}{run 3} & \textbf{0.9793} & \textbf{0.9995} & \textbf{0.9819} & \textbf{0.9996} & \textbf{0.9991} & \textbf{0.9791} \\
 
    %THIRD PARTICIPANT
    \midrule
    \multirow{3}{*}{P3} & \multicolumn{1}{|l}{run 1} & 0.9196 & \textbf{0.9985} & 0.9205 & 0.9986 & \textbf{0.9971} & 0.9169 & \multicolumn{1}{|l}{run 1} & 0.9015 & 0.9977 & 0.9037 & 0.9979 & 0.9957 & 0.9007 \\

    & \multicolumn{1}{|l}{run 2} & \textbf{0.930}0 & 0.9982 & \textbf{0.9333} & \textbf{0.9988} & \textbf{0.9971} & \textbf{0.9276} & \multicolumn{1}{|l}{run 2} & \textbf{0.9216} & \textbf{0.9985} & \textbf{0.9238} & \textbf{0.9987} & \textbf{0.9974} & \textbf{0.9207} \\

    & \multicolumn{1}{|l}{run 3} &  0.9222 & 0.9982 & 0.9258 & 0.9986 & 0.9969 & 0.9199 &  \multicolumn{1}{|l}{run 3} & 0.9015 & 0.9979 & 0.9005 & 0.9985 & 0.9966 & 0.8987 \\
 
    %FOURTH PARTICIPANT
    \midrule
    \multirow{3}{*}{P4} & \multicolumn{1}{|l}{run 1} & \textbf{0.9274} & \textbf{0.9983} & \textbf{0.9284} & \textbf{0.9987} & \textbf{0.9971} & \textbf{0.9262} & \multicolumn{1}{|l}{run 1} & \textbf{0.9410} & 0.9987 & \textbf{0.9424} & \textbf{0.9990} & \textbf{0.9978} & 0.9406 \\

    & \multicolumn{1}{|l}{run 2} & 0.9268 & \textbf{0.9983} & 0.9277 & \textbf{0.9987} & \textbf{0.9971} & 0.9252 & \multicolumn{1}{|l}{run 2} & \textbf{0.9410} & \textbf{0.9988} & 0.9422 & \textbf{0.9990} & \textbf{0.9978} & \textbf{0.9409}\\

    & \multicolumn{1}{|l}{run 3} &  0.9067 & 0.9979 & 0.9059 & 0.9983 & 0.9963 & 0.9046 & \multicolumn{1}{|l}{run 3} & 0.9326 & 0.9984 & 0.9336 & 0.9988 & 0.9974 & 0.9323\\
 
    %FIFTH PARTICIPANT 
    \midrule 
    \multirow{3}{*}{P5} & \multicolumn{1}{|l}{run 1} &  \textbf{0.7537} & \textbf{0.9944} & \textbf{0.7539} & \textbf{0.9943} & \textbf{0.9892} & \textbf{0.7507} & \multicolumn{1}{|l}{run 1} & \textbf{0.7187} & \textbf{0.9937} & \textbf{0.7215} & \textbf{0.9940} & \textbf{0.9886} & \textbf{0.7160} \\

    & \multicolumn{1}{|l}{run 2} & 0.4362 & 0.9870 & 0.4412 & 0.9873 & 0.9754 & 0.4328 & \multicolumn{1}{|l}{} \\

    & \multicolumn{1}{|l}{run 3} &  0.7123 & 0.9927 & 0.7109 & 0.9930 & 0.9868 & 0.7044 & \multicolumn{1}{|l}{} \\

    \bottomrule
    \end{tabular}
    \end{adjustbox}
\end{table*}

%%% CLASSIFICATION MEASURES -- BLAST CLASSIFICATION
\begin{table*}[!h]
    \centering
    \caption{Summary of statistical measures by method and property type (only geometry vs. geometry and physicochemical properties) for the BLAST-based community decomposition of level $3$. Here: TPR = True Positive Rate, TNR = True Negative Rate, PPV = Positive Predictive Value, NPV = Negative Predictive Value, ACC = ACCuracy, F1 = $F_1$ score. For each task and for each measure, the best value for each method is in bold. The best among them is highlighted in red.  \label{table:summary_class_meas_FASTA}}
    
    \begin{adjustbox}{width=1\textwidth}
    \begin{tabular}{l  l c c c c c c  l c c c c c c}
    
    \toprule
    & \multicolumn{7}{c}{Geometry} & \multicolumn{7}{c}{Geometry and physicochemical properties} \\
    \cmidrule(l){2-8}
    \cmidrule(l){9-15}
    & method & TPR & TNR &  PPV & NPV  & ACC & F1 & method & TPR & TNR &  PPV & NPV  & ACC & F1 \\[1.5ex]

    %FIRST PARTICIPANT
    \midrule
    \multirow{3}{*}{P1} &
    \multicolumn{1}{|l}{run 1} & 0.9086 & 0.9949 & 0.9082 & 0.9959 & 0.9917 & 0.9069 & \multicolumn{1}{|l}{run 1} & 0.9961 & 0.9998 & 0.9962 & \textcolor{red}{\textbf{1.0000}} & 0.9997 & 0.9960\\

    & \multicolumn{1}{|l}{run 2} & 
    \textcolor{red}{\textbf{0.9890}} & \textcolor{red}{\textbf{0.9996}} & \textcolor{red}{\textbf{0.9891}} & 0.9996 & \textcolor{red}{\textbf{0.9993}} & \textcolor{red}{\textbf{0.9888}} & 
    \multicolumn{1}{|l}{run 2} & \textbf{0.9981} & \textcolor{red}{\textbf{1.0000}} & \textbf{0.9981} & \textcolor{red}{\textbf{1.0000}} & \textcolor{red}{\textbf{1.0000}} & \textbf{0.9980}\\

    & \multicolumn{1}{|l}{run 3} &  0.9844 & \textcolor{red}{\textbf{0.9996}} & 0.9869 & \textcolor{red}{\textbf{0.9997}} & \textcolor{red}{\textbf{0.9993}} & 0.9840 & \multicolumn{1}{|l}{run 3} & 0.8529 & 0.9904 & 0.8562 & 0.9931 & 0.9854 & 0.8452\\
 
    %SECOND PARTICIPANT
    \midrule
    \multirow{3}{*}{P2} &
    \multicolumn{1}{|l}{run 1} & 0.9760 & \textbf{0.9992} & 0.9766 & \textbf{0.9993} & \textbf{0.9985} & 0.9758 & \multicolumn{1}{|l}{run 1} & 0.9929 & 0.9997 & 0.9930 & 0.9999 & 0.9996 & 0.9928\\

    & \multicolumn{1}{|l}{run 2} & 0.9728 & 0.9991 & 0.9746 & 0.9991 & 0.9983 & 0.9723 & \multicolumn{1}{|l}{run 2} & 0.9909 & 0.9998 & 0.9909 & 0.9999 & 0.9997 & 0.9908\\
    
    & \multicolumn{1}{|l}{run 3} &  \textbf{0.9767} & 0.9985 & \textbf{0.9770} & 0.9989 & 0.9977 & \textbf{0.9764} & \multicolumn{1}{|l}{run 3} & \textcolor{red}{\textbf{0.9987}} & \textbf{0.9999} & \textcolor{red}{\textbf{0.9987}} & \textcolor{red}{\textbf{1.0000}} & \textbf{0.9999} & \textcolor{red}{\textbf{0.9987}} \\
 
    %THIRD PARTICIPANT
    \midrule
    \multirow{3}{*}{P3} &
    \multicolumn{1}{|l}{run 1} & 0.9689 & 0.9981 & 0.9690 & 0.9985 & 0.9970 & 0.9680 & \multicolumn{1}{|l}{run 1} & \textbf{0.9942} & 0.9996 & \textbf{0.9943} & \textbf{0.9999} & 0.9995 &  \textbf{0.9942}\\

    & \multicolumn{1}{|l}{run 2} & \textbf{0.9747} & 0.9980 & \textbf{0.9754} & \textbf{0.9989} & 0.9972 & \textbf{0.9740} & \multicolumn{1}{|l}{run 2} & 0.9916 & \textbf{0.9997} & 0.9921 & 0.9998 & \textbf{0.9996} & 0.9916\\

    & \multicolumn{1}{|l}{run 3} &  0.9721 & \textbf{0.9987} & 0.9738 & 0.9985 & \textbf{0.9976} & 0.9716 & \multicolumn{1}{|l}{run 3} & 0.9806 & 0.9981 & 0.9809 & 0.9994 & 0.9980 & 0.9803\\
 
    %FOURTH PARTICIPANT
    \midrule
    \multirow{3}{*}{P4} &
    \multicolumn{1}{|l}{run 1} & \textbf{0.9799} & \textbf{0.9987} & \textbf{0.9805} & \textbf{0.9991} & \textbf{0.9980} & \textbf{0.9798}  & \multicolumn{1}{|l}{run 1} & \textbf{0.9903} & \textbf{0.9995} & \textbf{0.9909} & \textbf{0.9996} & \textbf{0.9992} & \textbf{0.9904}\\

    & \multicolumn{1}{|l}{run 2} & 0.9793 & \textbf{0.9987} & 0.9801 & 0.9990 & 0.9979 & 0.9791 & \multicolumn{1}{|l}{run 2} & \textbf{0.9903} & \textbf{0.9995} & 0.9908 & \textbf{0.9996} & 0.9991 & \textbf{0.9904}\\

    & \multicolumn{1}{|l}{run 3} &  0.9734 & 0.9984 & 0.9738 & 0.9987 & 0.9974 & 0.9732 & \multicolumn{1}{|l}{run 3} & 0.9870 & 0.9993 & 0.9876 & 0.9994 & 0.9988 & 0.9871\\

    %FIFTH PARTICIPANT
    \midrule
    \multirow{3}{*}{P5} &
    \multicolumn{1}{|l}{run 1} & \textbf{0.9209} & \textbf{0.9953} & \textbf{0.9209} & \textbf{0.9958} & \textbf{0.9923} & \textbf{0.9197} & \multicolumn{1}{|l}{run 1} & \textbf{0.9501} & \textbf{0.9975} & \textbf{0.9506} & \textbf{0.9988} & \textbf{0.9968} & \textbf{0.9491}\\

    & \multicolumn{1}{|l}{run 2} & 0.7168 & 0.9852 & 0.7201 & 0.9853 & 0.9734 & 0.7156 & \multicolumn{1}{|l}{}\\

    & \multicolumn{1}{|l}{run 3} &  0.9047 & 0.9944 & 0.9031 & 0.9953 & 0.9910 & 0.9019 & \multicolumn{1}{|l}{}\\

    \bottomrule

    \end{tabular}
    \end{adjustbox}
\end{table*}

\subsection{Discussion}
The methods that participated in this SHREC contest are representative of various types of approaches to the 3D object retrieval problem, ranging from purely feature-based engineered methods, mainly based on features represented with histograms (P1, P3, and P4), to the combination of features and dimensional reduction techniques (P1 Run 3), to deep neural networks (P2) and transfer learning from deep graph convolutional networks (P5).

On the one hand, the retrieval performances are positive
for all methods, in either the PDB or BLAST classifications. On the other hand, the NDCG and ADR measures are specifically designed for interpreting a multi-level dataset classification as in this case, and thus offer a complementary evaluation of the classical retrieval measures (e.g., NN, FT, ST, precision-recall plots, etc.) and classification measures (TPR, TNR, PPV, NPV, ACC, confusion matrices, etc.). These performance indicators show that the highest ADR scores vary from 0.729 (geometric) to 0.809 (geometry and pysico-chemical properties) being 1 the best possible value for the ADR: this confirms that these approaches are good but not optimal. Similarly, the highest possible area under a NDCG curve equals 1, while the best scores in this contribution are around 0.9 (for the Task B).

In this SHREC contest, a training dataset was explicitly provided having with the ground truth based on the PDB classification. The surface distribution in the classes mirrored the distribution of classes in the test set, see Figure \ref{fig:barchart_distributions}; the number of conformations per PDB ranges from 2 to 160.
This highlights one of the difficulties that learning methods have faced, namely the presence of classes of very heterogeneous size, which makes prediction very difficult. The severity of the PDB classification is then mitigated by the BLAST one, but this classification has been used only for the interpretation of the results and not previously provided to the participants. 

A further difficulty for learning methods is the dataset design choice of using different proteins (and their conformations) between training set and test set. This was done to investigate the ability of 3D retrieval approaches to reason about and predict the conformations of a protein, even if not yet ``seen" by the training system. This probably motivates that the best overall performance for the PDB classification was obtained by a technique based on engineered features. This fact is further confirmed by the lower prediction ability of the same descriptor when combined with a dimensional reduction technique as demonstrated by the method P1 (run 3), which show a particular decrease when the geometry is enriched with physicochemical properties.
Conversely, when we consider the BLAST classification, i.e. proteins that share fairly long amino acids sequences are considered similar altogether their conformations, we see that learning-based methods improve their performance proportionally more than direct methods, such as the P1 (run 3) method. In our view, this reflects the fact that similarities between sequences are reflected in similarities of 3D structure, and with this classification comes greater homogeneity between the features of the ``extended'' classes.

Additional considerations can be derived from the geometric-only and mixed geometry and physicochemical properties comparison.
Not surprisingly, we notice an improvement in the performance of the various approaches when switching from runs purely geometry-driven (Task A) to runs that consider both geometry and physicochemical properties (Task B). Nevertheless, the direct comparison between the proposed runs is not always possible because for some participants the geometric method may vary between the 2 tasks. We notice that the most widely adopted solution is the introduction of a histogram for the physicochemical properties, that is then used as an additional feature vector whose outcome is combined with the dissimilarity scores given by the geometric description. Furthermore, from the experiments available to us, we note that the most significant improvements are seen for methods that are based on learning. This is particularly reflected in the methods P2 and P5. This suggests that physicochemical properties play an important role in the characterisation of a protein but, perhaps, more research is still needed to deeply understand their role and how best to integrate them into engineered descriptors; for instance, considering a joint description as currently proposed in P1, that adopts joint 3D histograms.

\section{Concluding remarks}
\label{sec:obs_conlusion}

In this paper, we have provided a detailed analysis and evaluation of state-of-the-art retrieval and classification algorithms
dealing with protein similarity assessment based on molecular surfaces, which we believe deserve attention from the research community. The introduction of physicochemical properties into the benchmark, represents an element of originality in the available benchmarks for structural biology and provides a more complete representation of the protein.
To enable the participation of learning-based methods, both a training and a test set were provided for this benchmark dataset. Moreover, we are aware that in some of the PDB codes we used, the underlying structures may correspond to mutations of the same protein, or be isoform, or share a common fold; for this reason, we performed a multi-level performance analysis, comparing the performance of the proposed methods to both a classification made according to the protein PDB code and an aggregation between proteins made by using BLASTP. 

Beyond the extensive analysis that has been carried out throughout the paper, we hope that the experimental results presented here may offer interesting hints for further investigation. For instance, a better and more informed definition of similarity can be preliminary to a better and more effective definition of the complementarity between binding biomolecules.

The benchmark, as well as the dissimilarity matrices that originated the results described in Section \ref{sec:results} and in the appendices, are available at \url{https://github.com/rea1991/SHREC2021}.

\section*{Acknowledgements}
The track organisers thank Dr. Michela Spagnuolo and Dr. Davide Boscaini for the fruitful discussions. Special thanks go to Ms. Daniela Bejan, for her help in using the software PyMol.

This project is co-funded by the project ``TEACUP: Metodi e TEcniche innovative per lo sviluppo di librerie per la modellazione, l'Analisi e il confronto CompUtazionale di Proteine", POR FSE, Programma Operativo Regione Liguria 2014-2020, No RLOF18ASSRIC/68/1.
The CNR-IMATI research is partially developed in the activities DIT.AD021.080.001 and DIT.AD009.091.001. 
This research was partially supported by TAILOR, a project funded by EU Horizon 2020 research and innovation programme under GA No 952215, and by Vietnam National University Ho Chi Minh City (VNU-HCM) under grant number DS2020-42-01.

%REFERENCES
\bibliographystyle{ieeetr}      % mathematics and physical sciences

\appendix

\newpage
\section{Confusion matrices (PDB-based community decomposition)\label{app:conf_mat_SCOPe}}
\label{Appendix}

\begin{figure}[htb!]
    \begin{center}
    \begin{tabular}{cc}
    \begin{tabular}{ccccc}
         %RUN 1
         \includegraphics[scale=0.375, trim={0.04cm 0 0.05cm 0}, clip]{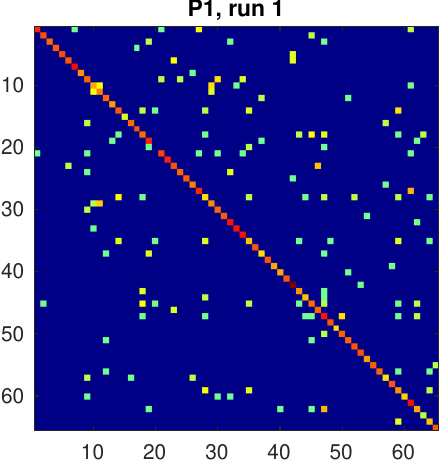}
         &
         \includegraphics[scale=0.375, trim={0.04cm 0 0.05cm 0}, clip]{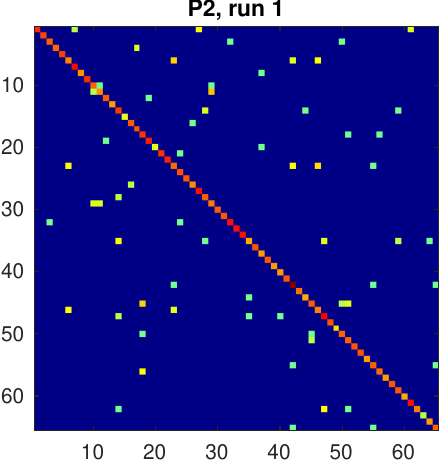}
         &
         \includegraphics[scale=0.375, trim={0.04cm 0 0.05cm 0}, clip]{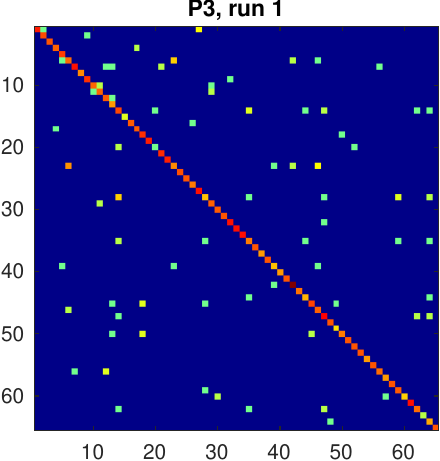}
         &
         \includegraphics[scale=0.375, trim={0.04cm 0 0.05cm 0}, clip]{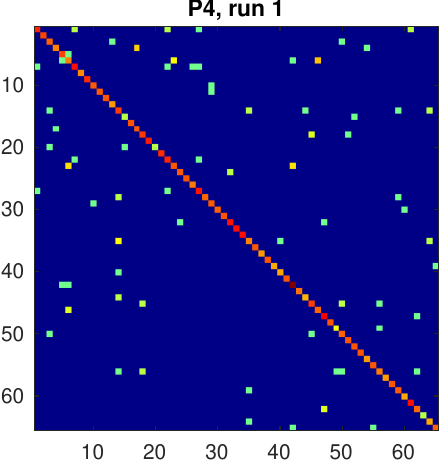}
         &
         \includegraphics[scale=0.375, trim={0.04cm 0 0.05cm 0}, clip]{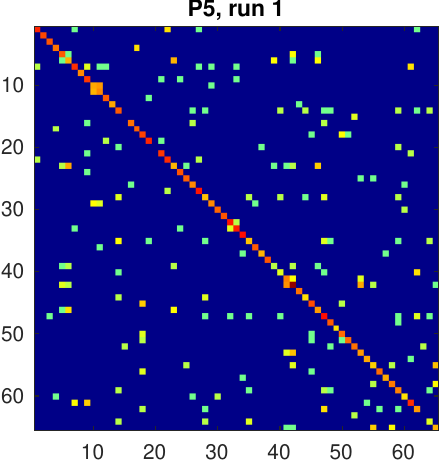}
         \\
         
         %RUN 2
         \includegraphics[scale=0.375, trim={0.04cm 0 0.05cm 0}, clip]{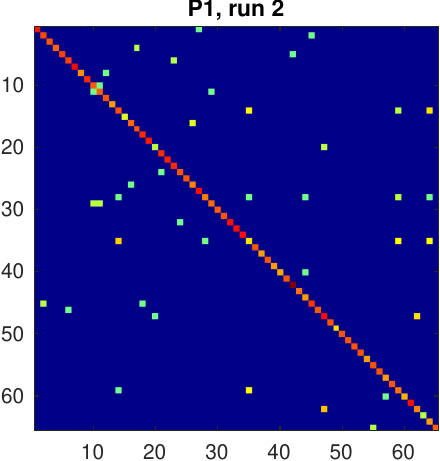}
         &
         \includegraphics[scale=0.375, trim={0.04cm 0 0.05cm 0}, clip]{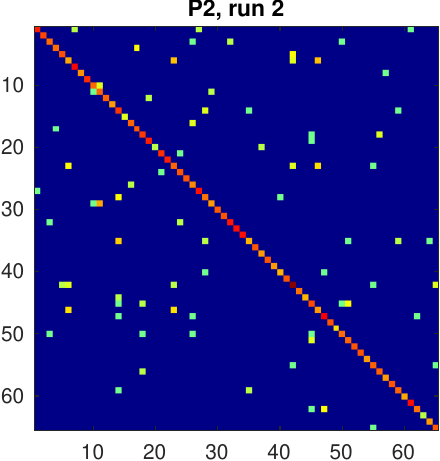}
         &
         \includegraphics[scale=0.375, trim={0.04cm 0 0.05cm 0}, clip]{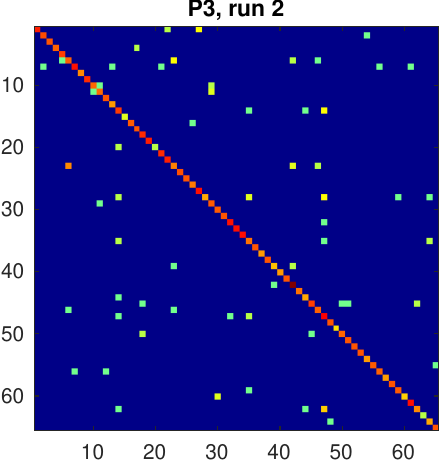}
         &
         \includegraphics[scale=0.375, trim={0.04cm 0 0.05cm 0}, clip]{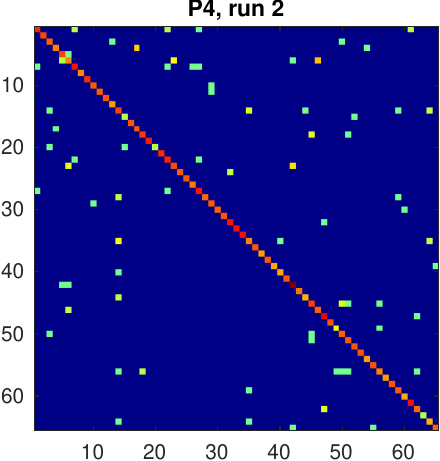}
         &
         \includegraphics[scale=0.375, trim={0.04cm 0 0.05cm 0}, clip]{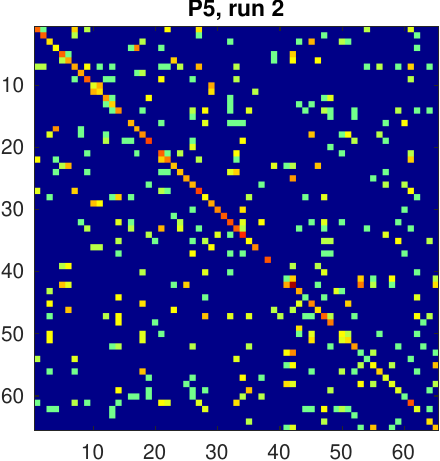}
         \\
         
         %RUN 3
         \includegraphics[scale=0.375, trim={0.04cm 0 0.05cm 0}, clip]{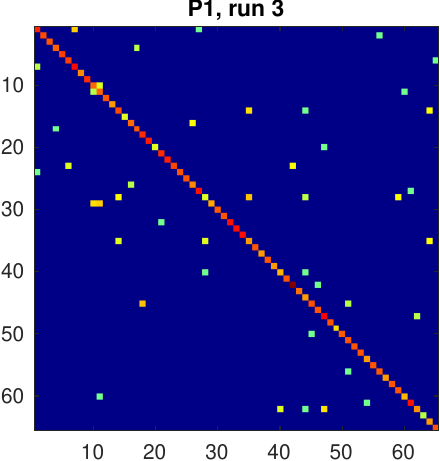}
         &
         \includegraphics[scale=0.375, trim={0.04cm 0 0.05cm 0}, clip]{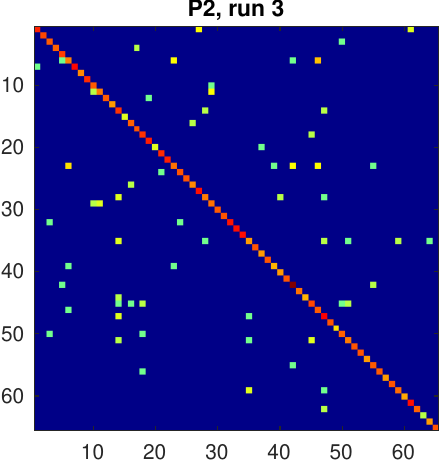}
         &
         \includegraphics[scale=0.375, trim={0.04cm 0 0.05cm 0}, clip]{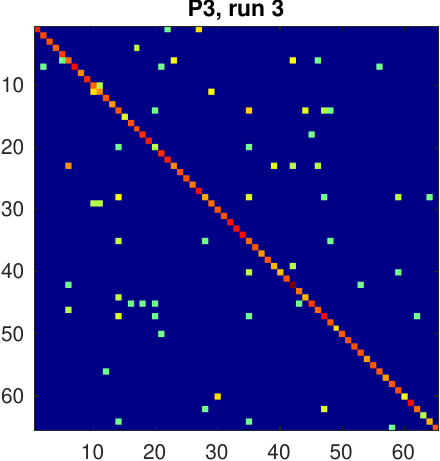}
         &
         \includegraphics[scale=0.375, trim={0.04cm 0 0.05cm 0}, clip]{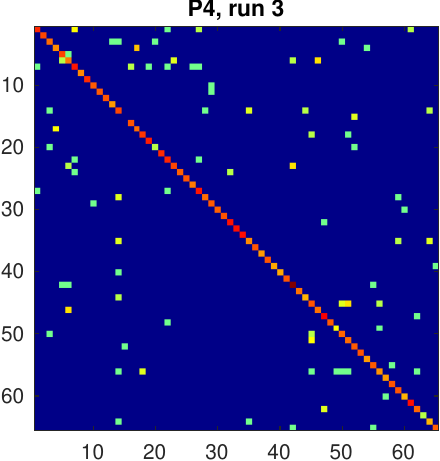}
         &
         \includegraphics[scale=0.375, trim={0.04cm 0 0.05cm 0}, clip]{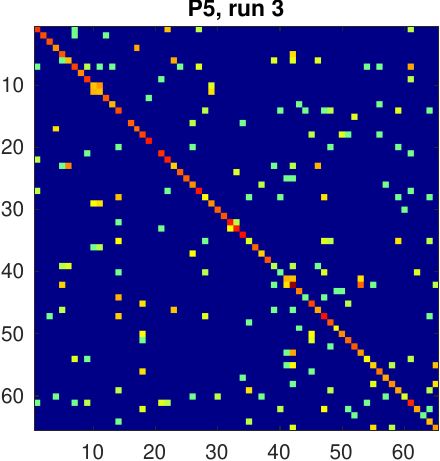}
         \\
    \end{tabular}
    &
    \begin{minipage}[c]{0.04\linewidth}
    \includegraphics[scale=0.6, trim={17cm 0 1.75cm 0}, clip]{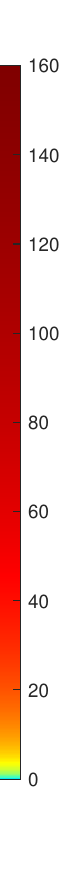}
    \end{minipage}
    \end{tabular}
    \end{center}
    \caption{Confusion matrices for task A (geometry only), with respect to the PDB-based community decomposition.}
    \label{fig:conf_mat_geom_SCOPe}
\end{figure}

\begin{figure}[htb!]
    \begin{center}
    \begin{tabular}{cc}
    \begin{tabular}{ccccc}
         %RUN 1
         \includegraphics[scale=0.375, trim={0.04cm 0 0.05cm 0}, clip]{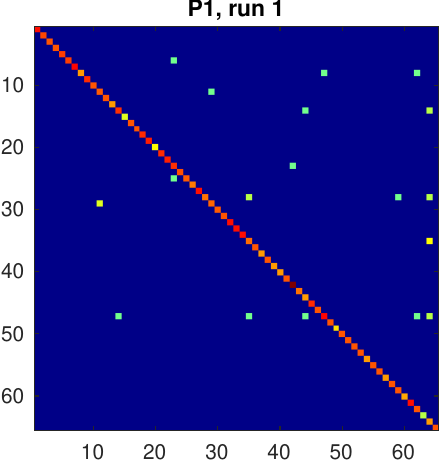}
         &
         \includegraphics[scale=0.375, trim={0.04cm 0 0.05cm 0}, clip]{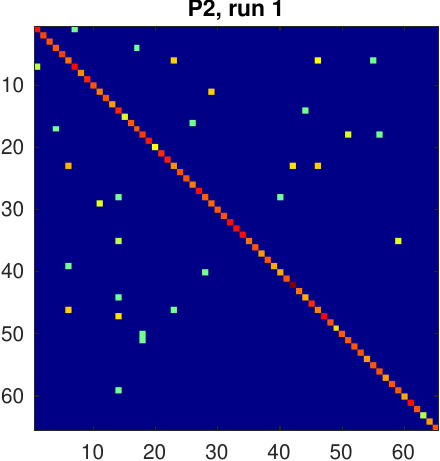}
         &
         \includegraphics[scale=0.375, trim={0.04cm 0 0.05cm 0}, clip]{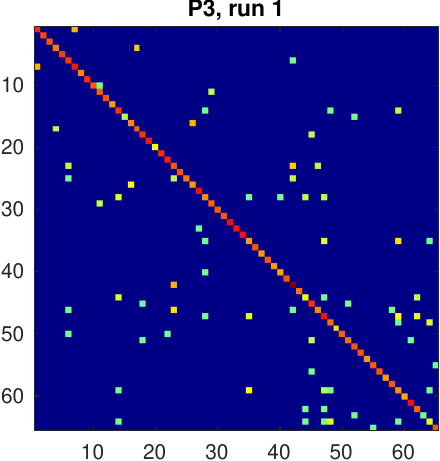}
         &
         \includegraphics[scale=0.375, trim={0.04cm 0 0.05cm 0}, clip]{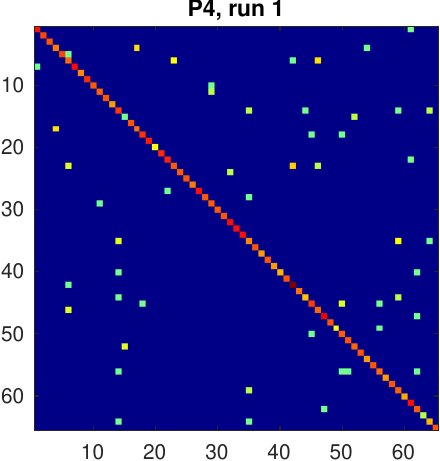}
         &
         \includegraphics[scale=0.375, trim={0.04cm 0 0.05cm 0}, clip]{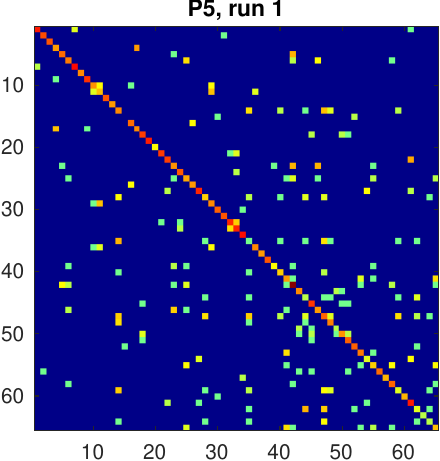}
         \\
         
         %RUN 2
         \includegraphics[scale=0.375, trim={0.04cm 0 0.05cm 0}, clip]{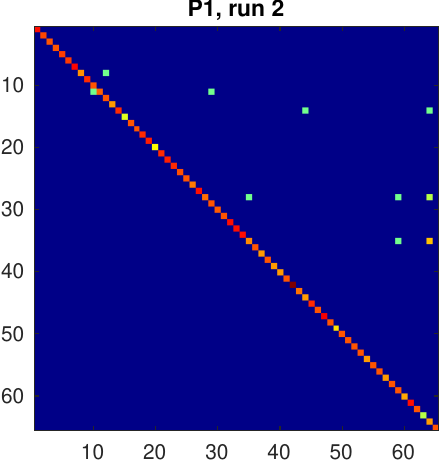}
         &
         \includegraphics[scale=0.375, trim={0.04cm 0 0.05cm 0}, clip]{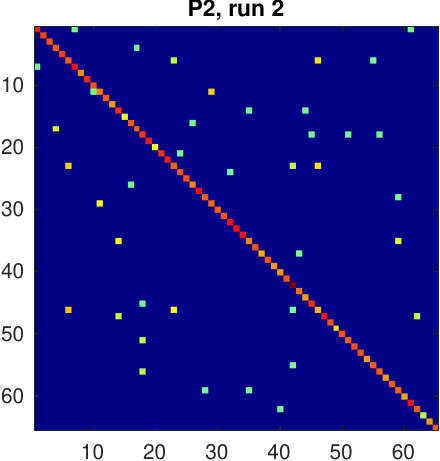}
         &
         \includegraphics[scale=0.375, trim={0.04cm 0 0.05cm 0}, clip]{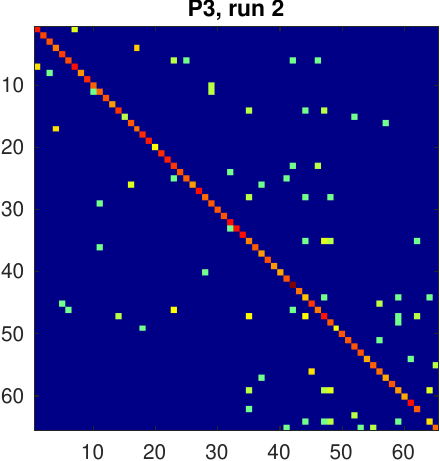}
         &
         \includegraphics[scale=0.375, trim={0.04cm 0 0.05cm 0}, clip]{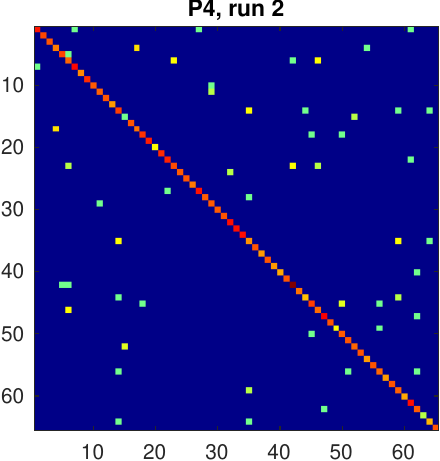}
         &
         
         \\
         
         %RUN 3
         \includegraphics[scale=0.375, trim={0.04cm 0 0.05cm 0}, clip]{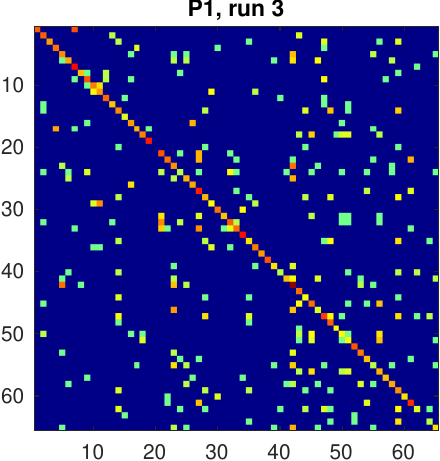}
         &
         \includegraphics[scale=0.375, trim={0.04cm 0 0.05cm 0}, clip]{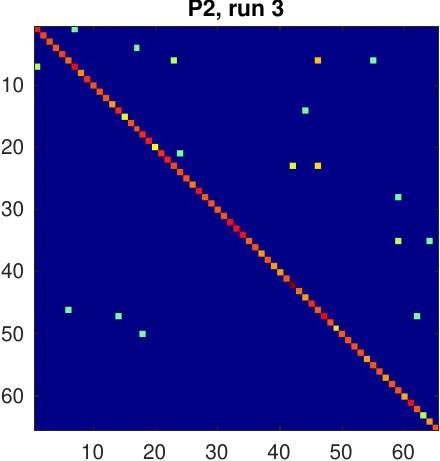}
         &
         \includegraphics[scale=0.375, trim={0.04cm 0 0.05cm 0}, clip]{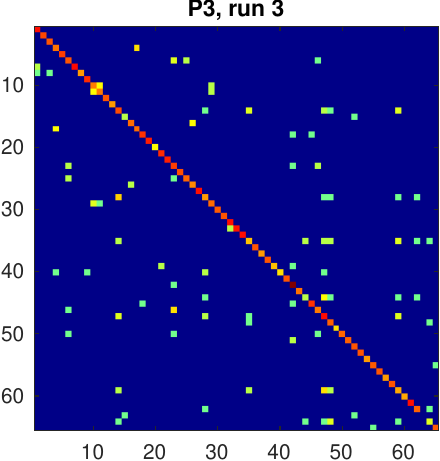}
         &
         \includegraphics[scale=0.375, trim={0.04cm 0 0.05cm 0}, clip]{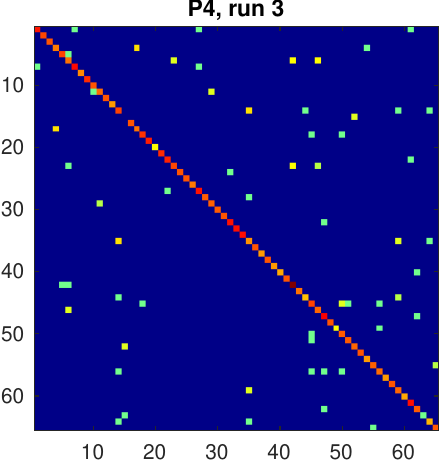}
         &
        
         \\
    \end{tabular}
    &
    \begin{minipage}[c]{0.04\linewidth}
    \includegraphics[scale=0.6, trim={17cm 0 1.75cm 0}, clip]{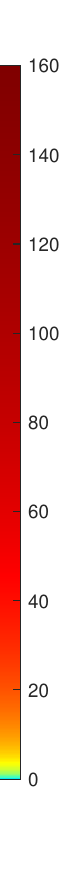}
    \end{minipage}
    \end{tabular}
    \end{center}
    \caption{Confusion matrices for task B (geometry and physicochemical properties), with respect to the PDB-based community decomposition.}
    \label{fig:conf_mat_geomchem_SCOPe}
\end{figure}

\newpage
\section{Confusion matrices (BLAST-based community decomposition of level 3)\label{app:conf_mat_FASTA_4levels}}

\begin{figure*}[htb!]
    \begin{center}
    \begin{tabular}{cc}
    \begin{tabular}{ccccc}
         %RUN 1
         \includegraphics[scale=0.375, trim={0.04cm 0 0.05cm 0}, clip]{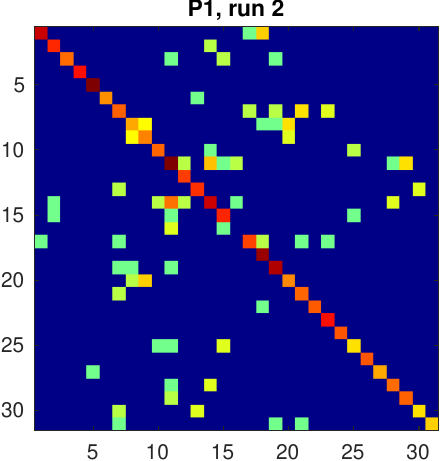}
         &
         \includegraphics[scale=0.375, trim={0.04cm 0 0.05cm 0}, clip]{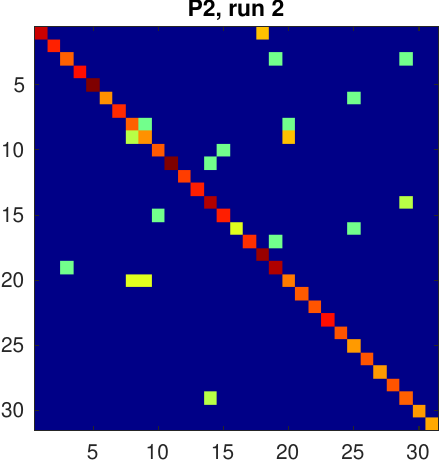}
         &
         \includegraphics[scale=0.375, trim={0.04cm 0 0.05cm 0}, clip]{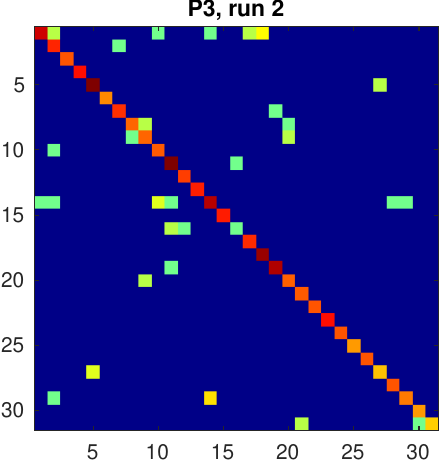}
         &
         \includegraphics[scale=0.375, trim={0.04cm 0 0.05cm 0}, clip]{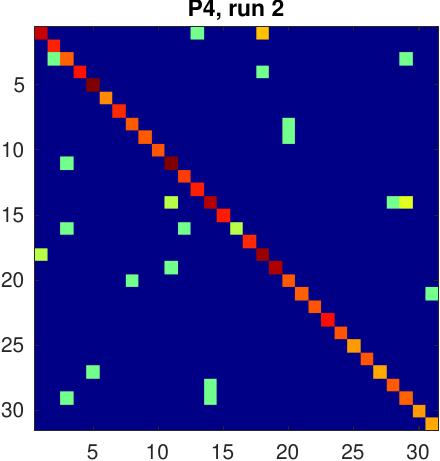}
         &
         \includegraphics[scale=0.375, trim={0.04cm 0 0.05cm 0}, clip]{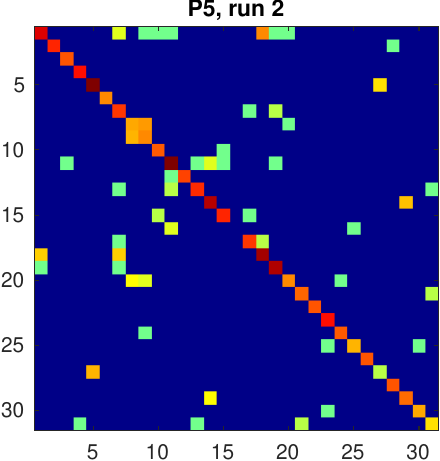}
         \\
         
         %RUN 2
         \includegraphics[scale=0.375, trim={0.04cm 0 0.05cm 0}, clip]{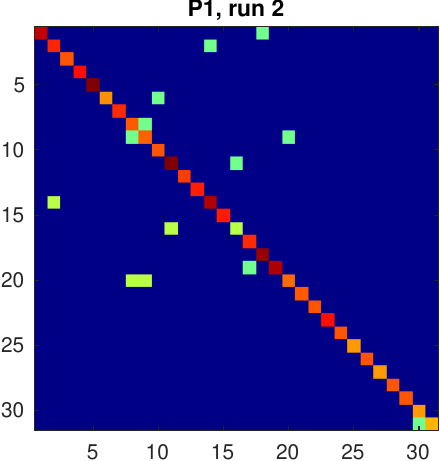}
         &
         \includegraphics[scale=0.375, trim={0.04cm 0 0.05cm 0}, clip]{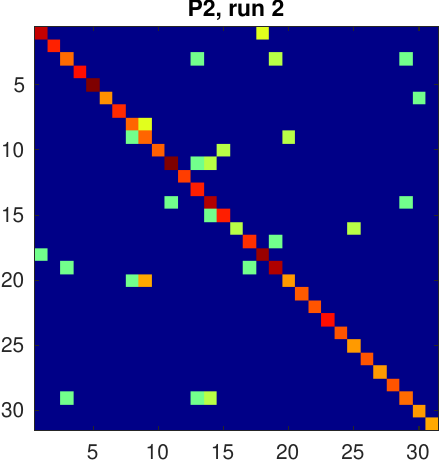}
         &
         \includegraphics[scale=0.375, trim={0.04cm 0 0.05cm 0}, clip]{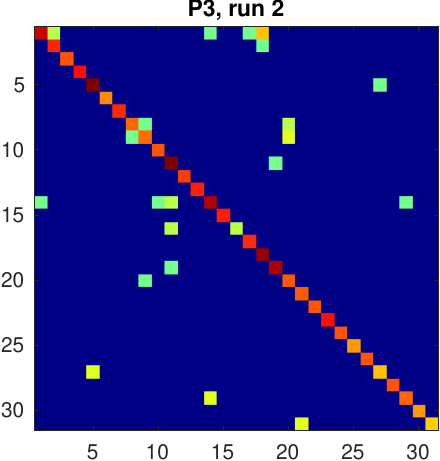}
         &
         \includegraphics[scale=0.375, trim={0.04cm 0 0.05cm 0}, clip]{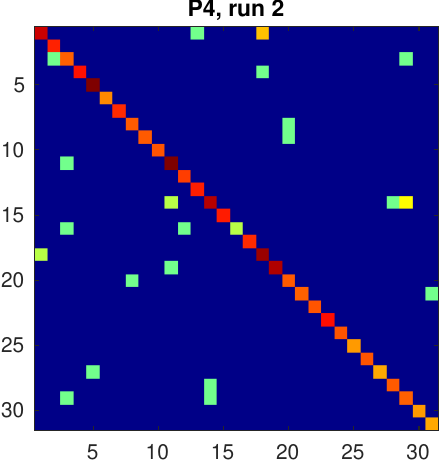}
         &
         \includegraphics[scale=0.375, trim={0.04cm 0 0.05cm 0}, clip]{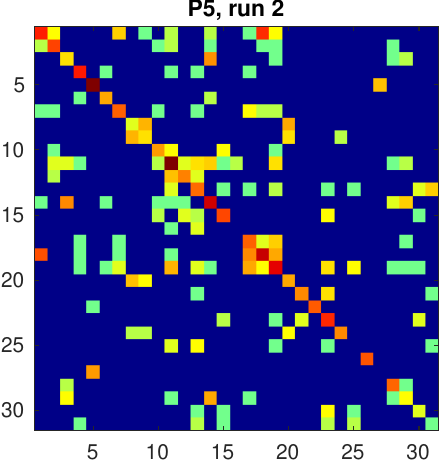}
         \\
         
         %RUN 3
         \includegraphics[scale=0.375, trim={0.04cm 0 0.05cm 0}, clip]{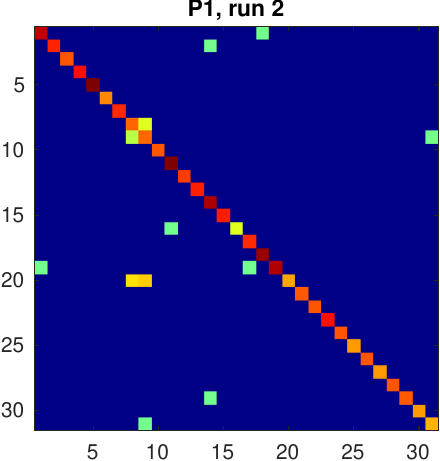}
         &
         \includegraphics[scale=0.375, trim={0.04cm 0 0.05cm 0}, clip]{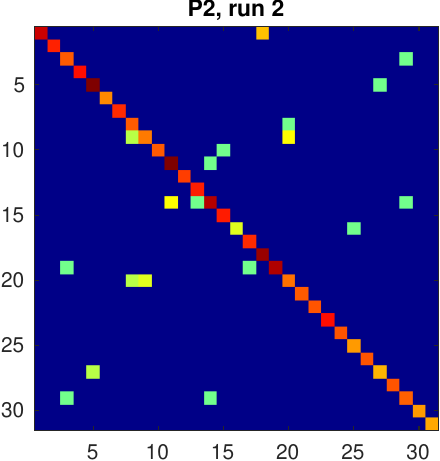}
         &
         \includegraphics[scale=0.375, trim={0.04cm 0 0.05cm 0}, clip]{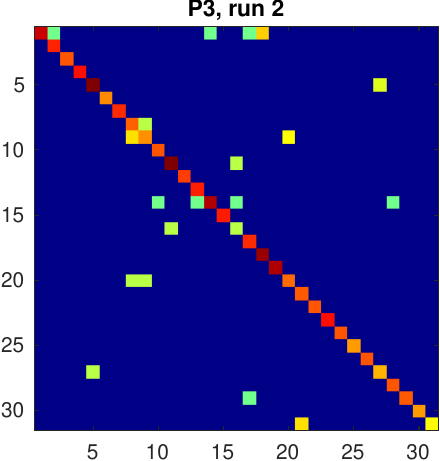}
         &
         \includegraphics[scale=0.375, trim={0.04cm 0 0.05cm 0}, clip]{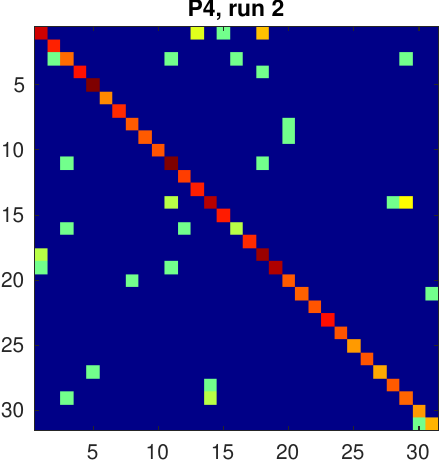}
         &
         \includegraphics[scale=0.375, trim={0.04cm 0 0.05cm 0}, clip]{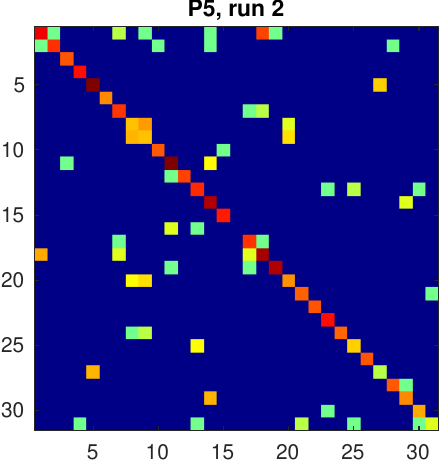}
         \\
    \end{tabular}
    &
    \begin{minipage}[c]{0.04\linewidth}
    \includegraphics[scale=0.6, trim={17cm 0 1.75cm 0}, clip]{colorbar_geom-eps-converted-to.pdf}
    \end{minipage}
    \end{tabular}
    \end{center}
    \caption{Confusion matrices for task A (geometry only), with respect to the BLAST-based community decomposition of level $3$.}
    \label{fig:conf_mat_geom_FASTA_4levels}
\end{figure*}

\begin{figure*}[htb!]
    \begin{center}
    \begin{tabular}{cc}
    \begin{tabular}{ccccc}
         %RUN 1
         \includegraphics[scale=0.375, trim={0.04cm 0 0.05cm 0}, clip]{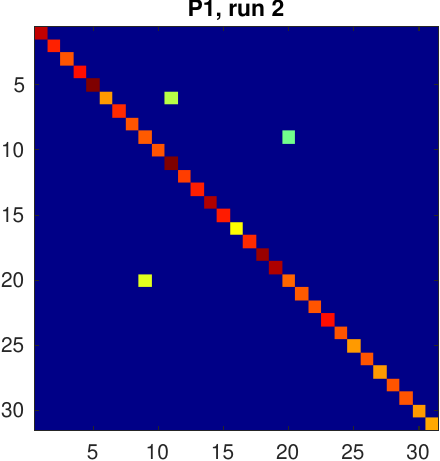}
         &
         \includegraphics[scale=0.375, trim={0.04cm 0 0.05cm 0}, clip]{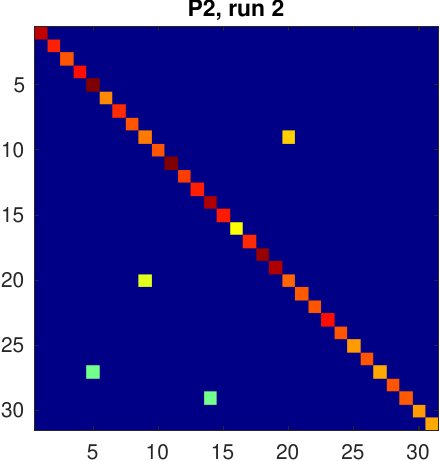}
         &
         \includegraphics[scale=0.375, trim={0.04cm 0 0.05cm 0}, clip]{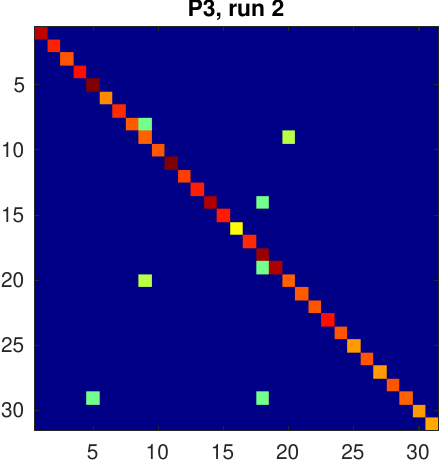}
         &
         \includegraphics[scale=0.375, trim={0.04cm 0 0.05cm 0}, clip]{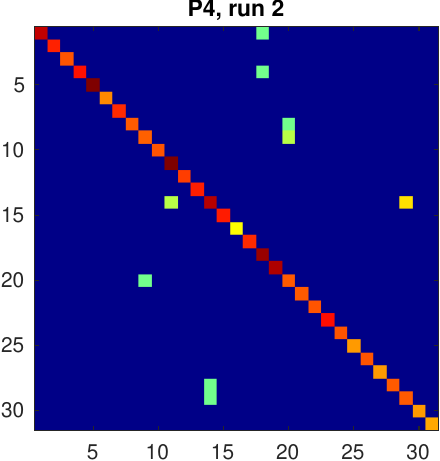}
         &
         \includegraphics[scale=0.375, trim={0.04cm 0 0.05cm 0}, clip]{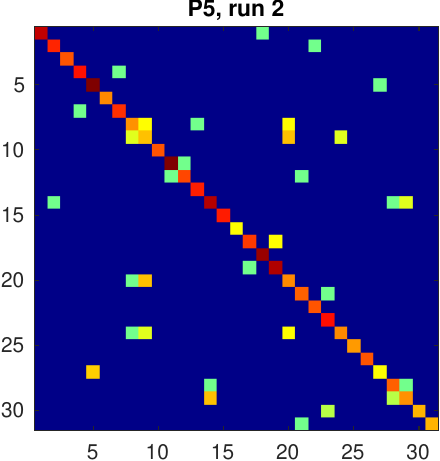}
         \\
         
         %RUN 2
         \includegraphics[scale=0.375, trim={0.04cm 0 0.05cm 0}, clip]{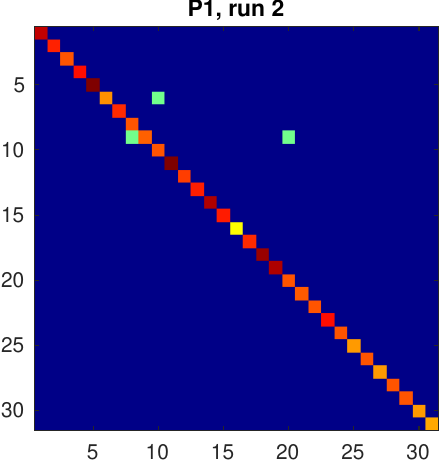}
         &
         \includegraphics[scale=0.375, trim={0.04cm 0 0.05cm 0}, clip]{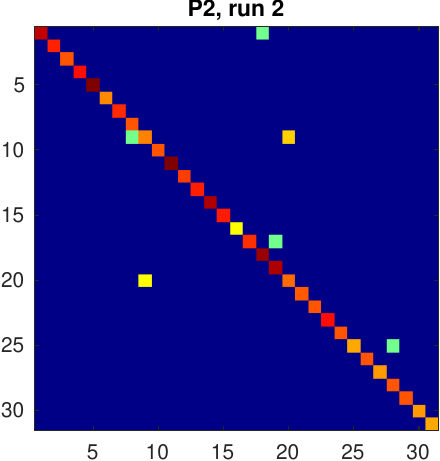}
         &
         \includegraphics[scale=0.375, trim={0.04cm 0 0.05cm 0}, clip]{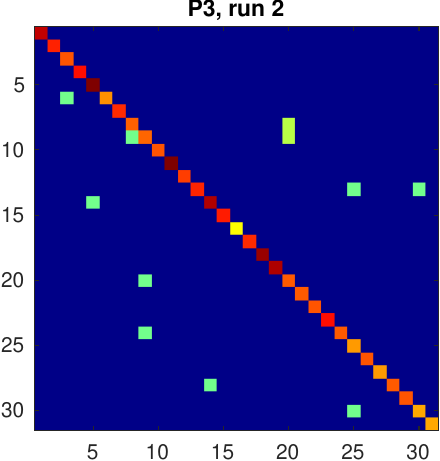}
         &
         \includegraphics[scale=0.375, trim={0.04cm 0 0.05cm 0}, clip]{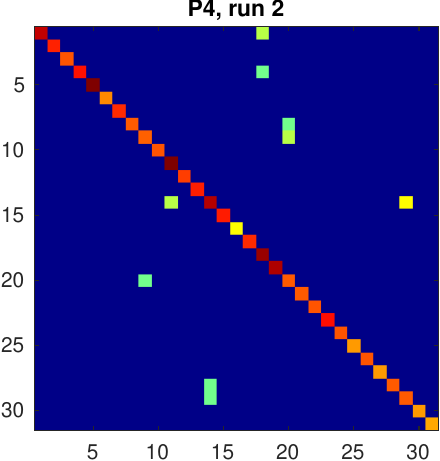}
         &
         
         \\
         
         %RUN 3
         \includegraphics[scale=0.375, trim={0.04cm 0 0.05cm 0}, clip]{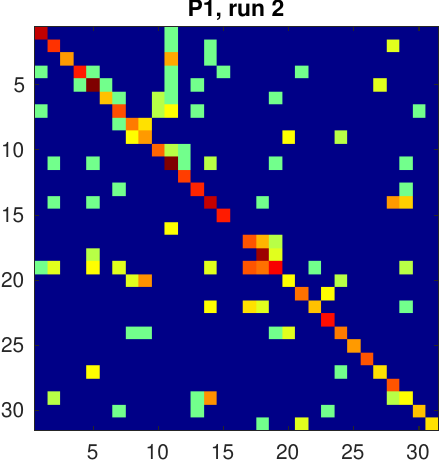}
         &
         \includegraphics[scale=0.375, trim={0.04cm 0 0.05cm 0}, clip]{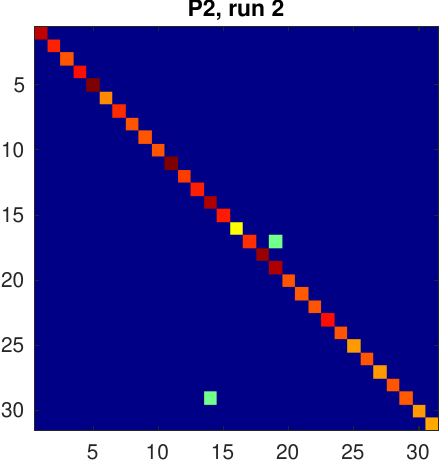}
         &
         \includegraphics[scale=0.375, trim={0.04cm 0 0.05cm 0}, clip]{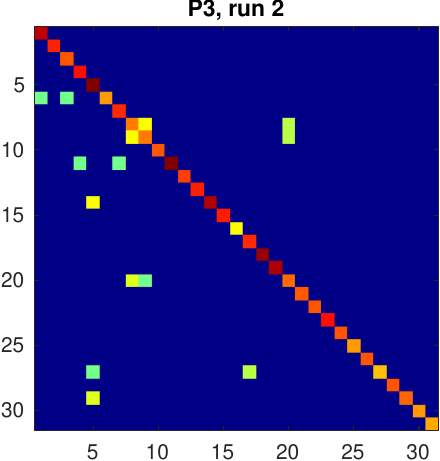}
         &
         \includegraphics[scale=0.375, trim={0.04cm 0 0.05cm 0}, clip]{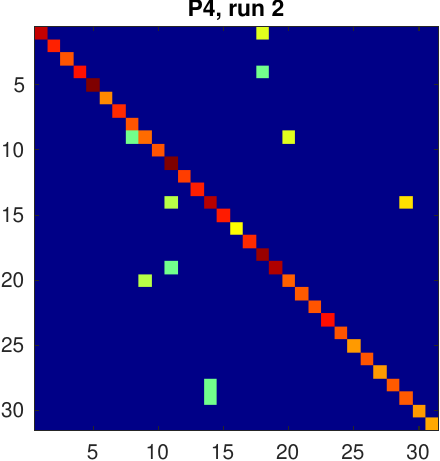}
         &
        
         \\
    \end{tabular}
    &
    \begin{minipage}[c]{0.04\linewidth}
    \includegraphics[scale=0.6, trim={17cm 0 1.75cm 0}, clip]{colorbar_geomchem-eps-converted-to.pdf}
    \end{minipage}
    \end{tabular}
    \end{center}
    \caption{Confusion matrices for task B (geometry and physicochemical properties), with respect to the BLAST-based community decomposition of level $3$.}
    \label{fig:conf_mat_geomchem_FASTA_4levels}
\end{figure*}

\section{Supplementary material: Performances with respect to a 3-level BLAST classification \label{sec:suppl_mat}}

\begin{figure*}[htb!]
    \begin{center}
    \begin{tabular}{cc}
    \begin{tabular}{ccccc}
         %RUN 1
         \includegraphics[scale=0.375, trim={0.04cm 0 0.05cm 0}, clip]{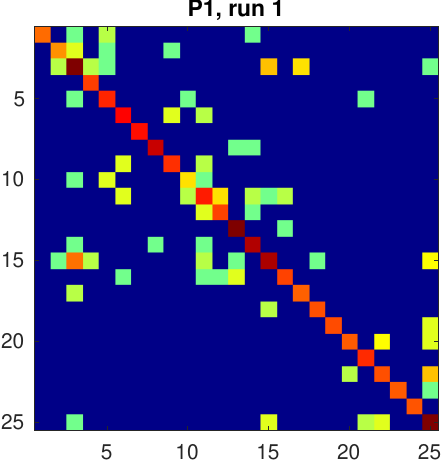}
         &
         \includegraphics[scale=0.375, trim={0.04cm 0 0.05cm 0}, clip]{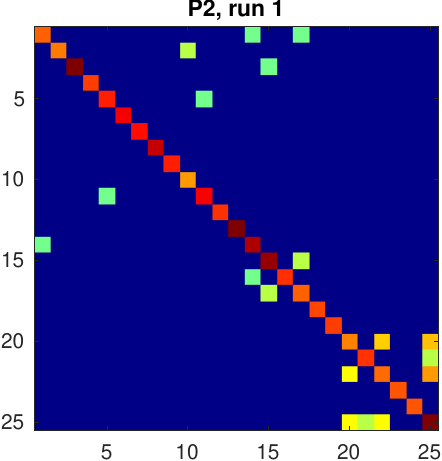}
         &
         \includegraphics[scale=0.375, trim={0.04cm 0 0.05cm 0}, clip]{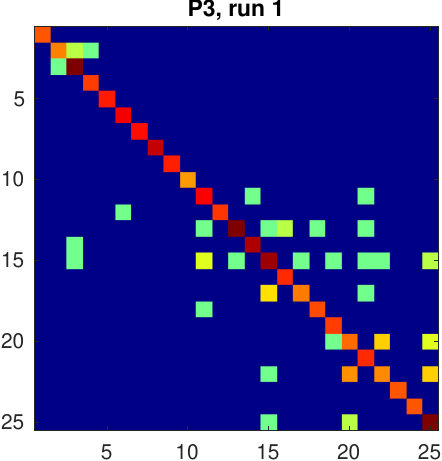}
         &
         \includegraphics[scale=0.375, trim={0.04cm 0 0.05cm 0}, clip]{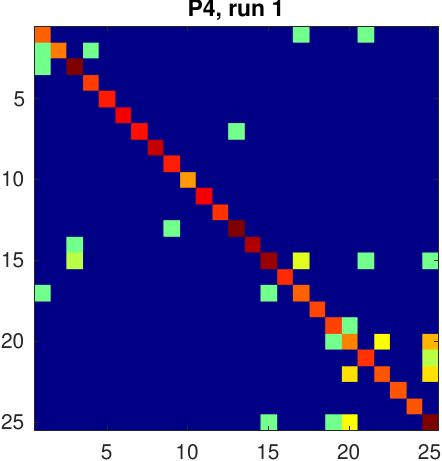}
         &
         \includegraphics[scale=0.375, trim={0.04cm 0 0.05cm 0}, clip]{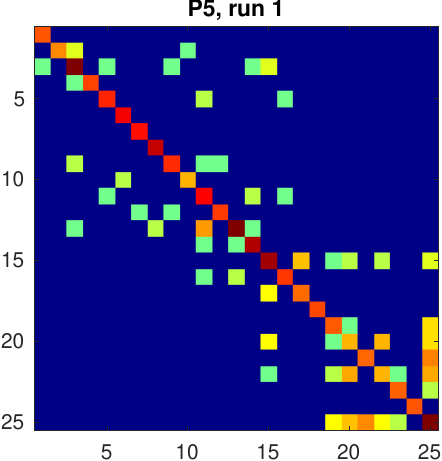}
         \\
         
         %RUN 2
         \includegraphics[scale=0.375, trim={0.04cm 0 0.05cm 0}, clip]{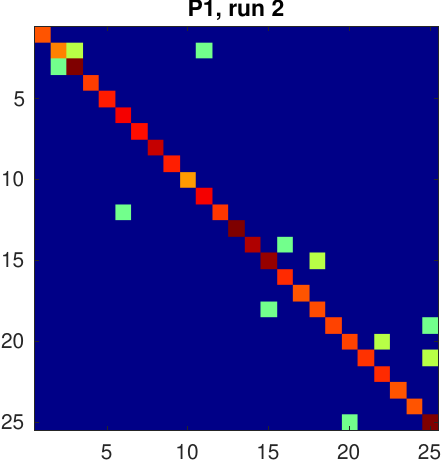}
         &
         \includegraphics[scale=0.375, trim={0.04cm 0 0.05cm 0}, clip]{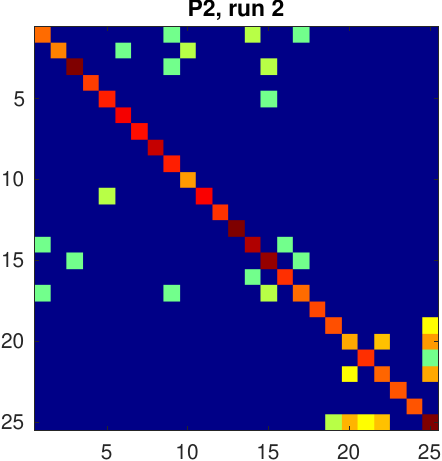}
         &
         \includegraphics[scale=0.375, trim={0.04cm 0 0.05cm 0}, clip]{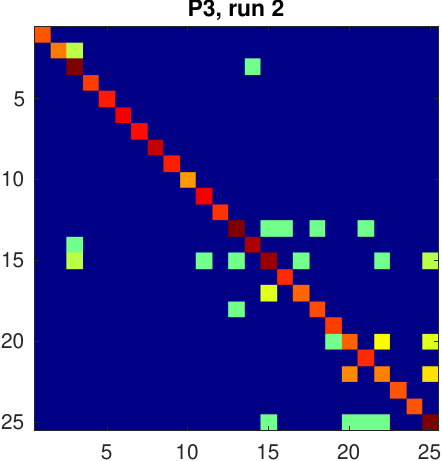}
         &
         \includegraphics[scale=0.375, trim={0.04cm 0 0.05cm 0}, clip]{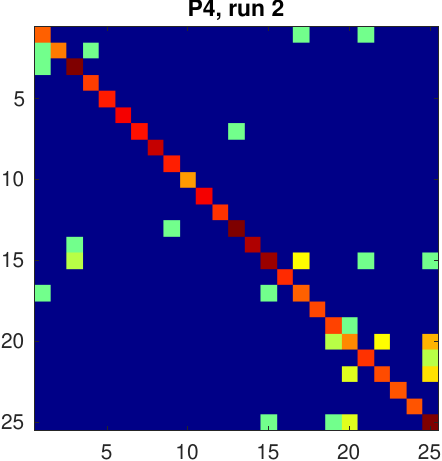}
         &
         \includegraphics[scale=0.375, trim={0.04cm 0 0.05cm 0}, clip]{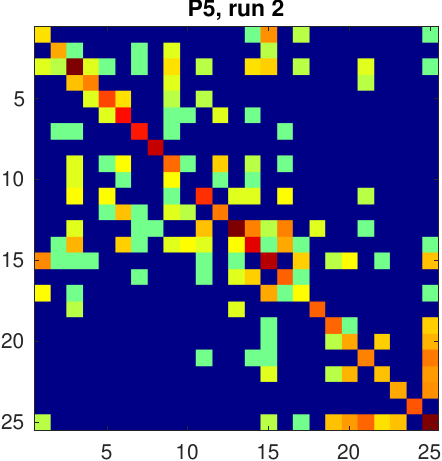}
         \\
         
         %RUN 3
         \includegraphics[scale=0.375, trim={0.04cm 0 0.05cm 0}, clip]{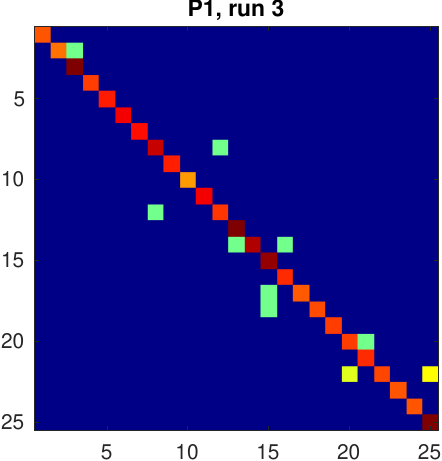}
         &
         \includegraphics[scale=0.375, trim={0.04cm 0 0.05cm 0}, clip]{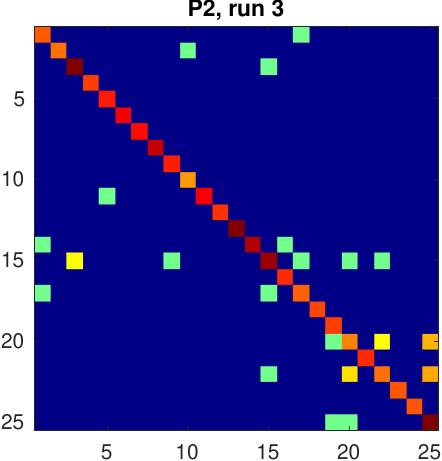}
         &
         \includegraphics[scale=0.375, trim={0.04cm 0 0.05cm 0}, clip]{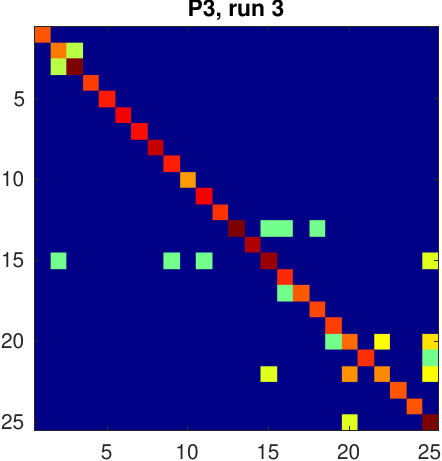}
         &
         \includegraphics[scale=0.375, trim={0.04cm 0 0.05cm 0}, clip]{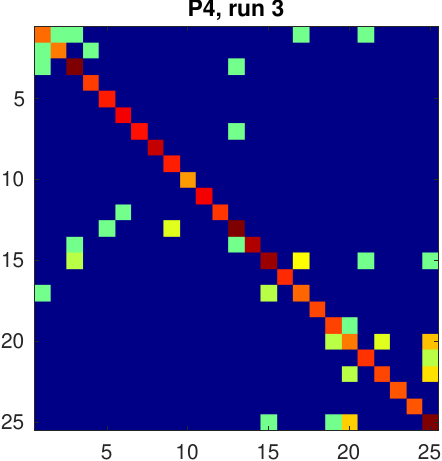}
         &
         \includegraphics[scale=0.375, trim={0.04cm 0 0.05cm 0}, clip]{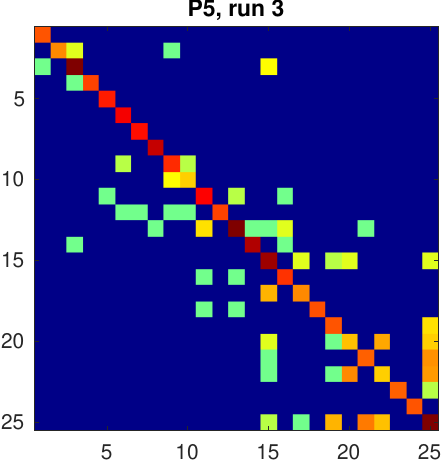}
         \\
    \end{tabular}
    &
    \begin{minipage}[c]{0.04\linewidth}
    \includegraphics[scale=0.6, trim={17cm 0 1.75cm 0}, clip]{colorbar_geom-eps-converted-to.pdf}
    \end{minipage}
    \end{tabular}
    \end{center}
    \caption{Confusion matrices for task A (geometry only), with respect to the BLAST-based community decomposition of level $2$.}
    \label{fig:conf_mat_geom_FASTA_3levels}
\end{figure*}

\begin{figure*}[htb!]
    \begin{center}
    \begin{tabular}{cc}
    \begin{tabular}{ccccc}
         %RUN 1
         \includegraphics[scale=0.375, trim={0.04cm 0 0.05cm 0}, clip]{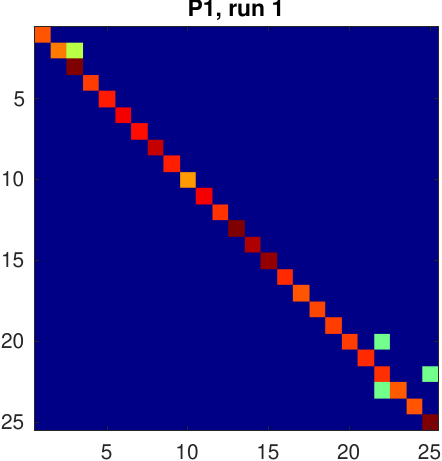}
         &
         \includegraphics[scale=0.375, trim={0.04cm 0 0.05cm 0}, clip]{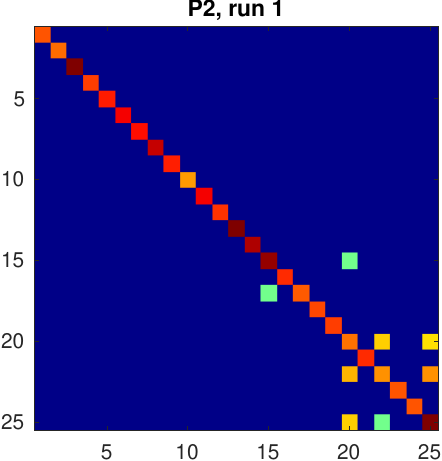}
         &
         \includegraphics[scale=0.375, trim={0.04cm 0 0.05cm 0}, clip]{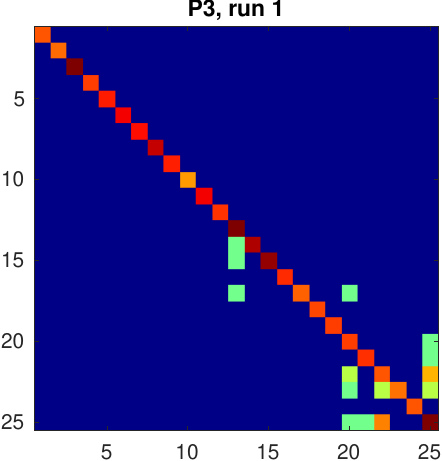}
         &
         \includegraphics[scale=0.375, trim={0.04cm 0 0.05cm 0}, clip]{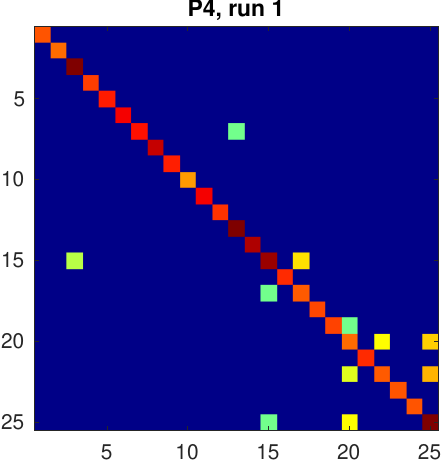}
         &
         \includegraphics[scale=0.375, trim={0.04cm 0 0.05cm 0}, clip]{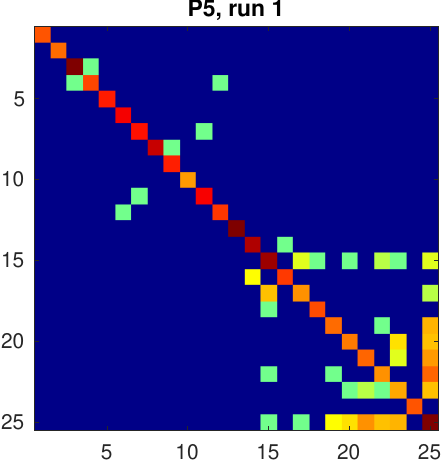}
         \\
         
         %RUN 2
         \includegraphics[scale=0.375, trim={0.04cm 0 0.05cm 0}, clip]{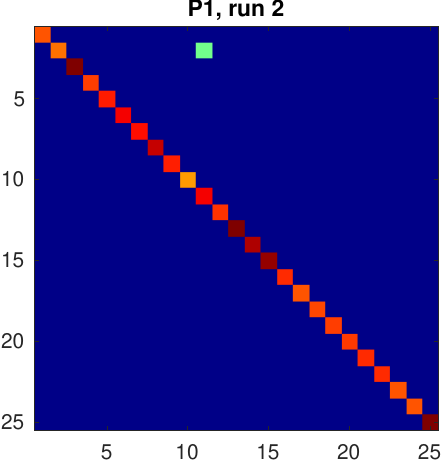}
         &
         \includegraphics[scale=0.375, trim={0.04cm 0 0.05cm 0}, clip]{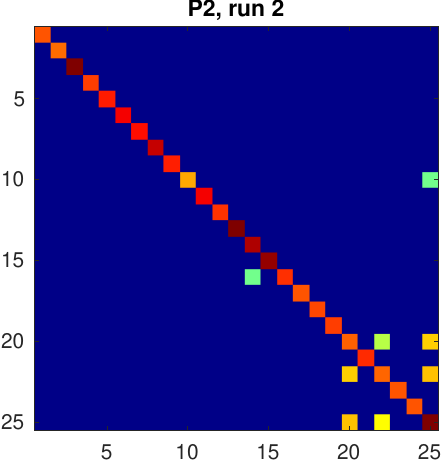}
         &
         \includegraphics[scale=0.375, trim={0.04cm 0 0.05cm 0}, clip]{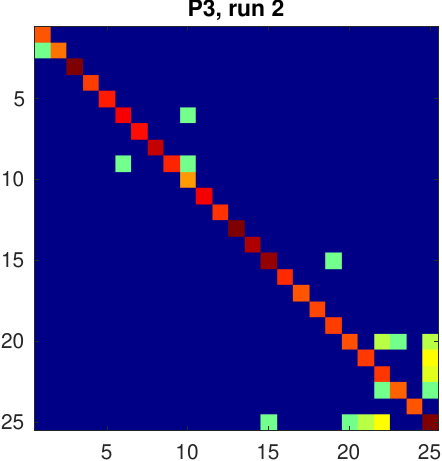}
         &
         \includegraphics[scale=0.375, trim={0.04cm 0 0.05cm 0}, clip]{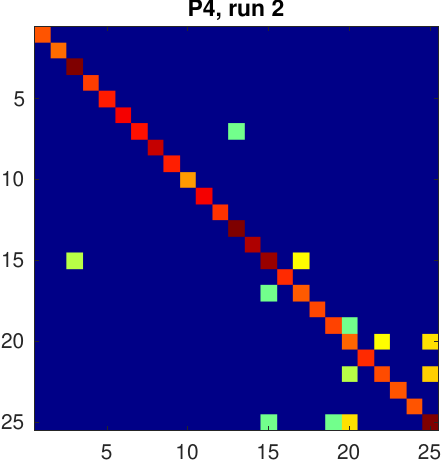}
         &
         
         \\
         
         %RUN 3
         \includegraphics[scale=0.375, trim={0.04cm 0 0.05cm 0}, clip]{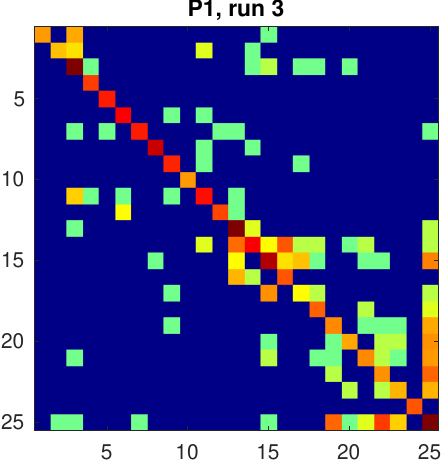}
         &
         \includegraphics[scale=0.375, trim={0.04cm 0 0.05cm 0}, clip]{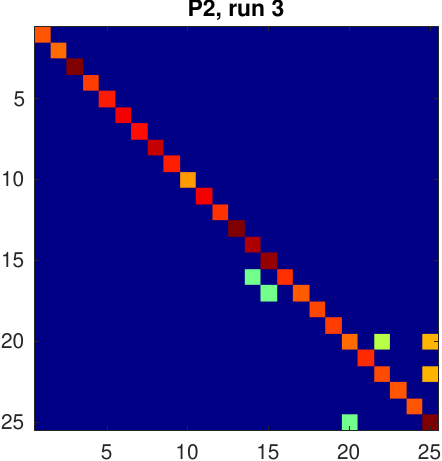}
         &
         \includegraphics[scale=0.375, trim={0.04cm 0 0.05cm 0}, clip]{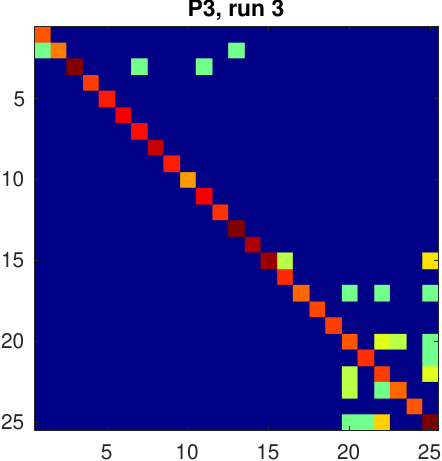}
         &
         \includegraphics[scale=0.375, trim={0.04cm 0 0.05cm 0}, clip]{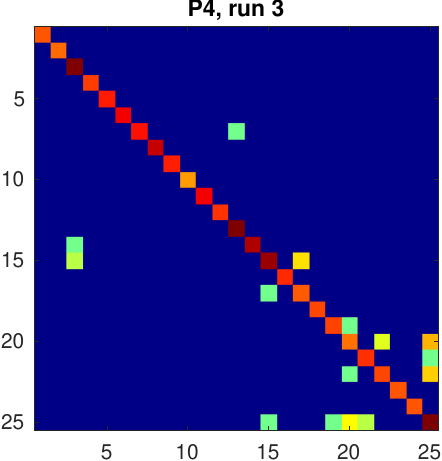}
         &
        
         \\
    \end{tabular}
    &
    \begin{minipage}[c]{0.04\linewidth}
    \includegraphics[scale=0.6, trim={17cm 0 1.75cm 0}, clip]{colorbar_geomchem-eps-converted-to.pdf}
    \end{minipage}
    \end{tabular}
    \end{center}
    \caption{Confusion matrices for task B (geometry and physicochemical properties), with respect to the BLAST-based community decomposition of level $2$.}
    \label{fig:conf_mat_geomchem_FASTA_3levels}
\end{figure*}

%%% CLASSIFICATION MEASURES -- BLAST CLASSIFICATION
\begin{table*}[!h]
    \centering
    \caption{Summary of statistical measures by method and property type (only geometry vs. geometry and physicochemical properties) for the BLAST-based community decomposition of level $2$. Here: TPR = True Positive Rate, TNR = True Negative Rate, PPV = Positive Predictive Value, NPV = Negative Predictive Value, ACC = ACCuracy, F1 = $F_1$ score. For each task and for each measure, the best value for each participant is in bold. The best among them is highlighted in red.  \label{table:summary_class_meas_FASTA_old}}
    
    \begin{adjustbox}{width=1\textwidth}
    \begin{tabular}{l  l c c c c c c  l c c c c c c}
    
    \toprule
    & \multicolumn{7}{c}{Geometry} & \multicolumn{7}{c}{Geometry and physicochemical properties} \\
    \cmidrule(l){2-8}
    \cmidrule(l){9-15}
    & method & TPR & TNR &  PPV & NPV  & ACC & F1 & method & TPR & TNR &  PPV & NPV  & ACC & F1 \\[1.5ex]

    %FIRST PARTICIPANT
    \midrule
    \multirow{3}{*}{P1} &
    \multicolumn{1}{|l}{run 1} & 0.9125 & 0.9922 & 0.9123 & 0.9940 & 0.9878 & 0.9113 & \multicolumn{1}{|l}{run 1} & 0.9968 & 0.9996 & 0.9968 & \textbf{\textcolor{red}{1.0000}} & 0.9996 & 0.9967\\

    & \multicolumn{1}{|l}{run 2} & 
    \textbf{\textcolor{red}{0.9903}} & \textbf{\textcolor{red}{0.9992}} & 0.9904 & 0.9995 & 0.9988 & \textbf{\textcolor{red}{0.9902}} & 
    \multicolumn{1}{|l}{run 2} & \textbf{\textcolor{red}{0.9993}} & \textbf{\textcolor{red}{1.0000}} & \textbf{\textcolor{red}{0.9994}} & \textbf{\textcolor{red}{1.0000}} & \textbf{\textcolor{red}{1.0000}} & \textbf{\textcolor{red}{0.9993}}\\

    & \multicolumn{1}{|l}{run 3} &  \textbf{\textcolor{red}{0.9903}} & 0.9990 & \textbf{\textcolor{red}{0.9905}} & \textbf{\textcolor{red}{0.9997}} & \textbf{\textcolor{red}{0.9989}} & 0.9900 & \multicolumn{1}{|l}{run 3} & 0.7991 & 0.9796 & 0.8081 & 0.9842 & 0.9689 & 0.7953\\
 
    %SECOND PARTICIPANT
    \midrule
    \multirow{3}{*}{P2} &
    \multicolumn{1}{|l}{run 1} & 0.9663 & \textbf{0.9967} & 0.9648 & 0.9978 & 0.9953 & 0.9652 & \multicolumn{1}{|l}{run 1} & 0.9747 & 0.9974 & 0.9733 & 0.9986 & 0.9966 & 0.9733\\

    & \multicolumn{1}{|l}{run 2} & 0.9495 & 0.9955 & 0.9484 & 0.9959 & 0.9927 & 0.9485 & \multicolumn{1}{|l}{run 2} & 0.9780 & \textbf{0.9977} & 0.9782 & 0.9981 & 0.9965 & 0.9778\\
    
    & \multicolumn{1}{|l}{run 3} &  \textbf{0.9701} & \textbf{0.9967} & \textbf{0.9680} & \textbf{0.9987} & \textbf{0.9960} & \textbf{0.9682} & \multicolumn{1}{|l}{run 3} & \textbf{0.9864} & 0.9976 & \textbf{0.9863} & \textbf{0.9996} & \textbf{0.9976} & \textbf{0.9854} \\
 
    %THIRD PARTICIPANT
    \midrule
    \multirow{3}{*}{P3} &
    \multicolumn{1}{|l}{run 1} & 0.9592 & 0.9969 & 0.9579 & 0.9977 & 0.9952 & 0.9575 & \multicolumn{1}{|l}{run 1} & 0.9767 & 0.9977 & 0.9784 & 0.9975 & 0.9960 & 0.9771\\

    & \multicolumn{1}{|l}{run 2} & 0.9682 & 0.9972 & 0.9675 & 0.9980 & 0.9959 & 0.9671 & \multicolumn{1}{|l}{run 2} & \textbf{0.9825} & \textbf{0.9983} & \textbf{0.9831} & \textbf{0.9986} & \textbf{0.9974} & \textbf{0.9825}\\

    & \multicolumn{1}{|l}{run 3} &  \textbf{0.9702} & \textbf{0.9974} & \textbf{0.9697} & \textbf{0.9983} & \textbf{0.9963} & \textbf{0.9687} & \multicolumn{1}{|l}{run 3} & 0.9760 & 0.9980 & 0.9774 & 0.9980 & 0.9966 & 0.9763\\
 
    %FOURTH PARTICIPANT
    \midrule
    \multirow{3}{*}{P4} &
    \multicolumn{1}{|l}{run 1} & 0.9682 & \textbf{0.9969} & 0.9675 & 0.9981 & 0.9957 & 0.9675  & \multicolumn{1}{|l}{run 1} & 0.9767 & 0.9974 & 0.9762 & \textbf{0.9986} & 0.9966 & 0.9760\\

    & \multicolumn{1}{|l}{run 2} & \textbf{0.9689} & \textbf{0.9969} & \textbf{0.9679} & \textbf{0.9982} & \textbf{0.9958} & \textbf{0.9678} & \multicolumn{1}{|l}{run 2} & \textbf{0.9786} & \textbf{0.9978} & \textbf{0.9784} & 0.9984 & \textbf{0.9968} & \textbf{0.9783}\\

    & \multicolumn{1}{|l}{run 3} &  0.9631 & 0.9966 & 0.9627 & 0.9973 & 0.9948 & 0.9626 & \multicolumn{1}{|l}{run 3} & 0.9754 & 0.9971 & 0.9750 & 0.9981 & 0.9960 & 0.9748\\

    %FIFTH PARTICIPANT
    \midrule
    \multirow{3}{*}{P5} &
    \multicolumn{1}{|l}{run 1} & \textbf{0.8996} & \textbf{0.9915} & \textbf{0.8996} & 0.9916 & \textbf{0.9855} & \textbf{0.8986} & \multicolumn{1}{|l}{run 1} & \textbf{0.9132} & \textbf{0.9906} & \textbf{0.9125} & \textbf{0.9928} & \textbf{0.9861} & \textbf{0.9120}\\

    & \multicolumn{1}{|l}{run 2} & 0.7155 & 0.9792 & 0.7250 & 0.9778 & 0.9623 & 0.7175 & \multicolumn{1}{|l}{}\\

    & \multicolumn{1}{|l}{run 3} &  0.8983 & 0.9913 & 0.8965 & \textbf{0.9918} & \textbf{0.9855} & 0.8963 & \multicolumn{1}{|l}{}\\

    \bottomrule

    \end{tabular}
    \end{adjustbox}
\end{table*}

%  NDCG --- BLAST CLASSIFICATION
\begin{figure*}[t!]
    \begin{center}
    \begin{tabular}{|cc|cc|}
    \hline
        \rowcolor{blue!12}\multicolumn{2}{|c|}{Geometry} & \multicolumn{2}{c|}{Geometry and physicochemical properties} \\
        \hline
         \includegraphics[scale=0.45, trim={0 0 0 0}, clip]{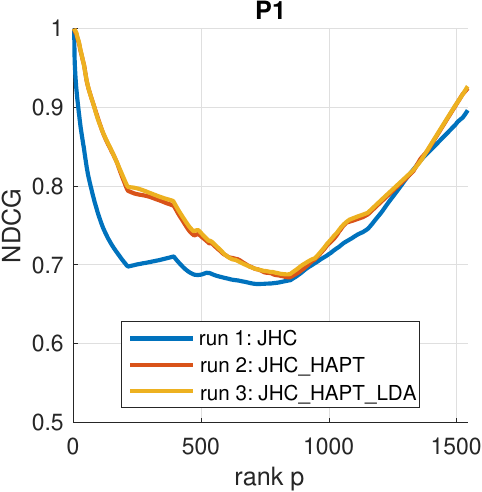}
         &
         \includegraphics[scale=0.45, trim={0 0 0 0}, clip]{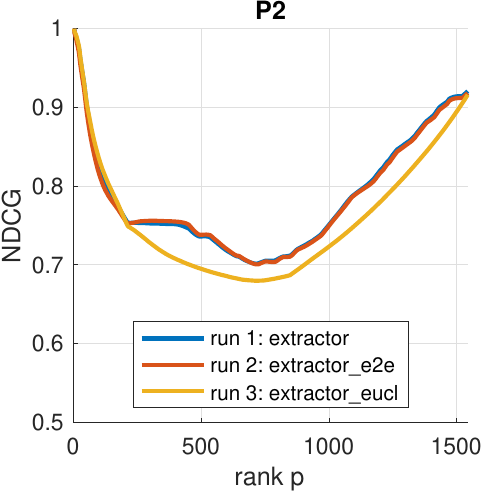}
         &
         \includegraphics[scale=0.45, trim={0 0 0 0}, clip]{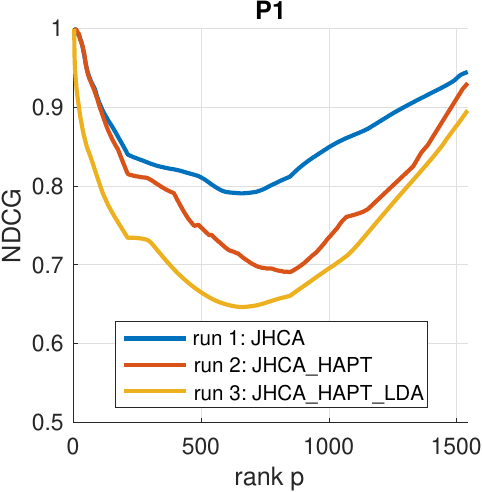}
         &
         \includegraphics[scale=0.45, trim={0 0 0 0}, clip]{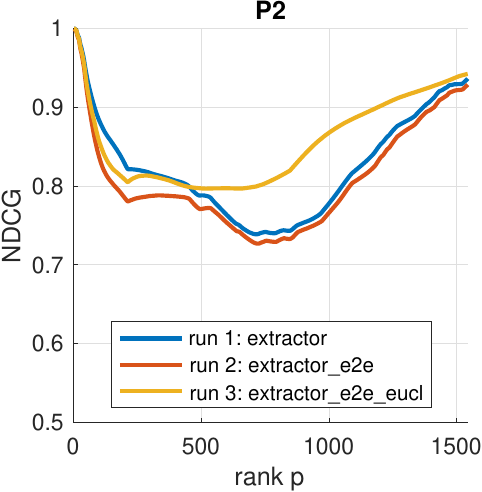}
         \\
         
         \includegraphics[scale=0.45, trim={0 0 0 0}, clip]{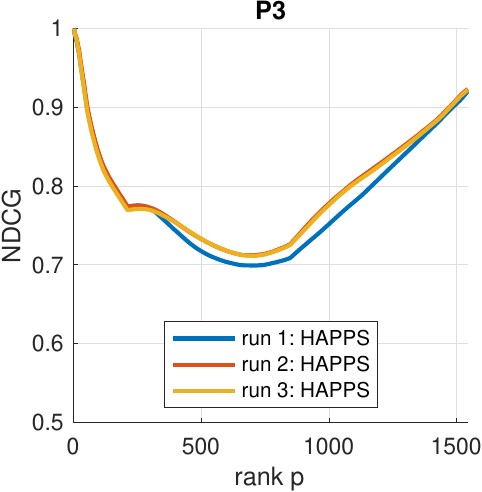}
         &
         \includegraphics[scale=0.45, trim={0 0 0 0}, clip]{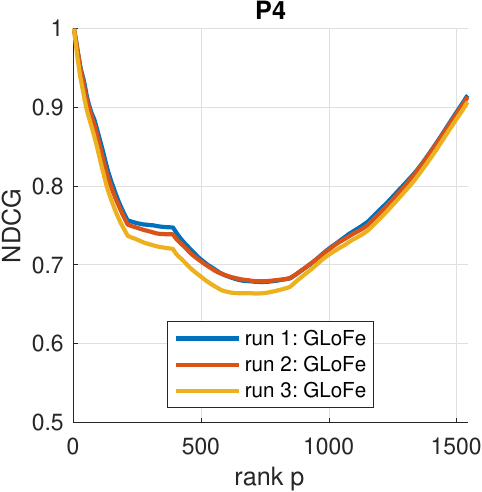}
         &
         \includegraphics[scale=0.45, trim={0 0 0 0}, clip]{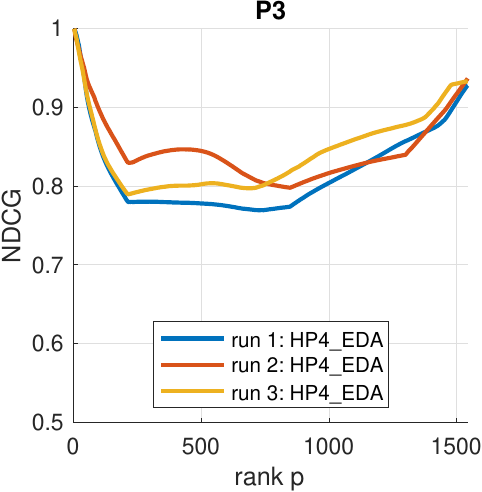}
         &
         \includegraphics[scale=0.45, trim={0 0 0 0}, clip]{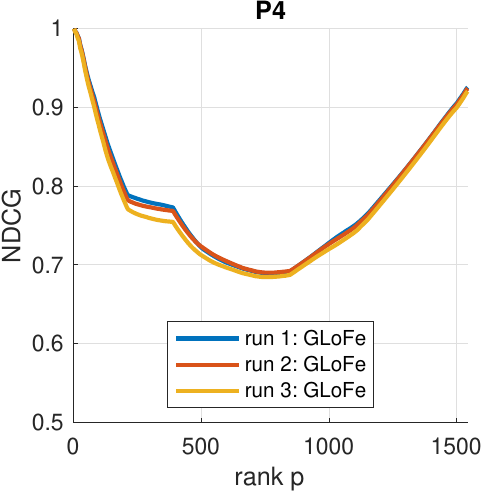}
         \\
         \includegraphics[scale=0.45, trim={0 0 0 0}, clip]{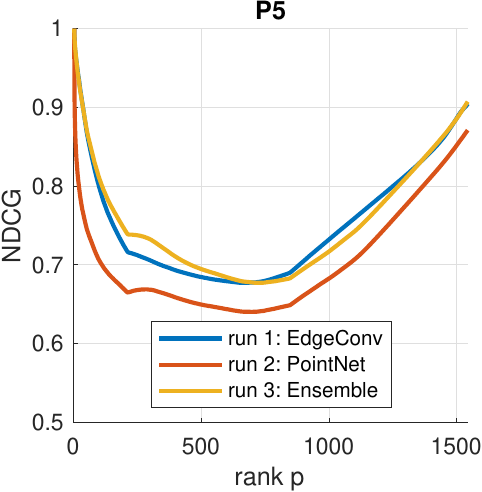}
         &
         \includegraphics[scale=0.45, trim={0 0 0 0}, clip]{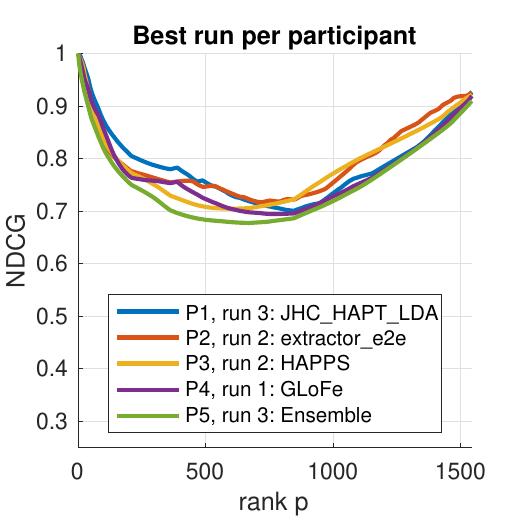}
         &
         \includegraphics[scale=0.45, trim={0 0 0 0}, clip]{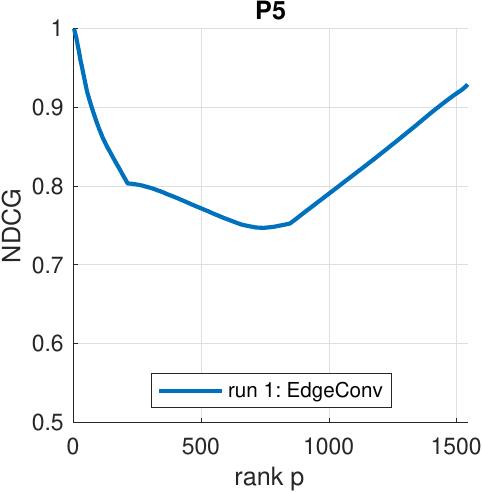}
         &
         \includegraphics[scale=0.45, trim={0 0 0 0}, clip]{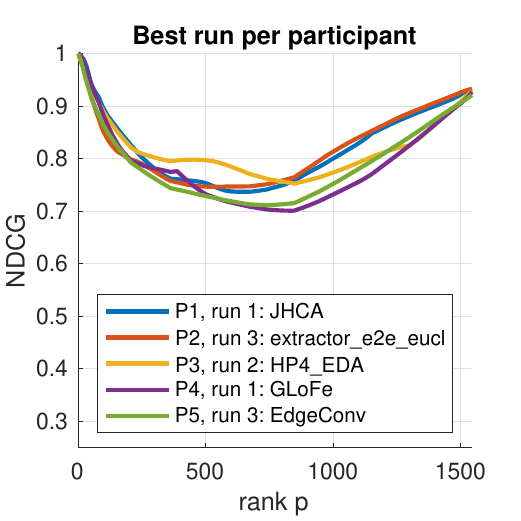}\\
         \hline
    \end{tabular}
    \end{center}  
    \caption{Normalized discounted cumulated gain (NDCG) for Task A (geometry only) and Task B (geometry and physicochemical properties), with respect to the $3$-level BLAST classification.}
    \label{fig:NDCG_geom_geomchem_FASTA_old}
\end{figure*}

% ADR MEASURE --- BLAST CLASSIFICATION
\begin{table*}[!t]
    \centering
    \caption{Summary of average dynamic recalls (ADRs) for the $3$-level BLAST-based classification. For each task, the best ADR for each participant is in bold. The best among them is highlighted in red. \label{table:summary_ADR_FASTA_old}}
    \begin{tabular}{l c c c c c  l c c c c c}
    
    \toprule
    \multicolumn{6}{c}{Geometry} & \multicolumn{6}{c}{Geometry and physicochemical properties} \\
    \cmidrule(l){1-6}
    \cmidrule(l){7-12}
    & P1 & P2 & P3 & P4 & P5 & & P1 & P2 & P3 & P4 & P5 \\[1.5ex]

    \midrule
    \multicolumn{1}{l}{run 1} &   0.677 &  \textbf{0.735} & 0.741  & \textbf{0.728} & 0.686 & \multicolumn{1}{|l}{run 1} & \textbf{0.829} & 0.800 & 0.800 & \textbf{0.756} & \textbf{0.779}  \\

    \multicolumn{1}{l}{run 2} & 0.749 & 0.728 & \textbf{0.751}  & 0.723 & 0.608 & \multicolumn{1}{|l}{run 2} & 0.771 & 0.768 & \textbf{\textcolor{red}{0.849}} & 0.751 & - \\

    \multicolumn{1}{l}{run 3} &  \textbf{\textcolor{red}{0.755}} & 0.715 & 0.746 & 0.708 & \textbf{0.698} & \multicolumn{1}{|l}{run 3} & 0.674 & \textbf{0.806} & 0.810 & 0.742 & - \\

    \bottomrule

    \end{tabular}
\end{table*}

\end{document}